\documentclass[11pt]{iopart}
\pdfoutput=1

\usepackage{lineno}

\usepackage{cite}
\usepackage[english]{babel}
\usepackage[utf8]{inputenc}
\usepackage[T1]{fontenc}
\usepackage{subfigure}
\usepackage{multirow}

\usepackage{amssymb}
\usepackage{graphicx}
\usepackage[colorlinks=true, allcolors=blue]{hyperref}

\begin{document}

\title[Infrared emission from dust and the spectral features of extragalactic gamma-ray sources]{Infrared emission from dust and the spectral features of extragalactic gamma-ray sources}
\author{D. R. de Matos Pimentel$^1$, E. Moura-Santos$^1$}

\address{$^1$ Instituto de F\'isica, Universidade de S\~ao Paulo, Rua do Mat\~ao trav. R 187, 05508-090, S\~ao Paulo - SP, Brazil}
\eads{douglas.pimentel@usp.br, emoura@if.usp.br}

\begin{abstract}
In this work, we investigate the role of emission by dust at infrared wavelengths in the absorption of gamma radiation from distant extragalactic sources, especially blazars. We use an existing EBL model based on direct starlight emission at UV/visible and secondary radiation due to dust (PAHs (polycyclic aromatic hydrocarbons), small and large grains) at IR due to partial absorption of the stellar component. The relative contribution of each grain type to the total EBL energy density was determined from a combined fit to the Markarian 501 ($z\sim 0.034$) SED in flare state, where both the parameters of the intrinsic source spectrum (with or without curvature) and the dust fractions were allowed to vary. By separating the attenuation due to each EBL component, the importance of individual grain types to the opacity of the extragalactic medium for the TeV emission of a blazar like Markarian 501 could be better understood. Using a nested log-likelihood ratio test, we compared null hypotheses represented by effective 1- and 2-grain models against a 3-grain alternative scenario. When the temperatures of the grains are fixed a priori, the 1-grain scenario with only PAHs can be excluded at more than 5$\sigma$ ($p=2.9\times 10^{-8}$), irrespective of the curvature in the intrinsic spectrum. The effective 3-grain EBL model with the tuned fractions was finally used to fit the SEDs of a sample of extragalactic gamma-ray sources (dominated by blazars). Such a sample is still dominated by starlight attenuation, therefore, no statistically significant improvement in the quality of fits was observed when the tuned fractions are used to account for the EBL attenuation and the intrinsic spectrum parameters are allowed to vary during the fit. The potential of this kind of analysis when the next generation of IACTs, represented by the Cherenkov Telescope Array (CTA), starts observations is enormous. The newly discovered AGNs at a broad range of redshifts should break many of the degeneracies currently observed.
\end{abstract}


\maketitle

\section{Introduction}
\label{intro}

The spectral energy distribution (SED) of gamma-ray sources is a valuable piece of information in order to understand the details of their different non-thermal emission processes. Given our current understanding of the quantum nature of matter and radiation, we do expect that part of the high energy photons emitted by extragalactic sources should be absorbed due to the interaction with low energy radiation fields, such as those contributing to the Cosmic Microwave Background (CMB) and the Extragalactic Background Light (EBL). In the standard cosmological model, being a relic of the Big Bang, the former was created with a blackbody spectrum. Precise measurements of the CMB temperature across the sky have allowed us to build a consistent picture of the energy content of the universe, including evidences of the presence of dark matter and dark energy (if one assumes that General Relativity holds true also at cosmological scales). Far more complicated, however, are the spectral features of the latter. The two main contributions to the EBL radiation field are direct star light (therefore, expected to peak at UV/visible wavelengths in comoving coordinates) and dust re-emission when grains are heated by part of the stellar emission, reaching, in turn, maximal spectral intensity at IR wavelengths. Accordingly, this radiation field is believed to have started being emitted at the end of the Dark Ages, when the first gravitationally bounded and nuclear fusion powered objects are formed and have since evolved tightly bound to the star formation rate and the cosmological expansion. It is clear, therefore, that understanding such a radiation is essential to have a full picture of the universe evolution in a regime which is very different from the linear perturbations employed in the CMB case. In addition to that, low energy radiation fields dictate the opacity level of the extragalactic medium to high energy radiation. At TeV energies, $e^+/e^-$ pair production is expected to reduce the mean free path of gamma-rays from extragalactic sources down to a few hundreds of Mpc \cite{Gould:1967zza,Gould:1967zzb}, and even though current estimates of the EBL photon number density, especially at mid-IR, are uncertain, they usually point to bolometric intensities between 50 to 70 nW m$^{-2}$ sr$^{-1}$ (i.e., about 5\% of the CMB intensity) and non-negligible attenuation effects. However, direct measurements of the EBL are hard to perform. They do require instruments with absolute calibration, so the sky brightness can be measured against a well established reference. Moreover, the careful subtraction of foregrounds like dust particle emission and other galactic components is required, as well as corrections of atmospheric effects like the zodiacal light \cite{Hauser:2001xs}. Constraints on the EBL intensity are also obtained from the integrated galaxy-counts, which uses deep field data from space- and ground-based facilities. This method has shown good agreement when compared with direct measurements of the Cosmic Infrared Background (CIB).

The operation of arrays of air imaging Cherenkov telescopes (IACTs), with their large effective collection areas (especially at stereoscopic configurations) and enhanced sensitivity across a broad energy range, has open up the possibility of disentangling intrinsic spectral features of powerful extragalactic sources from the EBL attenuation effects in the GeV-TeV energy range, by means of precise measurements of the SED of gamma-ray sources at different redshifts. The Cherenkov Telescope Array (CTA) \cite{Acharya:2017ttl}, in turn, represents the new generation of IACTs for gamma-ray astronomy and it is expected to bring both qualitative and quantitative changes to this scenario with its factor 10 improvement in sensitivity, fine angular and energy resolutions, allowing for more precise spectrum measurements and the discovery a whole new population of TeV blazars at
high redshifts.

Hence, many recent works are based on the extraction of information on the EBL by combining the measured attenuated spectra provided by IACTs or on-orbit satellites with some data-driven modeling of the EBL spectrum. A few different approaches can be found in the literature, to know, procedures where i) direct galaxy observations in the form of their luminosity functions are employed \cite{Primack:1998wn,Somerville:2011kq,Gilmore:2011ks}; ii) information on the cosmic star formation rate is used \cite{Malkan:1997yd,Stecker:2005qs,Franceschini:2008tp}. We focus on the last procedure and make use of the simplifying yet powerful assumption that the EBL spectral energy density can be modeled as the sum of four contributions (one stellar and three from dust), all having a Planck spectrum \cite{Finke:2009xi}. Here we focus on the role of dust in the attenuation of the spectra of AGNs. In particular, the mid-IR emission (at comoving coordinates) is usually believed to contribute significantly to the opacity of the extragalactic medium to gamma-rays with energies up to around 10 TeV. In this paper, we do quantify more precisely this point by studying the case of Markarian 501 (Mkn\,501).

The outline of paper is the following: in section \ref{model}, we briefly review the main assumptions behind the construction of an EBL model with blackbody spectra for both the stellar and dust contributions; the role of dust in the attenuation of the flux of an extragalactic source like Mkn\,501 is studied in section \ref{pah_and_mkn}; in this section, the SED of Mkn\,501 during the flare of 1997 \cite{aharonian1999time} is used in order to perform a combined fit of the intrinsic spectrum, as well as the relative contributions of different dust grains to the EBL; A nested likelihood ratio test is also performed to assess the importance of different grain sizes; in section \ref{global}, the dust fractions coming out of this combined fit are then used to feed an effective EBL model which, in turn, is employed to fit the intrinsic spectra of an extended sample of extragalactic TeV-emitters. Global properties of the fits are then analysed. Conclusions are finally presented in section \ref{conclusions}.

\section{Modeling the EBL spectrum}
\label{model}

Having stars and cosmic dust in the interstellar medium (ISM) as its main contributing sources, the EBL spectrum shows nowadays a maximum intensity in the wavelength range 0.1 $\mu$m - 1000 $\mu$m. In \cite{Finke:2009xi}, the authors have indeed modeled the EBL spectral density as the sum of these two main components\footnote{Since we will be interested in gamma-rays with energies $E<20$ TeV, the CMB attenuation can be ignored. For a source at $z=0.034$, its contribution to the optical depth is less than $6.1\times 10^{-18}$.}. Each star is assumed to have a metallicity similar to the sun and to emit as a blackbody, whose effective temperature depends on its luminosity and radius, the evolution of which is followed through the H-R diagram using an approach described in \cite{eggleton1989distribution}, where suitable parameterizations are found in order to describe the main stages of stellar evolution: the main sequence, the Hertzsprung gap and the giant branch. The applicability of the blackbody assumption is tested by comparing the spectra of simple stellar populations (SSPs) of stars of different ages with high resolution ones from detailed stellar structure codes \cite{Bruzual:2003tq}. In the wavelength range of interest for EBL studies ($\gtrsim 0.1\mu$m), there is a good agreement between these two types of SSPs.

Therefore, the stellar emissivity (i.e., the luminosity per unit comoving volume in W Mpc$^{-3}$)  is given by
\begin{equation}
\fl  \epsilon j_c^{\scriptsize{\textrm{star}}}(\epsilon;z)=m_e c^2 \epsilon^2 f_{esc}(\epsilon) \int\limits_{m_{\scriptsize{\textrm{min}}}}^{m_{\scriptsize{\textrm{max}}}}dm\, \xi(m)\int\limits_z^{z_{\scriptsize{\textrm{max}}}}dz_1\left|\frac{dt}{dz_1}\right|\psi_c(z_1) \dot{N}(\epsilon;m,t),
  \label{eq:j_star}
\end{equation}
where $\epsilon=h\nu/m_e c^2$ is the dimensionless energy of a photon of frequency $\nu$, $f_{esc}(\epsilon)$ is the function which describes the fraction of radiation that is able to escape from the environment of the star (i.e., without being absorbed by gas or dust), $\xi(m)$ is the normalized initial mass function (IMF) \footnote{The authors assume the initial mass function (IMF) obtained by Baldry and Glazebrook \cite{baldry03} which, for masses $m > 0.5M_\odot$, is in good agreement with the one due to Salpeter \cite{Salpeter:1955it}.} , $\psi_c(z)$ is the comoving star formation rate in units of M$_{\odot}$ yr$^{-1}$ Mpc$^{-3}$, $\dot{N}(\epsilon;m,t)$ is the number of emitted photons per unit energy per unit time for a single star and $t(z,z_1)$ its age at redshift $z$ after birth at redshift $z_1$. The cosmological evolution is governed by $dt/dz$, which for a $\Lambda$CDM model in a flat universe (zero spatial curvature) is given by
\begin{equation}
  \left|\frac{dt}{dz}\right|=\frac{1}{H_0(1+z)\sqrt{\Omega_r(1+z)^4 + \Omega_m(1+z)^3 + \Omega_{\Lambda}}}
\end{equation}
where $H_0=70$ km$^{-1}$ s$^{-1}$ Mpc$^{-1}$ is the Hubble constant and $\Omega_r$, $\Omega_m$ and $\Omega_{\Lambda}$ are the dimensionless density parameters of radiation, non-relativistic matter and dark energy, respectively. For the redshifts of interest in EBL studies, one can safely ignore the contribution from radiation ($\Omega_r\simeq 0$). We should take also $\Omega_m=0.3$ and $\Omega_\Lambda=0.7$.

The stellar emission is dominant between UV and near-IR, but part of these photons is absorbed by gas and dust. The function $f_{esc}(\epsilon)$ previously introduced allows us to get the direct starlight absorbed fraction. In \cite{Finke:2009xi}, it is assumed that all radiation with energy above 13.6 eV is completely absorbed by interstellar and intergalactic HI gas. Below this cutoff, a fraction of the photons is absorbed by ISM dust, whose grains are heated up and finally reemited at longer wavelengths (IR). Similar to the stellar contribution, a Planck spectral distribution is also assumed for dust, but now with three different temperatures associated to different grain types: small hot grains, large warm grains and polycyclic aromatic hydrocarbon (PAH) molecules. The dust emissivity is then given by
\begin{equation}
  \epsilon \, j_c^{\scriptsize{\textrm{dust}}}(\epsilon;z)=\frac{15}{\pi^4}\int d\epsilon' \left[\frac{1}{f_{esc}(\epsilon')} - 1\right] j_c^{\scriptsize{\textrm{star}}} (\epsilon';z)\sum_{n=1}^{3}\frac{f_n}{\Theta_n^4}\frac{\epsilon^4}{\exp(\epsilon/\Theta_n)-1}
  \label{eq:j_dust}
\end{equation}
where the sum runs over the three grain types, $f_n$ is the relative contribution of each dust component, $\Theta_n=k_B T_n/m_e c^2$ is the dimensionless temperature, $k_B$ is the Boltzmann constant and $T_n$ is the temperature of the grain in K. It is worth mentioning that the term $1/f_{esc}(\epsilon)$ does not diverge since $j_c^{star}(\epsilon;z)$ is proportional to $f_{esc}(\epsilon)$ as can be seen in Eq. \ref{eq:j_star}.
\begin{table*}[ht]
\begin{center}
\begin{tabular}{cccc}
\hline
\hline
Dust component & $n$ & $f_n$ & $\Theta_n$ [$10^{-9}$] \\
\hline
PAH & 1 & 0.25 & 76 \\
Small grains & 2 & 0.05 & 12 \\
Large grains & 3 & 0.70 & 7 \\
\hline
\hline
\end{tabular}
\end{center}
\caption{Summary of the dust parameters used in the calculations of this paper.}
\label{table:dust_pars}
\end{table*}

The applicability of the blackbody spectrum assumption for dust is difficult to assess. Unlike the stellar case, the physics of dust absorption and emission is hard to model. From the observational point of view, PAHs have been detected and characterized for decades by astronomers, showing a complex  emission and absorption spectra at mid-IR due to their many vibrational and rotational normal modes \cite{Krugel2003}.
\begin{figure}
\begin{center}
  \includegraphics[width=10cm]{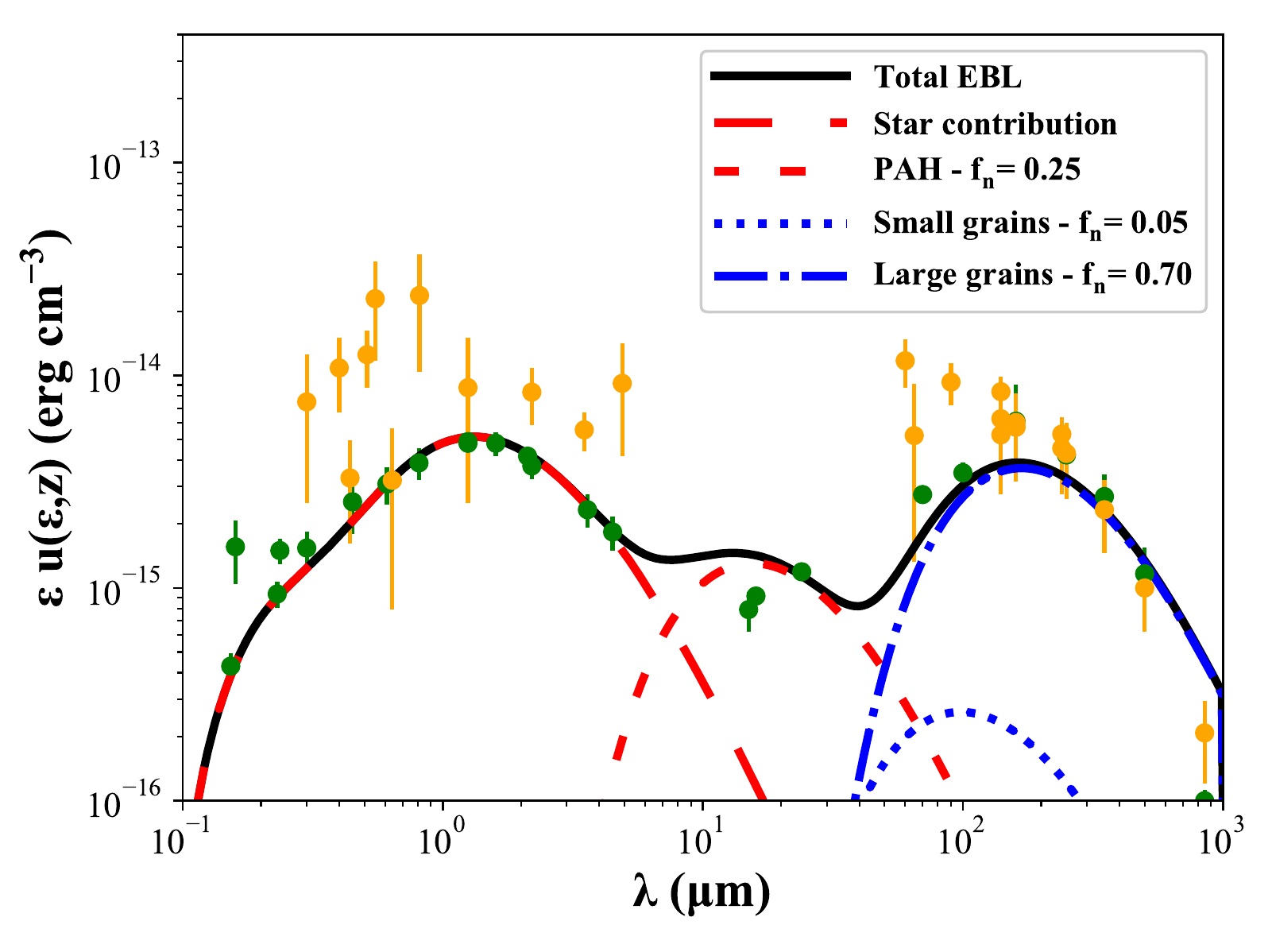}
  \caption{EBL energy density in comoving coordinates as measured by an observer at redshift $z=0$. In addition to the total (black) density, individual contributions are also shown: stellar (long-dashed red), small hot grains (dotted blue), large warm grains (dot-dashed blue) and PAHs (dashed red). Green and yellow points corresponding to lower and upper limits, respectively, and were obtained from \cite{biteau2015extragalactic}.}
  \label{fig:ebl_spectrum}
\end{center}
\end{figure}
With the stellar and dust emissivities from Eq. \ref{eq:j_star} and \ref{eq:j_dust}, it is straightforward to get the EBL energy density $u$ at a certain redshift $z$, for this function should satisfy a Boltzmann equation with a diluting term due to the expansion of the universe and an injection term proportional to the total emissivity (star+dust) \cite{Peebles1971}. In comoving coordinates, one can write
\begin{equation}
  \epsilon \, u_{c}(\epsilon;z) = \int\limits_{z}^{z_{\scriptsize{\textrm{max}}}}d{z_1}\frac{\epsilon'j_{c}(\epsilon';z_1)}{(1+z_1)}\left|\frac{dt}{dz_1}\right|
\end{equation}
where $\epsilon'=(1+z_1)\epsilon$ is the photon energy at redshift $z_1$, $j_c$ is the emissivity of the sources in comoving coordinates.

Five combinations of IMF and SFR were considered in \cite{Finke:2009xi}, the details of which can be found in \cite{Razzaque:2008te}. The combination showing the best agreement with data of UV/optical emissivities is chosen by the authors and was also the one we have selected in this work. In order to get a good agreement with their EBL energy density, some changes in the values of the fractions had to be made, by increasing in 10\% the contribution of large grains with respect to the nominal value of \cite{Finke:2009xi} and a corresponding decrease in the fraction of PAHs (table \ref{table:dust_pars} summarizes the values used in this paper). The final EBL comoving energy density as a function of wavelength can be seen in figure \ref{fig:ebl_spectrum}, where the contribution of each component is also shown separately. One can see the dominance of PAHs at mid-IR ($\sim 10\,\mu$m).

\section{Dust emission and the spectrum of Markarian 501}
\label{pah_and_mkn}

\begin{figure}[h]
  \centering
  \includegraphics[width=10cm]{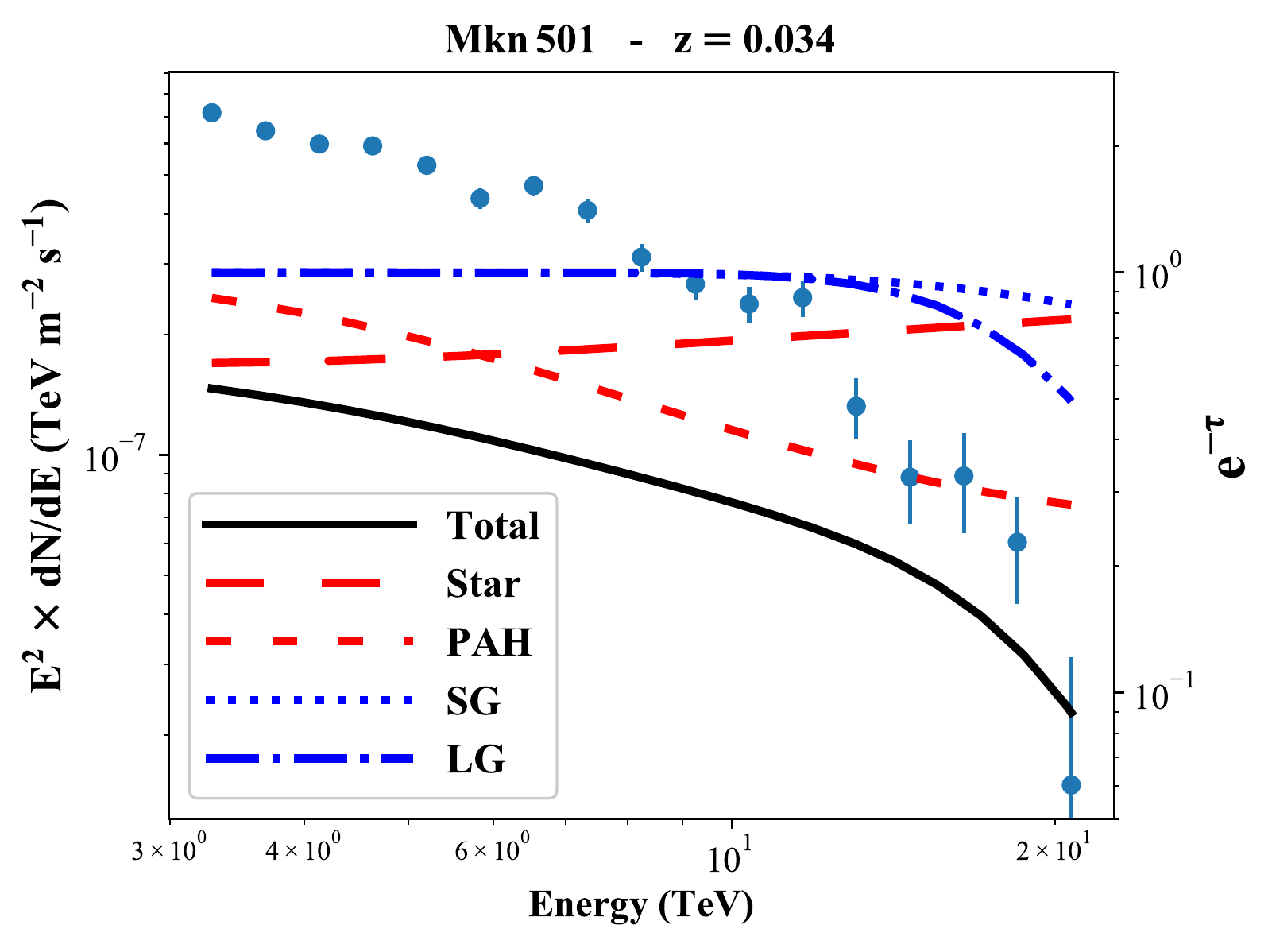}
  \caption{Mkn\,501 SED for the flare of in 1997. Also we plot the attenuation factor according to Finke et al. EBL model for a source located at $z=0.034$ where the contribution from stars and individual dust components are also shown.}
  \label{fig:mkn501}
\end{figure}

Extragalactic gamma-ray sources in the GeV-TeV  energy range present in the current available catalogs have direct star light as the main source of EBL attenuation. However, IACTs with good sensitivity around $\gtrsim 1$ TeV have also been measuring photon fluxes from a few sources in a redshift range where the dust component can play an important role in the attenuation process. We show that an example of this kind of source is Mkn\,501. An exceptional flare of this AGN was detected in 1997 by HEGRA \cite{aharonian1999time} and its SED characterized in \cite{aharonian2001reanalysis}. Figure \ref{fig:mkn501} shows the spectrum of Mkn\,501 superimposed to the attenuation factor from the extragalactic medium, as predicted by the EBL model of Finke et al., for a source at the location of Mkn\,501 ($z\sim 0.034$).

One can see for this particular source that, as the contribution from direct starlight to the opacity of the extragalactic medium decreases slowly and steadily for energies above $\sim 1$ TeV, the dust contribution increases monotonically up to $\sim 20$ TeV. Moreover, for energies $10\lesssim E \lesssim 20$ TeV, the optical depth is dominated, in the Finke et al. model, by the PAH component with the contribution from large and small grains rising fast above 10 TeV.

For the EBL model adopted in this paper, once the IMF and SFR are defined, the stellar contribution do not have any extra free parameters. On the other hand, for the dust component, in addition to an assumption on the escape probability ($f_{esc}$) of starlight, the relative contributions of different grain sizes ($f_{n}$) and their temperatures $\Theta_n$ also need to be determined. In \cite{Finke:2009xi}, in particular, the authors assume redshift-independent $f_{esc}$, $f_n$ and $\Theta_n$, choosing the last two variables so as to fit IR luminosity data at low redshifts ($z=0$ and $z=0.1$) from several observations \cite{blanton:2003galaxy,cole:20012df,kochanek:2001k,budavari:2005ultraviolet,tresse:2007cosmic,sawicki:2006keck,dahlen:2007evolution,smith:2009luminosity,magnelli:20090,takeuchi:2006iso,babbedge:2006luminosity,huang:2007local,le:2005infrared,flores:199915,cirasuolo:2007evolution,faucher:2008flat,reddy:2008multiwavelength,caputi:2007infrared,perez:2005spitzer}. The escape fraction is taken from \cite{Driver:2008sd} as a piecewise power-law at several wavelength ranges. Here, we should keep the same escape fraction function, as well as the temperatures of the three different grains. The relative grain  contributions, however, will be determined using a fit to the SED of Mkn\,501 in flare state in order to study the potential of this kind of observation to constrain both source intrinsic spectrum and EBL parameters.
\begin{figure*}[ht]
\begin{center}
\subfigure{
  \includegraphics[width=.49\textwidth]{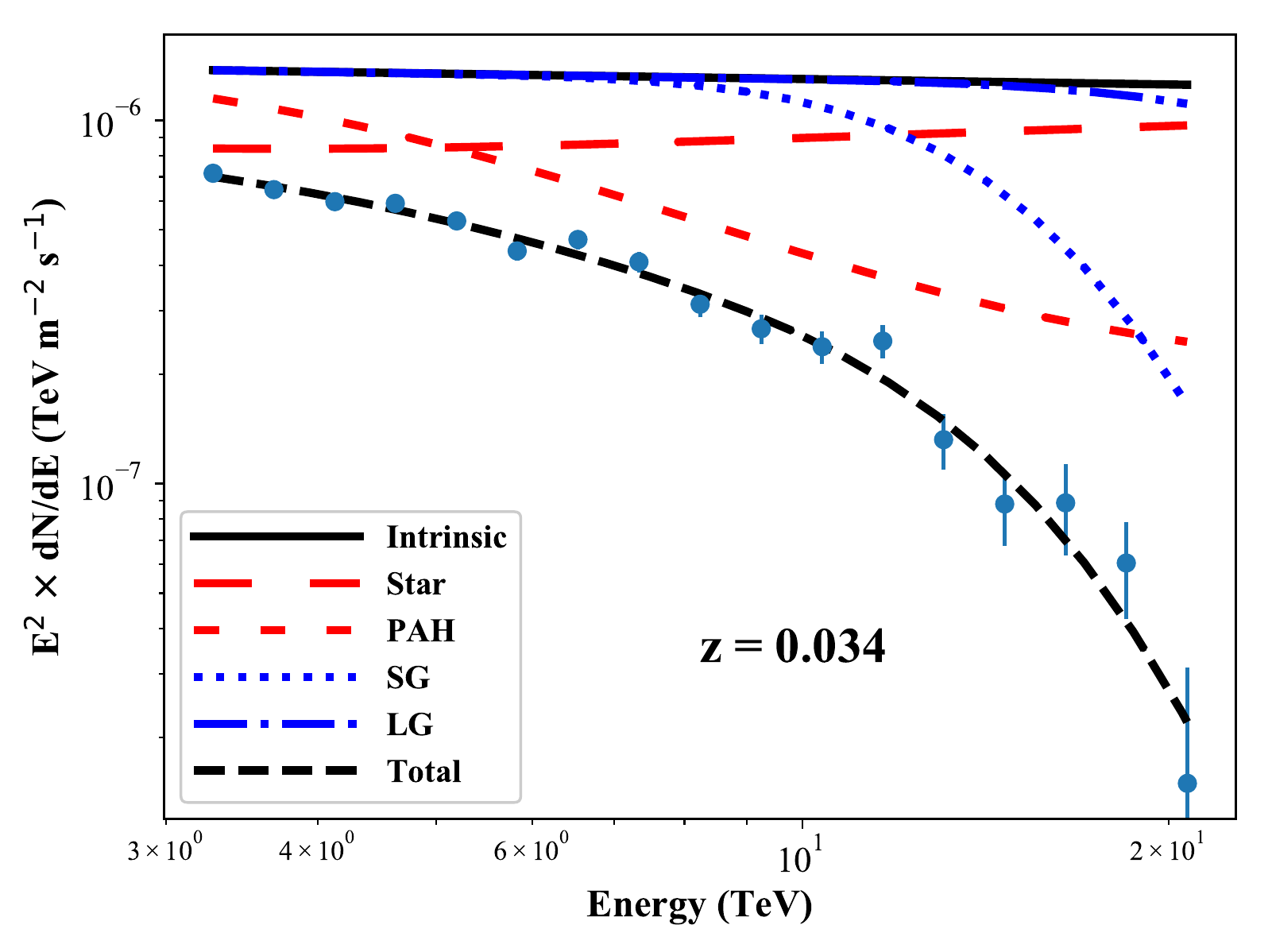}
}%
\subfigure{
  \includegraphics[width=.49\textwidth]{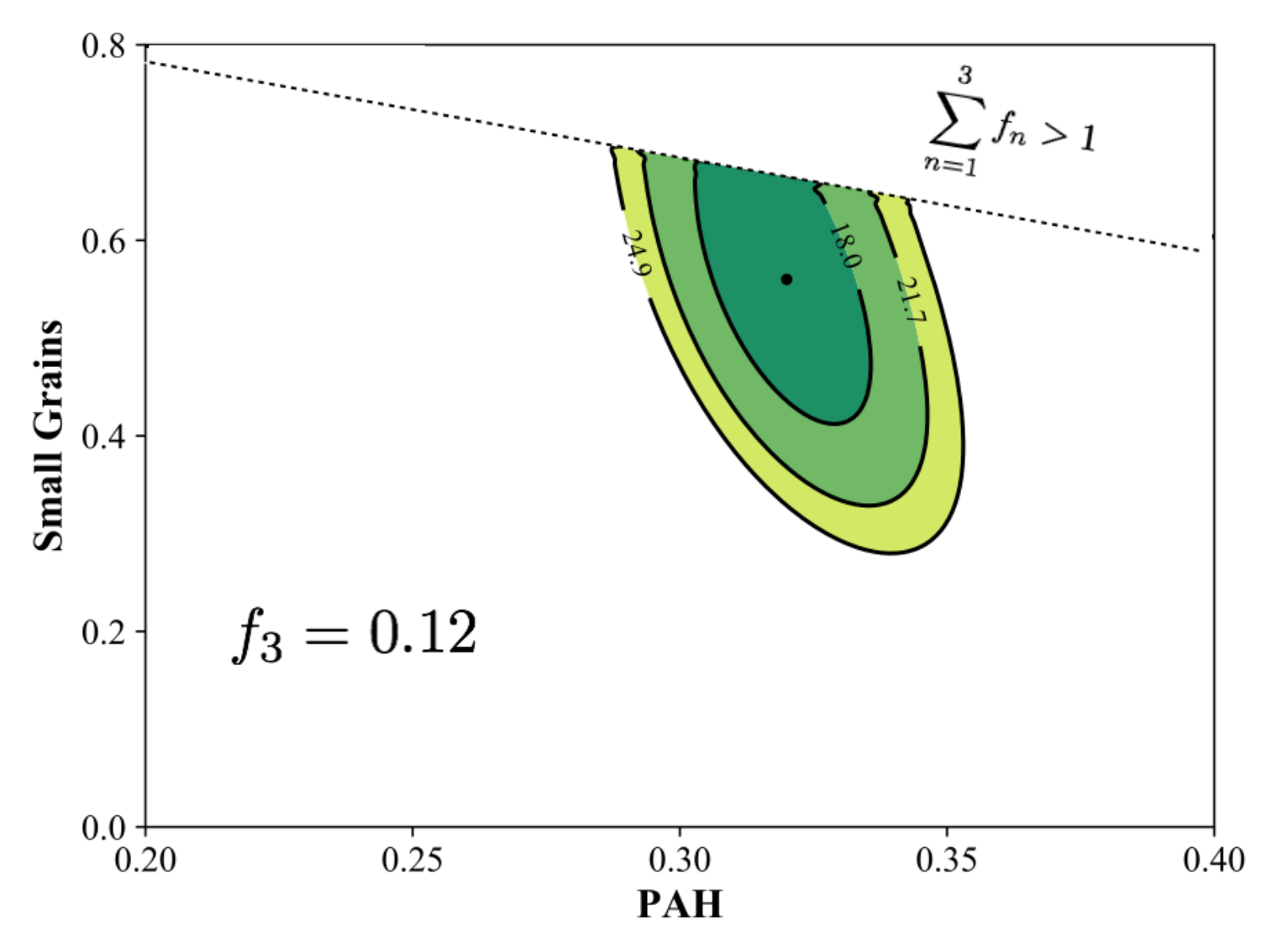}
} \\ 
\subfigure{
  \includegraphics[width=.49\textwidth]{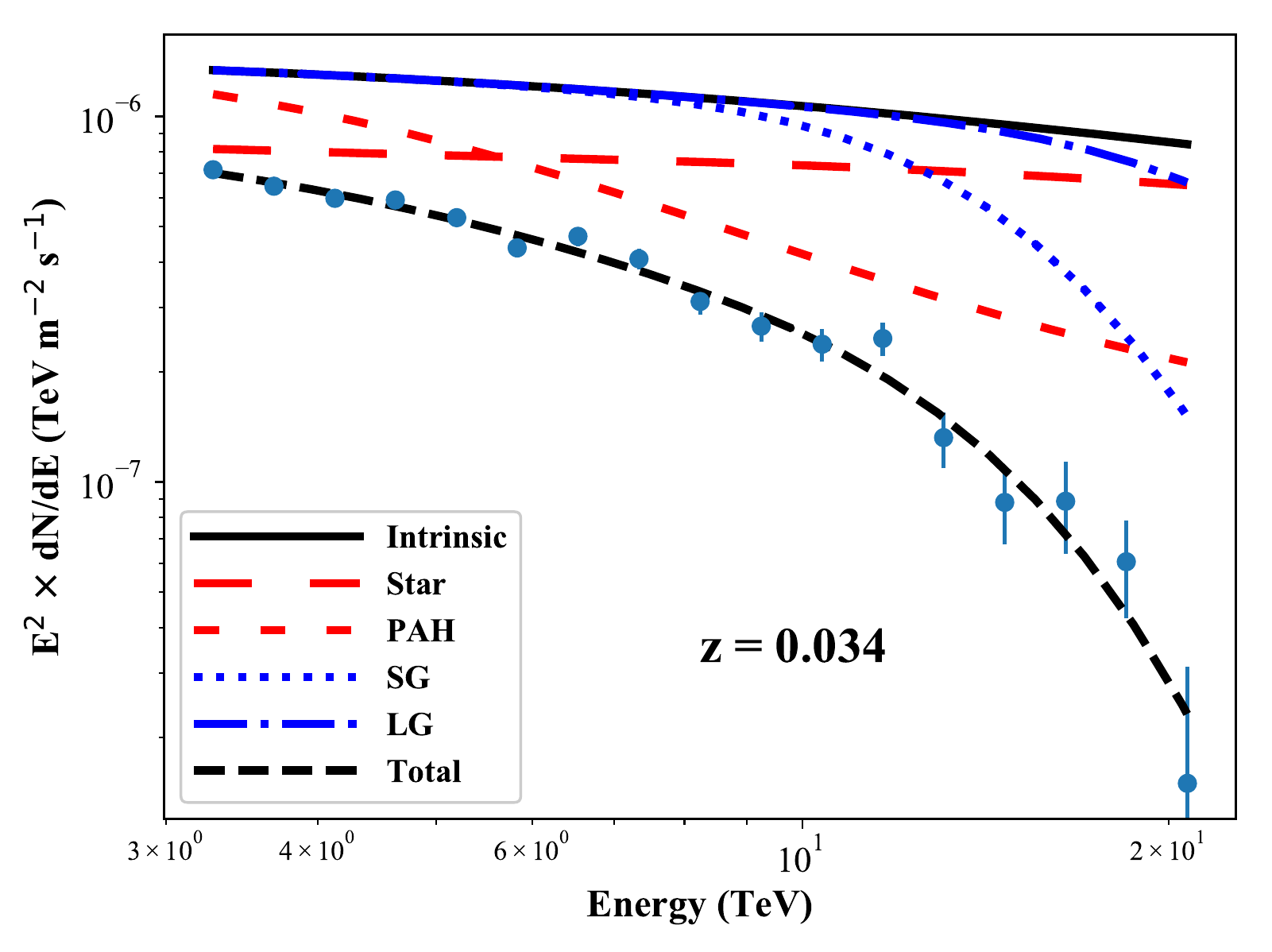}
}%
\subfigure{
  \includegraphics[width=.49\textwidth]{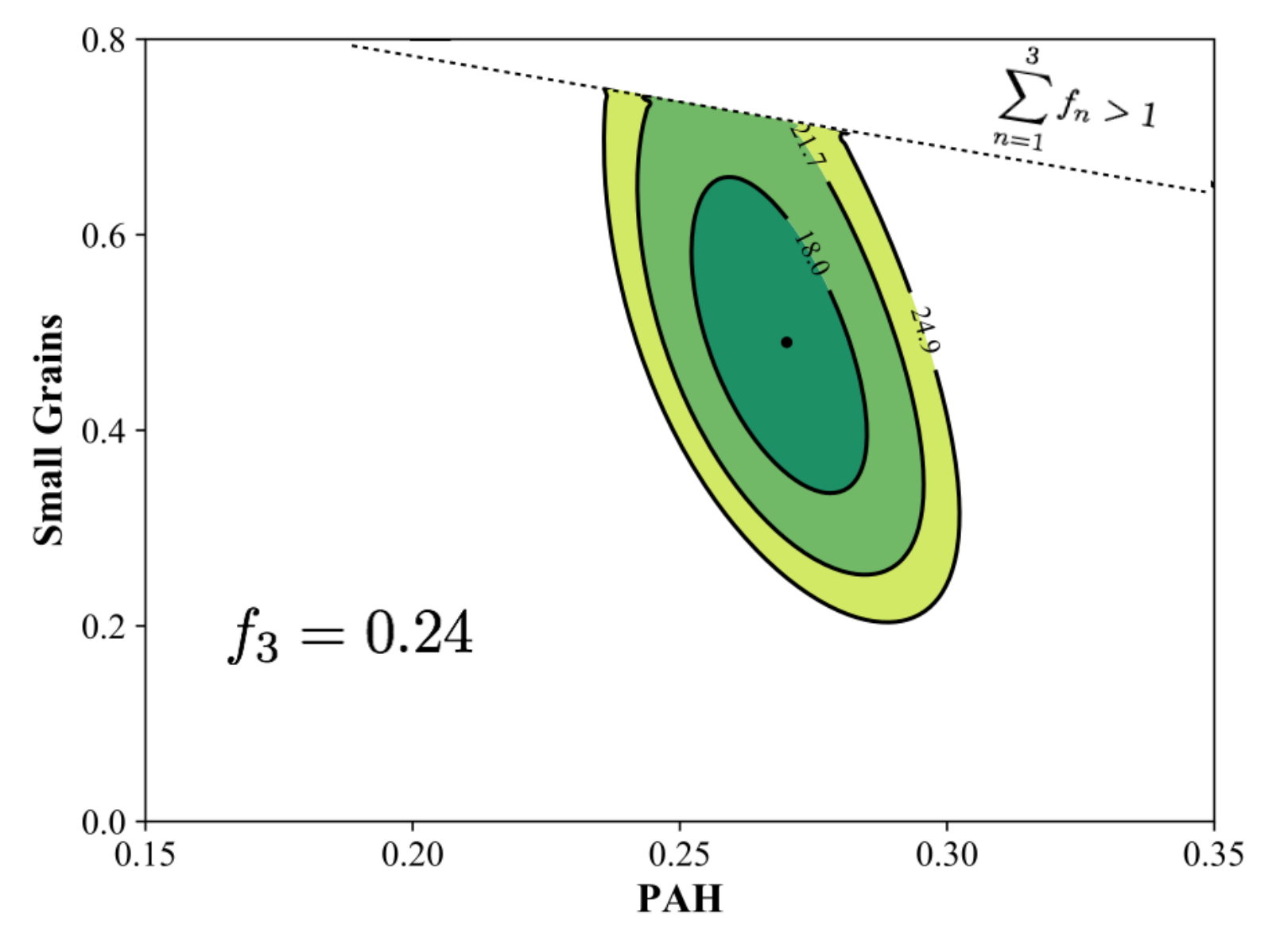}
} \\ 
\subfigure{
  \includegraphics[width=.49\textwidth]{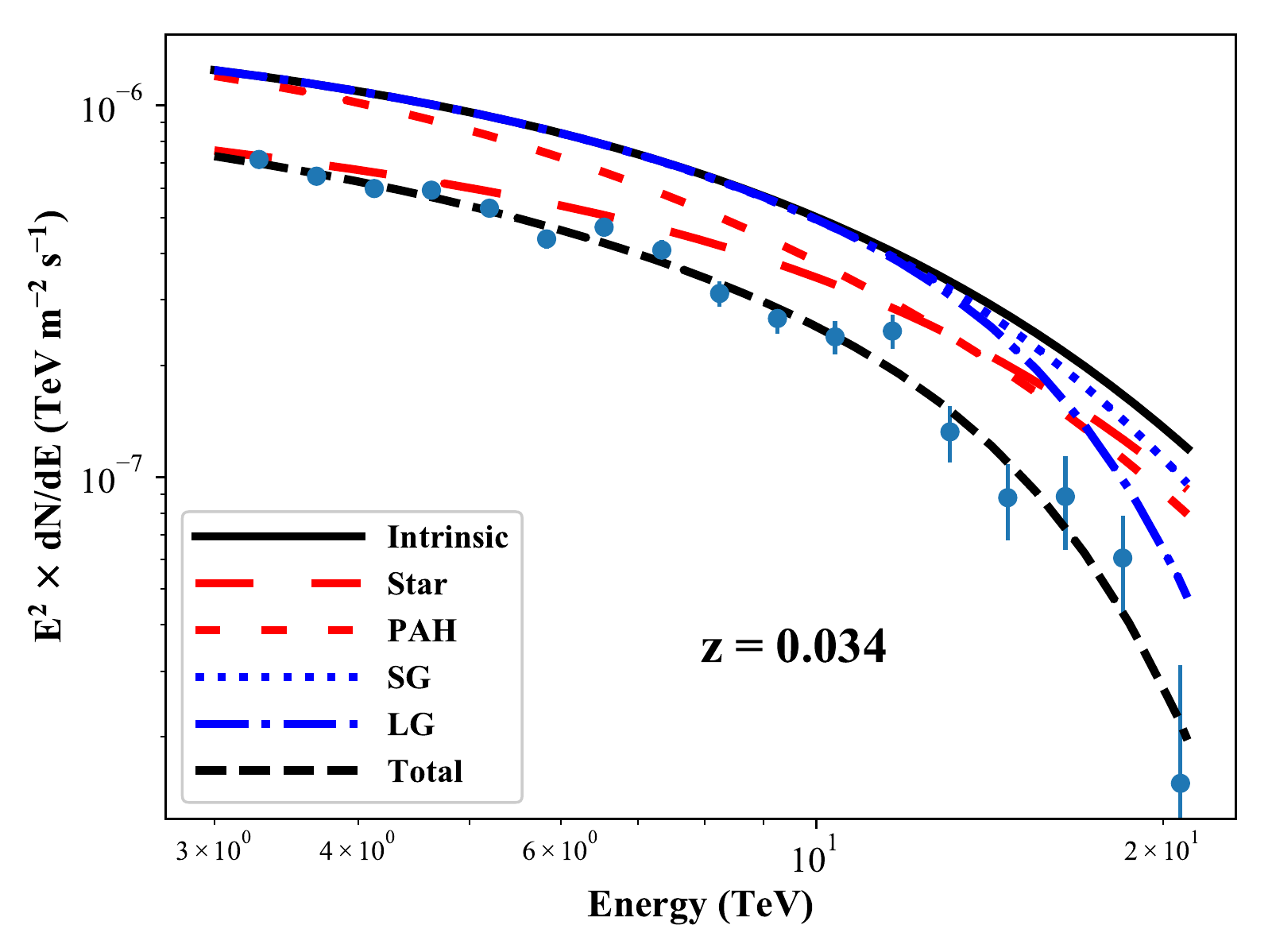}
}%
\subfigure{
  \includegraphics[width=.49\textwidth]{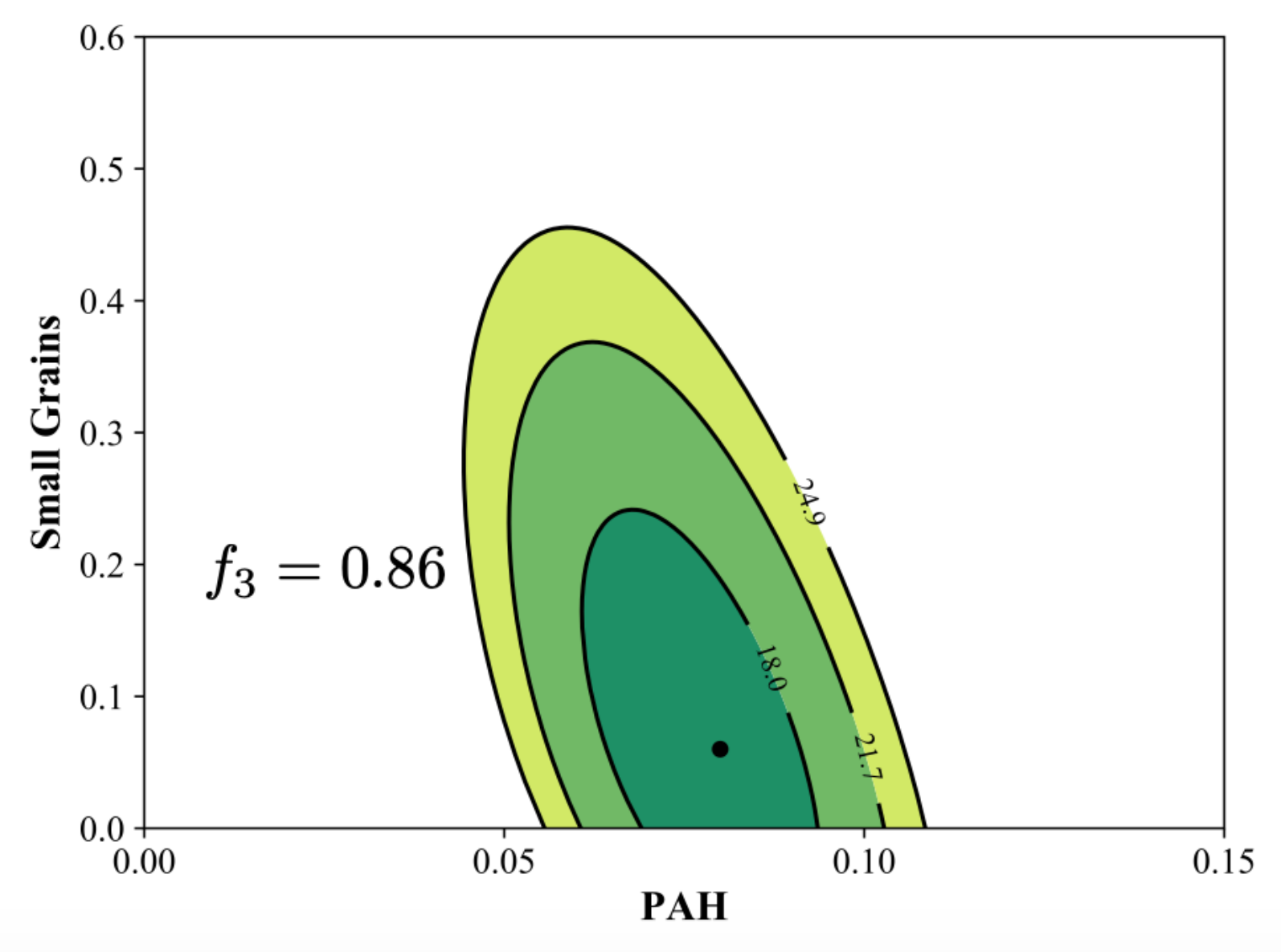}
}%
\end{center}
\caption{Combined fits of the intrinsic spectrum and relative contributions of dust grains using Mkn\,501 measured SED. Three different parameterizations are used for the intrinsic spectrum: power-law (top), log-parabola (middle) and power-law with an exponential cutoff (bottom). Left: measured SED superimposed to the convolution of the best-fit intrinsic spectrum with the attenuation factors of each EBL component, as well as the total attenuation. Right:  $\chi^2$ contours in the 2D space of dust parameters $f_{\tiny{\textrm{PAH}}} \times f_{\tiny{\textrm{SG}}}$ at confidence levels of 68\%, 95\% and 99\%.}
\label{fig:bestfit_mkn501_pl_lp_plexp}
\end{figure*}

The solution to the radiative transfer equation for the opaque extragalactic medium leads to the usual relationship between emitted (intrinsic) and detected (observed) fluxes $\Phi_{0}(E)$ and $\Phi(E)$:
\begin{equation}
  \Phi(E)=e^{-(\tau_{star}+\tau_{dust})}\Phi_{0}(E)
  \label{eq:flux_params}
\end{equation}
where $\tau_{star}(z)$ and $\tau_{dust}(z,\{f_n\})$ are the optical depths due to starlight and dust, respectively. The intrinsic spectrum will be modeled, generically, by a set of parameters $\{\alpha_i\}$. Here, we adopt three kinds of intrinsic spectra, to know, a single power-law, a log-parabola and also a power-law with an exponential cutoff. Since this last function has an energy dependent curvature, it is more likely for a combined blazar spectrum+EBL fit to converge in this case to a solution where part of the flux drop at the high energy part of the measured SEDs is absorbed already at the intrinsic source spectrum, instead of being created by EBL attenuation. We can, therefore, write explicitly
$$
  \Phi_{0}(E)=
\left\{
  \begin{array}{ll}
    N_0\,\left(\frac{E}{E_0}\right)^{-\Gamma} &  \qquad \textrm{(power-law - PL)} \\
    N_0\,\left(\frac{E}{E_0}\right)^{-a-b\log(E/E_0)}  & \qquad \textrm{(log-parabola - LP)} \\
    N_0\,\left(\frac{E}{E_0}\right)^{-\Gamma}e^{-E/E_{\textrm{\tiny{cut}}}} &  \qquad \textrm{(power-law with exponential cutoff - PLC)} \\
  \end{array}
\right.
  \label{eq:intr_models}
$$
where $E_0=1.0$ TeV is the reference energy (notice that $E_0$ is not fitted, but rather fixed to minimize the correlations between the free parameters), $N_0$ is a flux normalization factor, $\Gamma$ is the spectral index of the power-law, $a$ and $b$ are, respectively, the spectral index and curvature for the log-parabola and $E_{\tiny\textrm{cut}}$ is the exponential energy cutoff. Imposing the normalization condition for the relative grain fractions ($\sum_{n=1}^{3} f_n=1)$, the fits will be performed with either 4 (power-law) or 5 (log-parabola and power-law with exponential cutoff) free parameters. Assuming Gaussianity for the uncertainties in the flux measurements of Mkn\,501, we perform a $\chi^2$ minimization. The best fits for the SED as well as the contours in the 2D parameter space of dust properties are shown in figure \ref{fig:bestfit_mkn501_pl_lp_plexp}. In this figure, we have separated the attenuation effects of each EBL component, so the plots show the convolution of the best-fit intrinsic spectrum with the attenuation factors of each EBL component, as well as the total attenuation. We have to stress, however, that there is an important systematic uncertainty in the best-fit fractions associated to the lack of knowledge on the correct intrinsic spectrum. In the same figure we see, for example, that the absence of curvature in the power-law intrinsic spectrum leads to larger contributions of PAH and small grains compared to the other models. On the other hand, when the source spectrum has an exponential cutoff and, therefore, an energy dependent curvature, the fit converges to a solution where the shape of the measured SED at the high energy tail is essentially defined by the intrinsic curvature and some extra attenuation due mainly to large grains. Notice the importance of PAH in giving the attenuated spectrum of Mkn\,501 the correct energy dependence in the region just below 10 TeV. Also, for the power-law and log-parabola cases, the different energy dependences of the attenuation due to small and large grains lead the fit to prefer large values of small grain fractions in order to describe the very end of the SED and, in turn, to a somehow inverted hierarchy of contributions between small and large dust grains coming out of the fit, when power-law and log-parabola are used as the source spectrum. It is generally believed that small hot grains should amount to a fraction of 10\%, at most, of the so called ``solid grains'' (small plus large) in the interstellar medium \cite{Krugel2003}, that is, excluding what is usually called molecular dust (PAH). Therefore, we have also performed a fit where an upper-bound is imposed on the mentioned fraction, i.e. \begin{equation}
\tilde{f}_{\tiny{\textrm{sg}}} = \frac{f_{\tiny{\textrm{sg}}}}{f_{\tiny{\textrm{sg}}}+f_{\tiny{\textrm{lg}}}} \le 0.1
\end{equation}
and the results in this case are included in tables \ref{table:results_eblmodels+pl} (power-law) and \ref{table:results_eblmodels+log-par} (log-parabola). We can see that the fit tries to get as much EBL attenuation as possible using small grains in order to reproduce the strong suppression at the end of Mkn 501 SED. The best-fits, in both cases, saturate the bound on $\tilde{f}_{\tiny{\textrm{sg}}}$. But in the absence of curvature in the intrinsic spectrum, as in the power-law case, the $\tilde{f}$-bounded fit is much worse than the unbounded one: $\chi^2$/ndof=37.6/13 (bounded) versus 15.7/13 (unbounded).
\begin{figure}[ht]
\centering
  \includegraphics[width=1.0\textwidth]{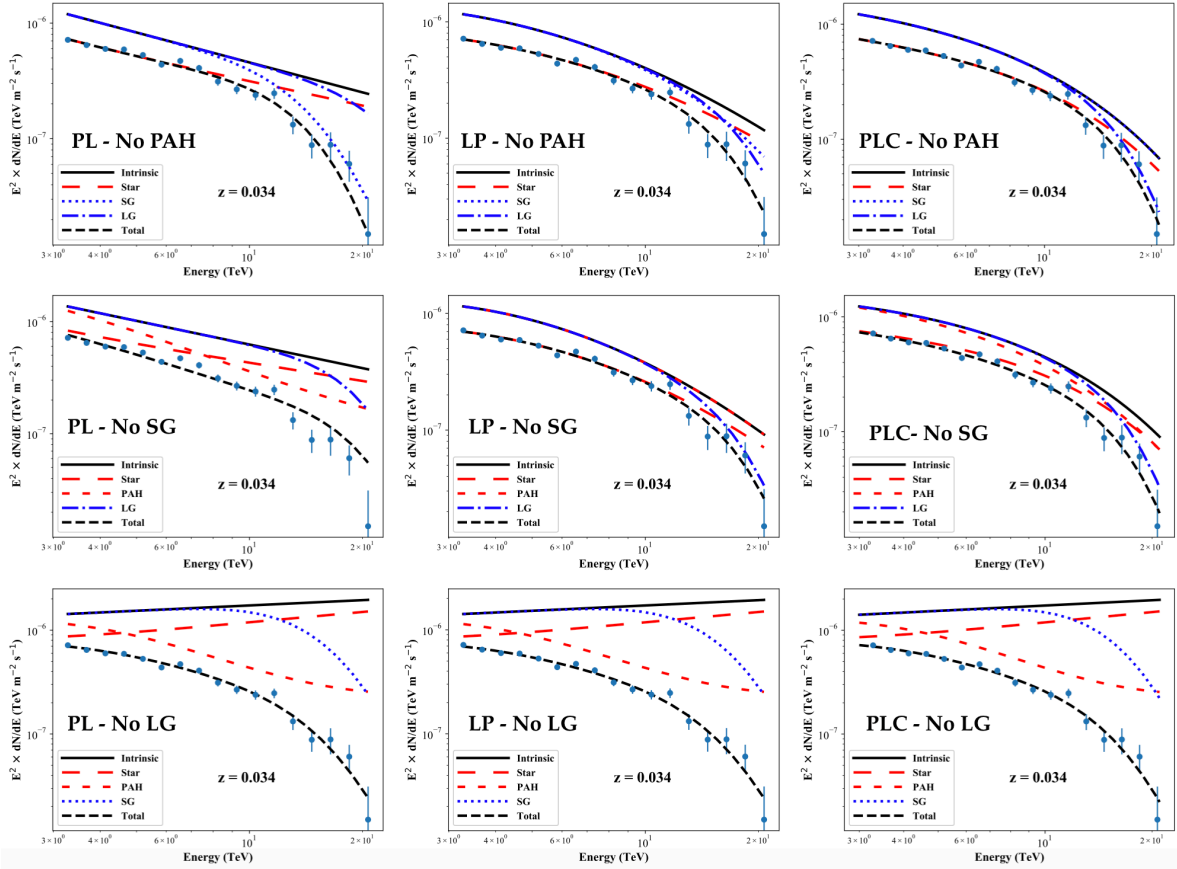}
  \caption{Best fit curves superimposed to measurements for an effective 2-grain model. The convolution of the intrinsic spectrum with the attenuation factors of individual grains is shown, as well as the total attenuation.}
  \label{fig:two_grains_models_bestfits}
\end{figure}

Some (if not all) of the degeneracies currently observed in the dust fraction parameter space are expected to be removed when a high quality multi-blazar sample is fit altogether, due to the increase in the number of degrees of freedom in such a fit. Even if different intrinsic spectra are used, the EBL attenuation at different redshifts should lead to stronger constraints on the dust fractions.
\begin{figure}[ht]
\centering
  \includegraphics[width=1.0\textwidth]{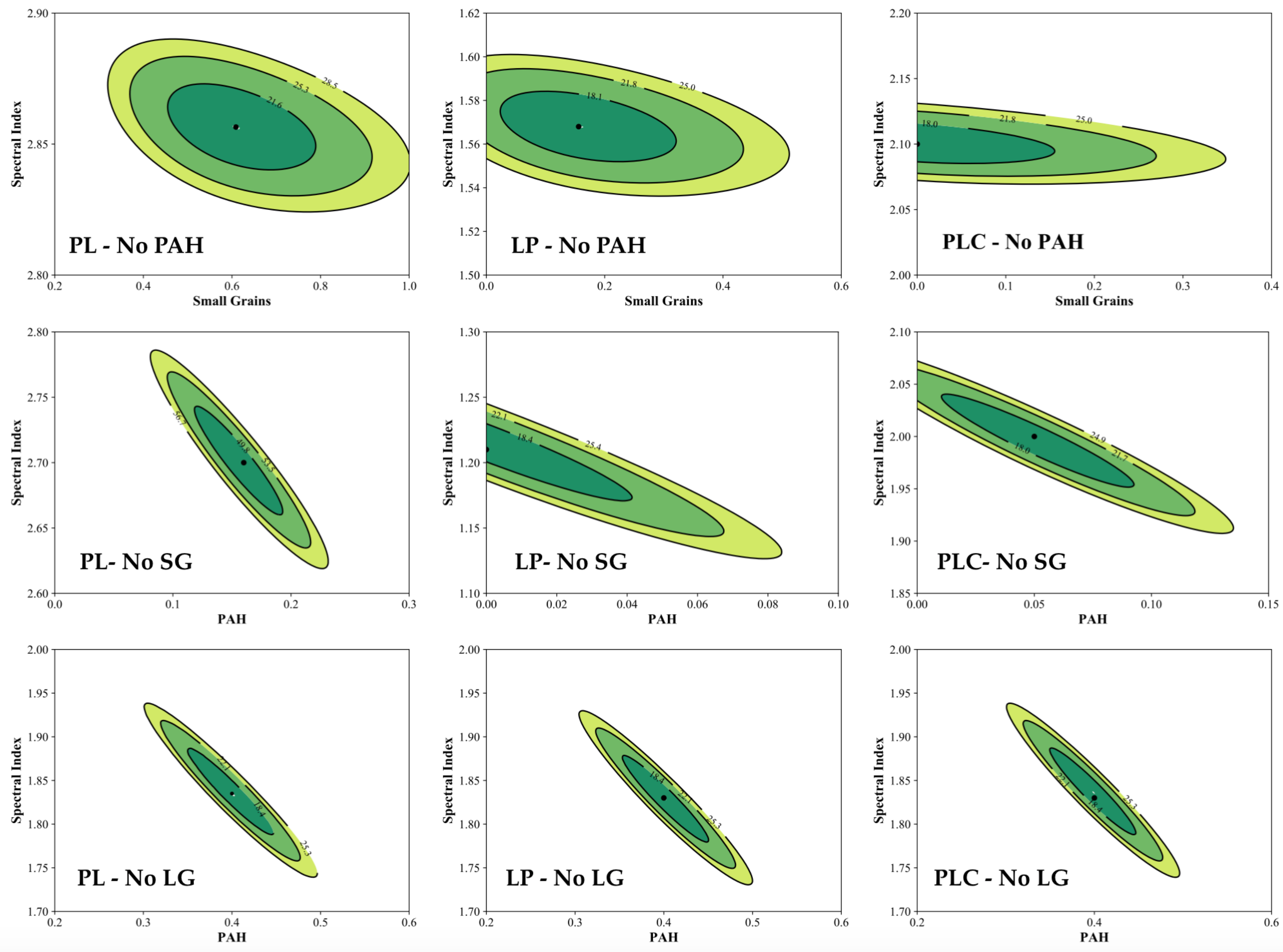}
  \caption{Confidence level curves at  68\%, 95\% and 99\% in a 2D parameter space with spectral index versus grain fraction: $\Gamma \times f_{\scriptsize{n}}$ (PL/PLC) or $a \times f_{\scriptsize{n}}$ (LP). The curves are for an effective 2-grain model.}
  \label{fig:two_grains_models_countous}
\end{figure}

In order to better understand the importance of individual grains in shaping the SED of Mkn\,501, we also performed the combined EBL-SED fit for effective 1- and 2-grain models. Figures \ref{fig:two_grains_models_bestfits} and \ref{fig:two_grains_models_countous} show the fit results for the 2-grain cases. Tables \ref{table:results_eblmodels+pl} (power-law), \ref{table:results_eblmodels+log-par} (log-parabola) and \ref{table:results_eblmodels+pl+expcut} (power-law with exponential cutoff) show additional information on the fits. One can see that when the intrinsic spectrum lacks curvature, the fit prefers to rely on small grains to model the spectrum shape at high energies. In the absence of this kind of dust (case PAH+large), the fit is the worse among all three effective 2-grain models. Caution should be exercised again when interpreting these results, because even though the relative contributions of the grains are varying, their temperatures are still fixed to the values of the nominal model, and figure \ref{fig:ebl_spectrum} shows that the grain temperature is a key parameter in shaping the EBL spectrum. We also see that curvature is able to compensate a big part of the dust attenuation, keeping the fit at reasonable quality. The contour plots show that the absence of either small or large grains introduce a strong correlation between the fractions of the remaining two grains. Notice that single grain models, with the temperatures given in table \ref{table:dust_pars}, do not provide good fits.
\begin{table*}[h!]
\setlength{\tabcolsep}{6pt}
\begin{center}
\begin{tabular}{cccccc}
\hline
\hline
\multicolumn{6}{ c }{power-law} \\
\hline
EBL model & $\chi^2$/ndof & $ \Gamma \pm \sigma $ & $f_{\tiny{\textrm{PAH}}} \pm \sigma $ & $f_{\tiny{\textrm{sg}}} \pm \sigma $ & $f_{\tiny{\textrm{lg}}} \pm \sigma $ \\
\hline
3 grains       & 15.7/13 & $ 2.05 \pm 0.39 $ & $0.32 \pm 0.15$ & $0.56 \pm 0.12$ & 0.12 \\
3 grains ($\tilde{f}_{\tiny{\textrm{sg}}}\le 0.1$)      & 37.6/13 & $ 2.75 \pm 0.29 $ & $0.12 \pm 0.12$ & $0.09 \pm 0.02$ & 0.79 \\
PAH+small  & 16.1/14 & $ 1.83 \pm 0.23 $ & $0.40 \pm 0.10$ & 0.60 & 0.00 \\
PAH+large   & 47.5/14 & $ 2.70 \pm 0.29 $ & $0.16 \pm 0.11$ & 0.00 & 0.84 \\
small+large & 19.3/14 & $ 2.86 \pm 0.06 $ & 0.00 & $0.61 \pm 0.14$ & 0.39 \\
PAH             & 98.0/15 & $ 0.68 \pm 0.04 $ & 1.00 & 0.00 & 0.00 \\
small           & 25.1/15 & $ 2.76 \pm 0.05 $ & 0.00 & 1.00 & 0.00 \\
large            & 48.9/15 & $ 3.09 \pm 0.04 $ & 0.00 & 0.00 & 1.00 \\
Finke et al.  & 41.6/15 & $ 2.44 \pm 0.04 $ & 0.25 & 0.05 & 0.70 \\
\hline
\hline
\end{tabular}
\end{center}
\caption{Summary of EBL+spectrum combined fits for a power-law intrinsic spectrum and the observed SED of Mkn\,501. Fractions without uncertainties were either kept fixed during the fit or obtained from fitted fractions by the normalization condition.}
\label{table:results_eblmodels+pl}
\end{table*}

\renewcommand{\arraystretch}{1.2}
\begin{table*}[h!]
\small
\setlength{\tabcolsep}{5pt}
\begin{center}
\begin{tabular}{ccccccc}
\hline
\hline
\multicolumn{7}{ c }{log-parabola} \\
\hline
EBL model & $\chi^2$/ndof & $ a \pm \sigma $ & $ b \pm \sigma $ & $f_{\tiny{\textrm{PAH}}} \pm \sigma $ & $f_{\tiny{\textrm{sg}}} \pm \sigma $ & $f_{\tiny{\textrm{lg}}} \pm \sigma $ \\
\hline
3 grains      & 15.7/12 & $ 1.96 \pm 0.48 $ & $ 0.16 \pm 0.60 $ & $ 0.27 \pm 0.25 $ & $0.49 \pm 0.28$ & 0.24 \\
3 grains ($\tilde{f}_{\tiny{\textrm{sg}}}\le 0.1$)      & 15.8/12 & $ 1.44 \pm 0.14 $ & $ 1.01 \pm 0.09$ & $0.00 \pm 0.14$ & $0.10 \pm 0.08$ & 0.90 \\
PAH+small & 16.1/13 & $ 1.83 \pm 0.23 $ & $ 0.00 \pm 0.23 $ & $ 0.40 \pm 0.10 $ & 0.60 & 0.00 \\
PAH+large  & 16.1/13 & $ 1.21 \pm 0.33 $ & $ 1.18 \pm 0.21 $ & $ 0.00 \pm 0.16 $ & 0.00 & 1.00 \\
small+large & 15.8/13 & $ 1.56 \pm 0.56 $ & $ 0.92 \pm 0.39 $ & 0.00 & $ 0.15 \pm 0.20 $ & 0.85 \\
PAH             & 62.4/14 & $ -0.99 \pm 0.29 $ & $ 1.00 \pm 0.17 $ & 1.00 & 0.00 & 0.00 \\
small           & 25.1/14 & $ 2.76 \pm 0.04 $ & $ 0.00 \pm 0.13 $ & 0.00 & 1.00 & 0.00 \\
large           & 16.2/14 & $ 1.21 \pm 0.33 $ & $ 1.18 \pm 0.21 $ & 0.00 & 0.00 & 1.00 \\
Finke et al.  & 18.8/14 & $ 0.93 \pm 0.32 $ & $ 0.95 \pm 0.20 $ & 0.25 & 0.05 & 0.70 \\
\hline
\hline
\end{tabular}
\end{center}
\caption{Summary of EBL+spectrum combined fit for a log-parabola intrinsic spectrum and the observed SED of Mkn\,501. Fractions without uncertainties were either kept fixed during the fit or obtained from fitted fractions by the normalization condition.}
\label{table:results_eblmodels+log-par}
\end{table*}
\renewcommand{\arraystretch}{1}

\renewcommand{\arraystretch}{1.2}
\begin{table*}[h!]
\small
\setlength{\tabcolsep}{5pt}
\begin{center}
\begin{tabular}{ccccccc}
\hline
\hline
\multicolumn{7}{ c }{power-law with exponential cutoff} \\
\hline
EBL model & $\chi^2$/ndof & $ \Gamma \pm \sigma $ & $ E_{\tiny{\textrm{cut}}} \pm \sigma \textrm{(TeV)} $ & $f_{\tiny{\textrm{PAH}}} \pm \sigma $ & $f_{\tiny{\textrm{sg}}} \pm \sigma $ & $f_{\tiny{\textrm{lg}}} \pm \sigma $ \\
\hline
3 grains      & 15.7/12 & $ 2.00 \pm 0.12 $ & $ 7.7 \pm 2.1 $ & $ 0.08 \pm 0.07 $ & $0.06 \pm 0.17$ & 0.86 \\
PAH+small & 16.1/13 & $ 1.83 \pm 0.24 $ & $ (0.1 \pm 9.9) \times 10^{7} $ & $ 0.40 \pm 0.10 $ & 0.60 & 0.00 \\
PAH+large  & 15.7/13 & $ 2.00 \pm 0.40 $ & $ 6.9 \pm 1.5 $ & $ 0.05 \pm 0.16 $ & 0.00 & 0.95 \\
small+large & 15.8/13 & $ 2.10 \pm 0.19 $ & $ 6.7 \pm 1.3 $ & 0.00 & $ 0.00 \pm 0.78 $ & 1.00 \\
PAH             & 50.4/14 & $ -0.31 \pm 0.16 $ & $ 7.6 \pm 1.2 $ & 1.00 & 0.00 & 0.00 \\
small           & 25.1/14 & $ 2.76 \pm 0.05 $ & $ (0.1 \pm 5.3) \times 10^{6} $ & 0.00 & 1.00 & 0.00 \\
large           & 15.8/14 & $ 2.10 \pm 0.19 $ & $ 6.7 \pm 1.3 $ & 0.00 & 0.00 & 1.00 \\
Finke et al.  & 16.6/14 & $ 1.62 \pm 0.18 $ & $ 8.3 \pm 1.9 $ & 0.25 & 0.05 & 0.70 \\
\hline
\hline
\end{tabular}
\end{center}
\caption{Summary of EBL+spectrum combined fit for a power law with an exponential cutoff intrinsic spectrum and the observed SED of Mkn\,501. Fractions without uncertainties were either kept fixed during the fit or obtained from fitted fractions by the normalization condition.}
\label{table:results_eblmodels+pl+expcut}
\end{table*}
\renewcommand{\arraystretch}{1}

\begin{figure*}[ht]
\begin{center}
\subfigure{
  \includegraphics[width=.33\textwidth]{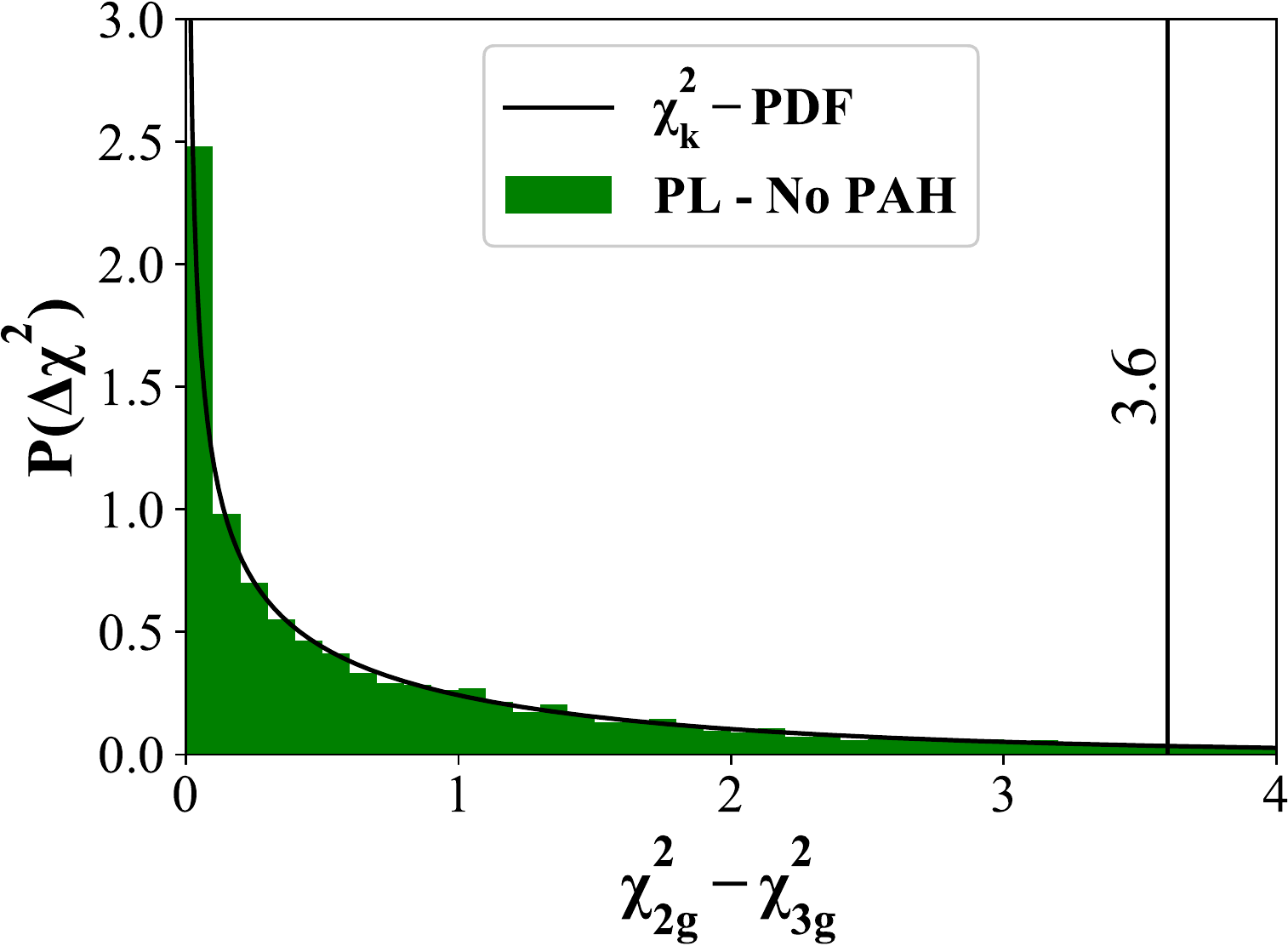}
}%
\subfigure{
  \includegraphics[width=.33\textwidth]{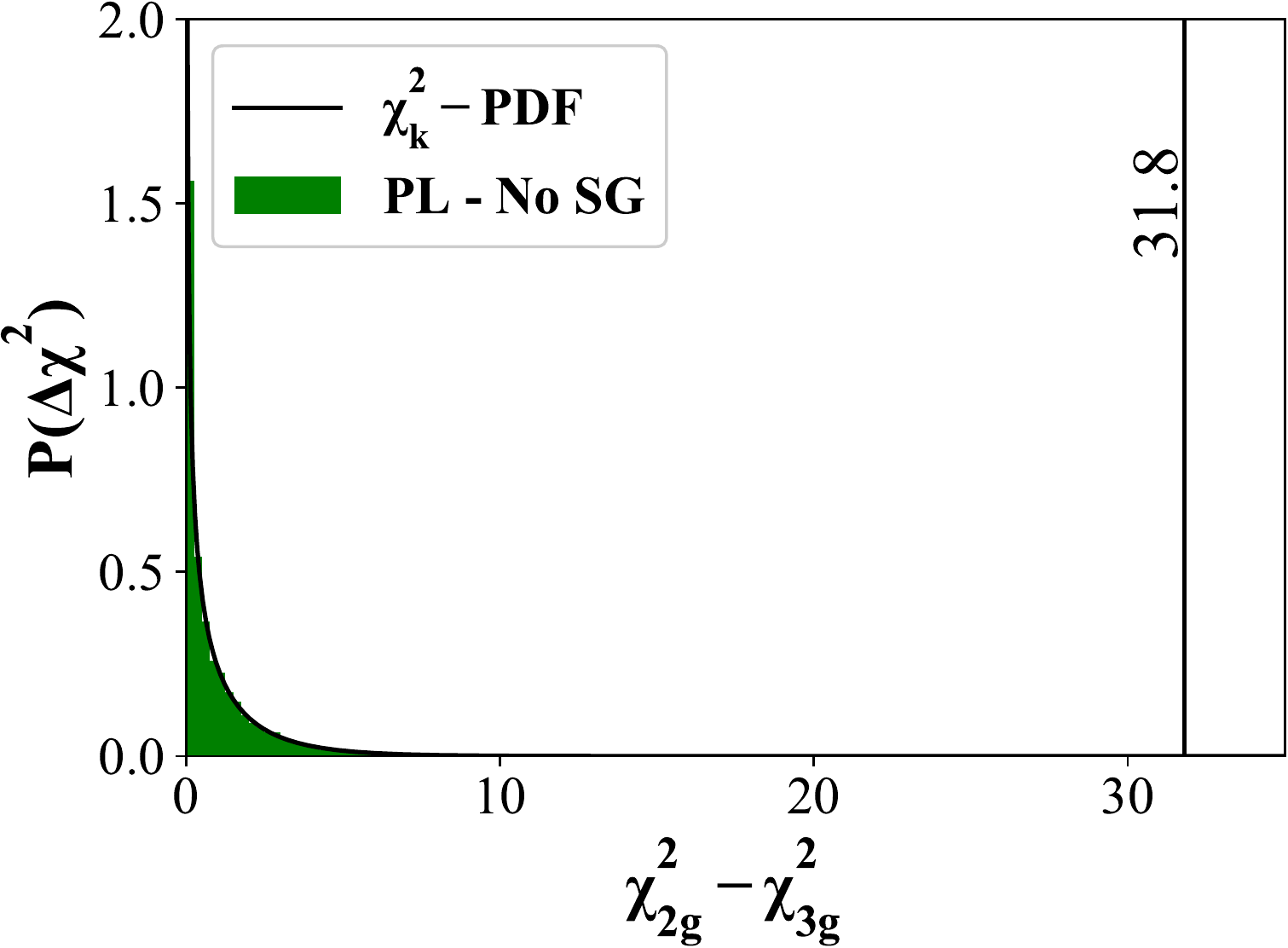}
}%
\subfigure{
  \includegraphics[width=.33\textwidth]{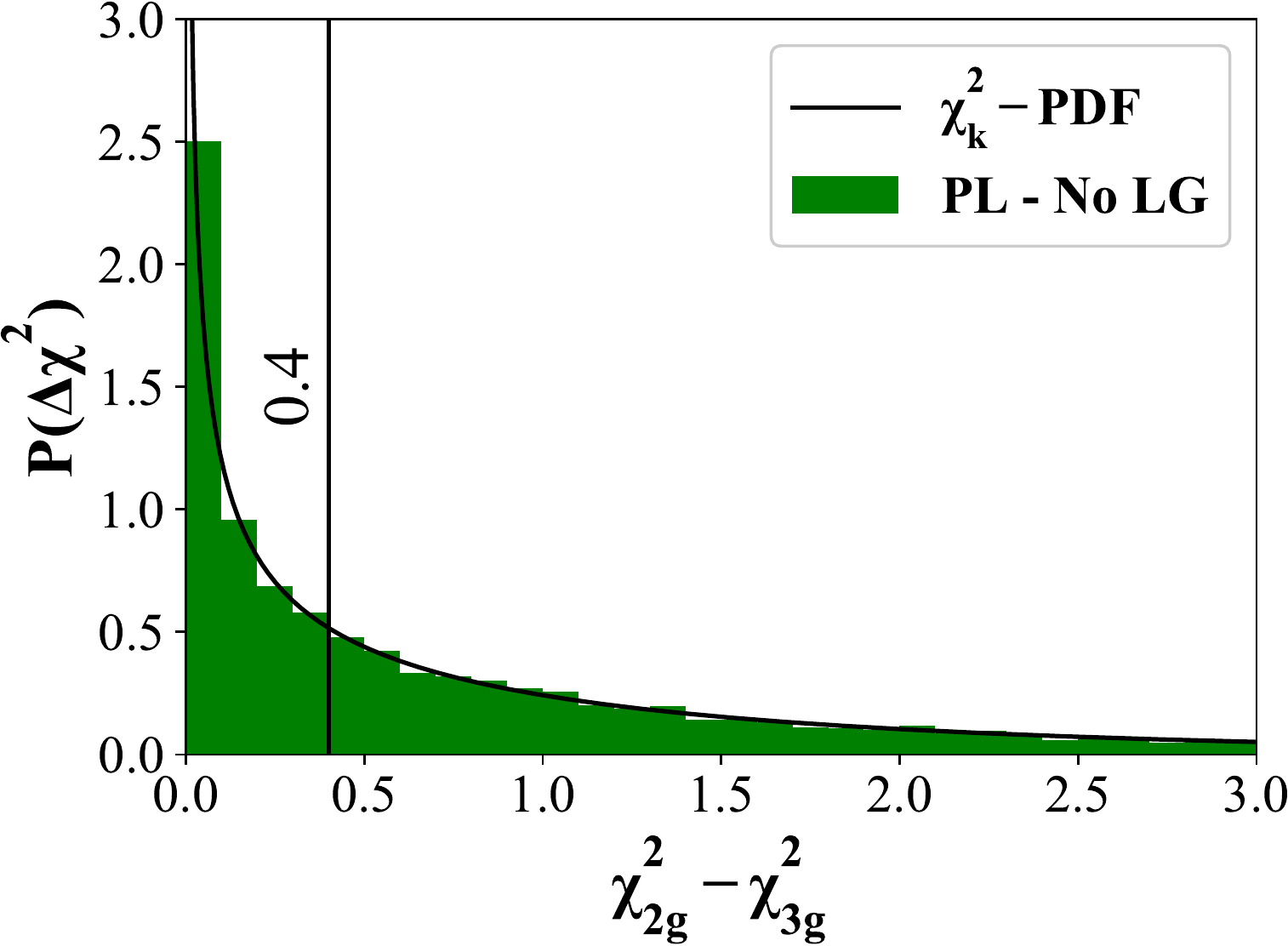}
} \\ 
\subfigure{
  \includegraphics[width=.33\textwidth]{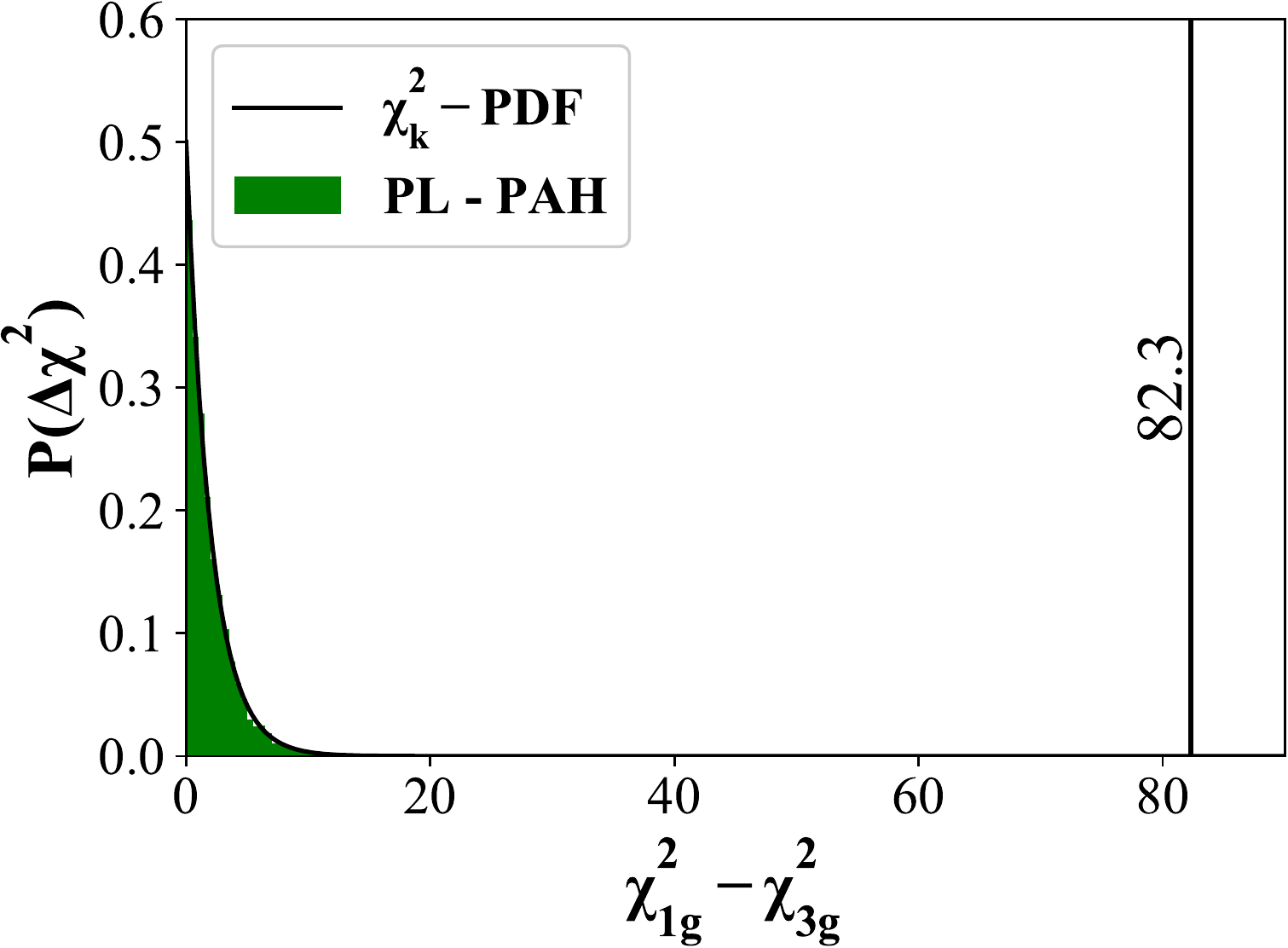}
}%
\subfigure{
  \includegraphics[width=.33\textwidth]{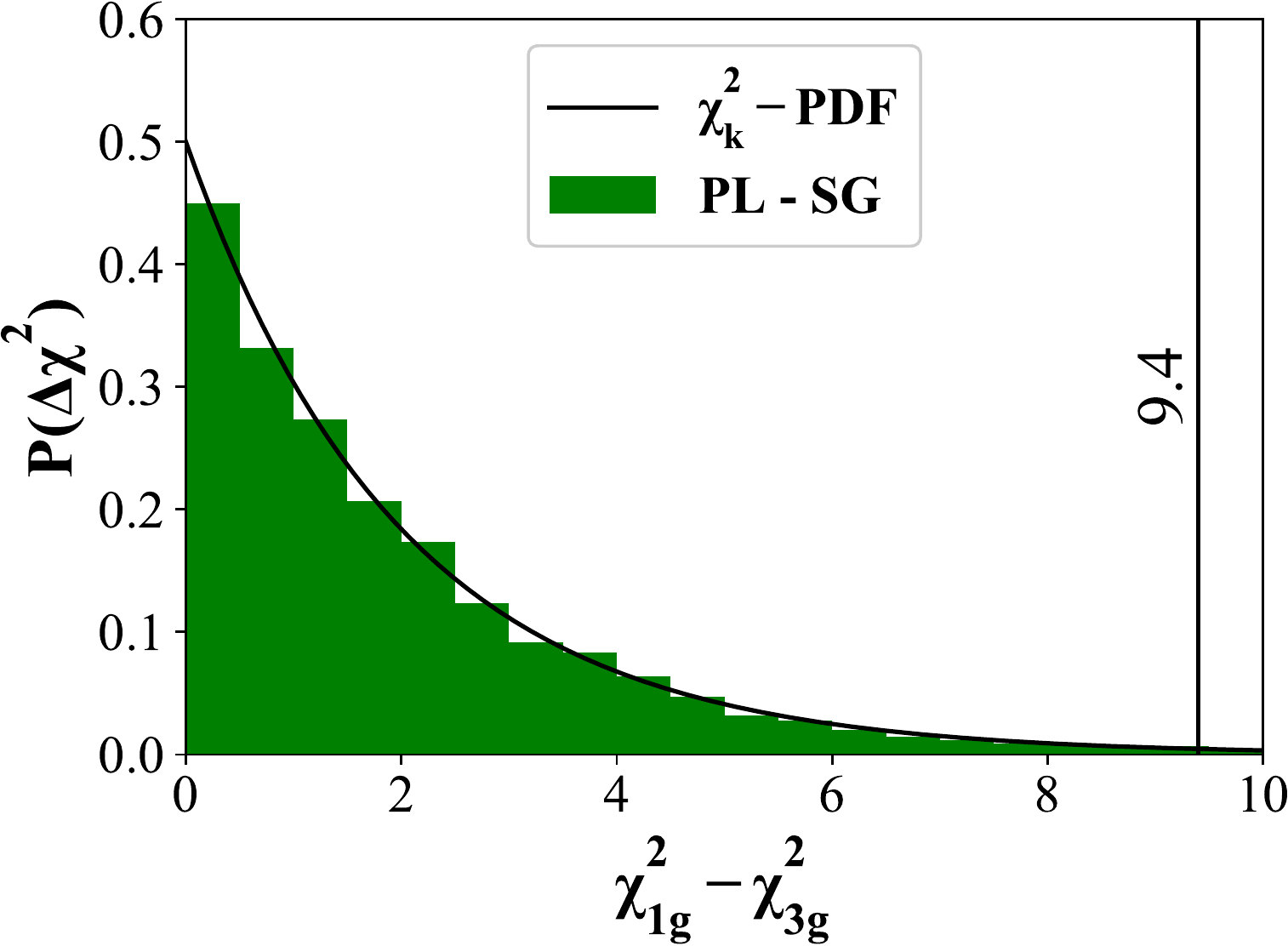}
}%
\subfigure{
  \includegraphics[width=.33\textwidth]{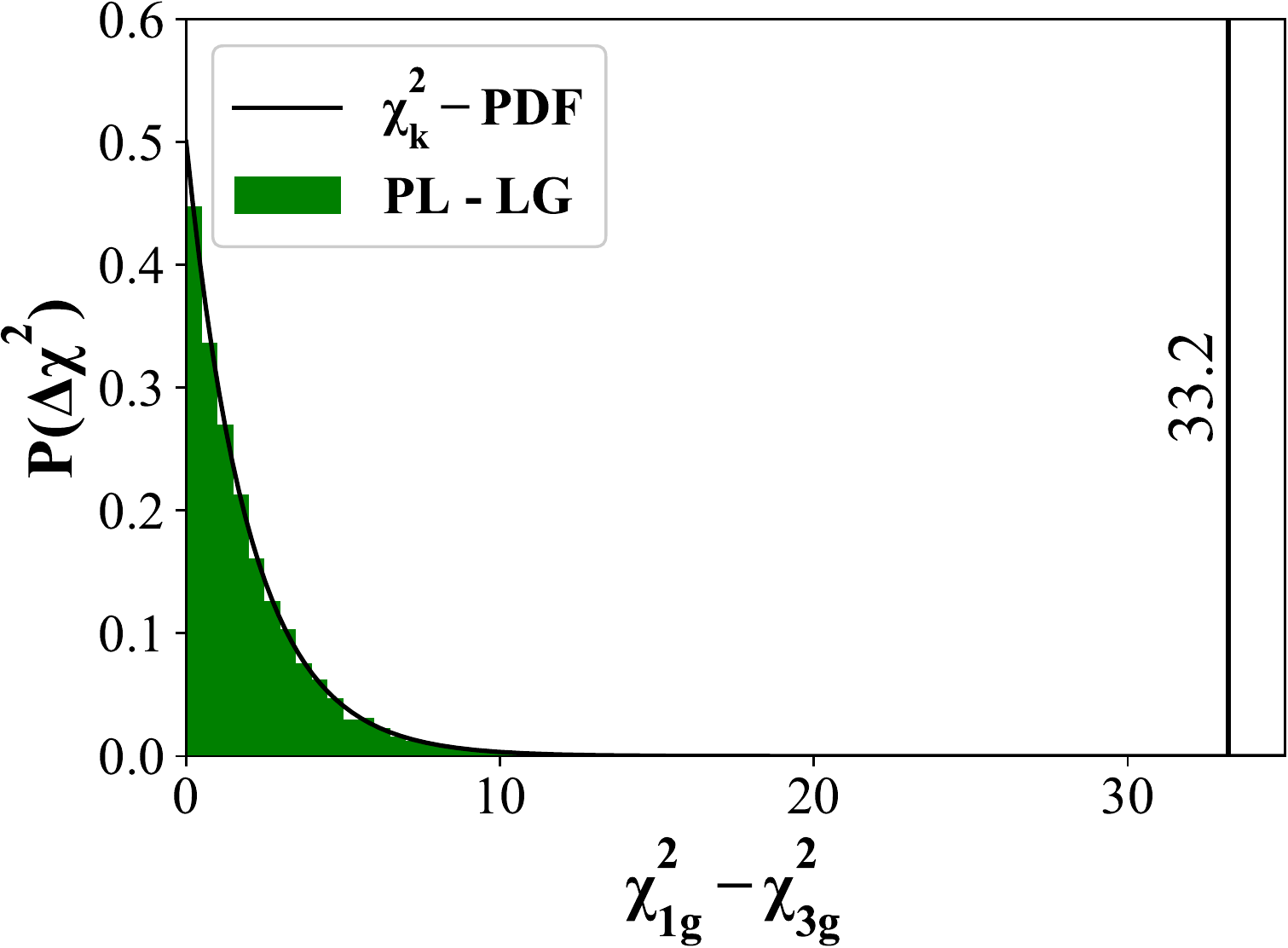}
} \\ 
\end{center}
  \caption{Distribution of test statistic ($-2\,\textrm{ln}({\cal L}_0/{\cal L}_1) = \Delta\chi^2$) for a power-law intrinsic spectrum, 1- and 2-grain models as null hypotheses ($H_0$) and 3-grain model as alternative hypothesis ($H_1$). Top: distribution for effective 2-grain models superimposed to the $\chi^{2}_{k}$ pdf with $k=1$ dof. Bottom: distribution for single grain models superimposed to the $\chi^{2}_{k}$ pdf with $k=2$ dofs. The vertical lines correspond to the test statistic value for the best fits of table \ref{table:results_eblmodels+pl}.}
  \label{fig:likeli_ratio_pl}
\end{figure*}

\begin{figure*}[ht]
\begin{center}
\subfigure{
  \includegraphics[width=.33\textwidth]{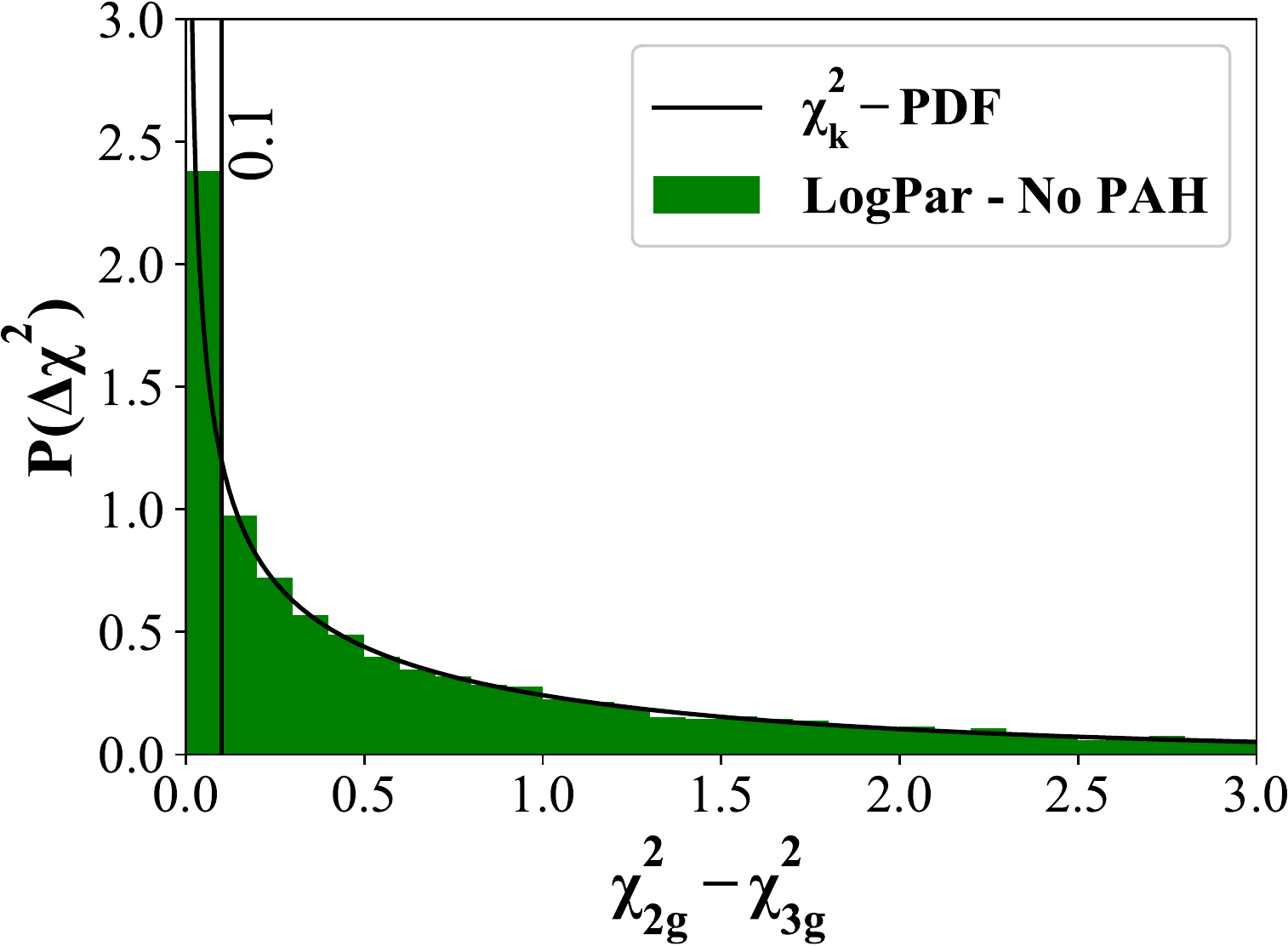}
}%
\subfigure{
  \includegraphics[width=.33\textwidth]{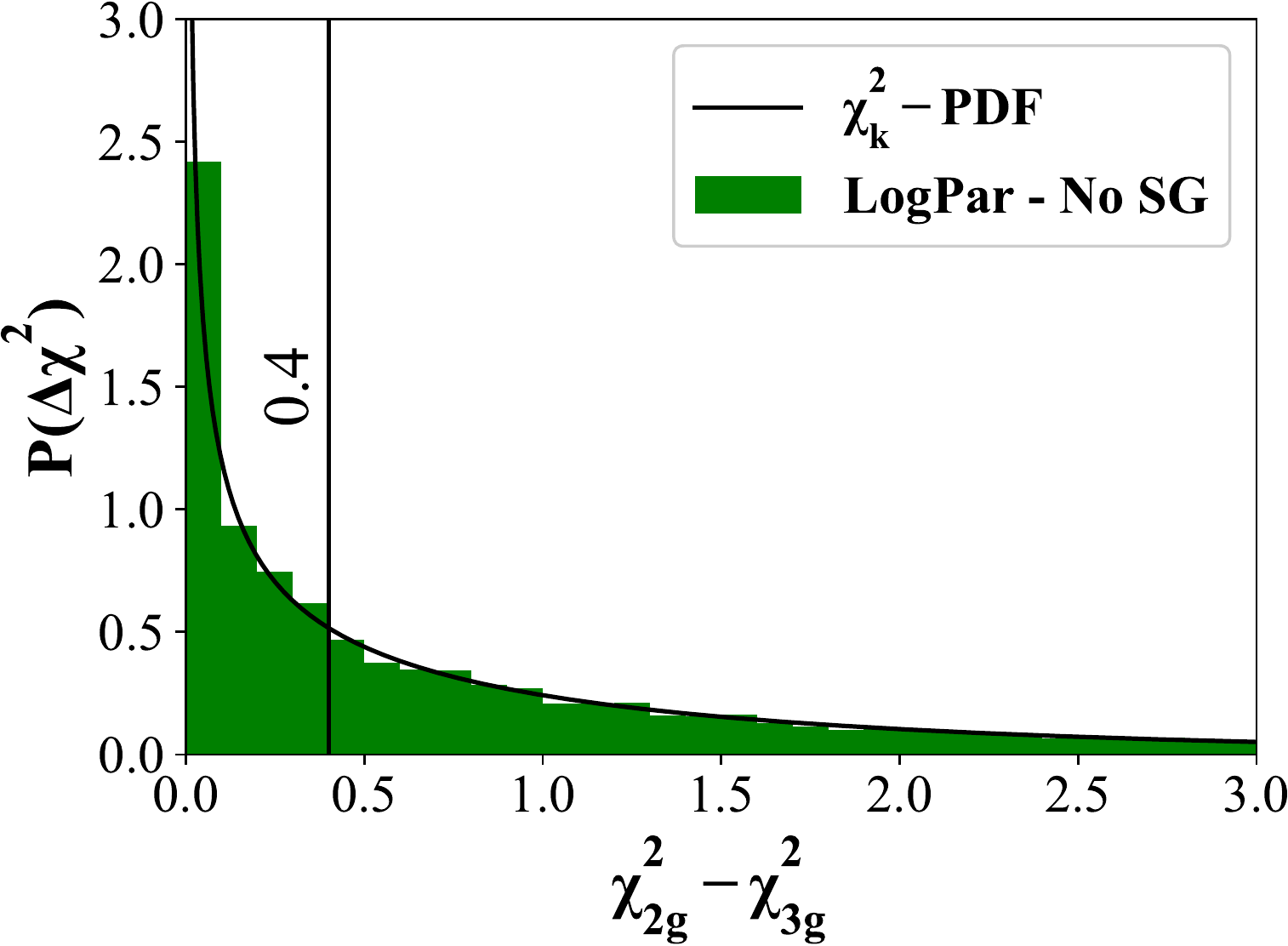}
}%
\subfigure{
  \includegraphics[width=.33\textwidth]{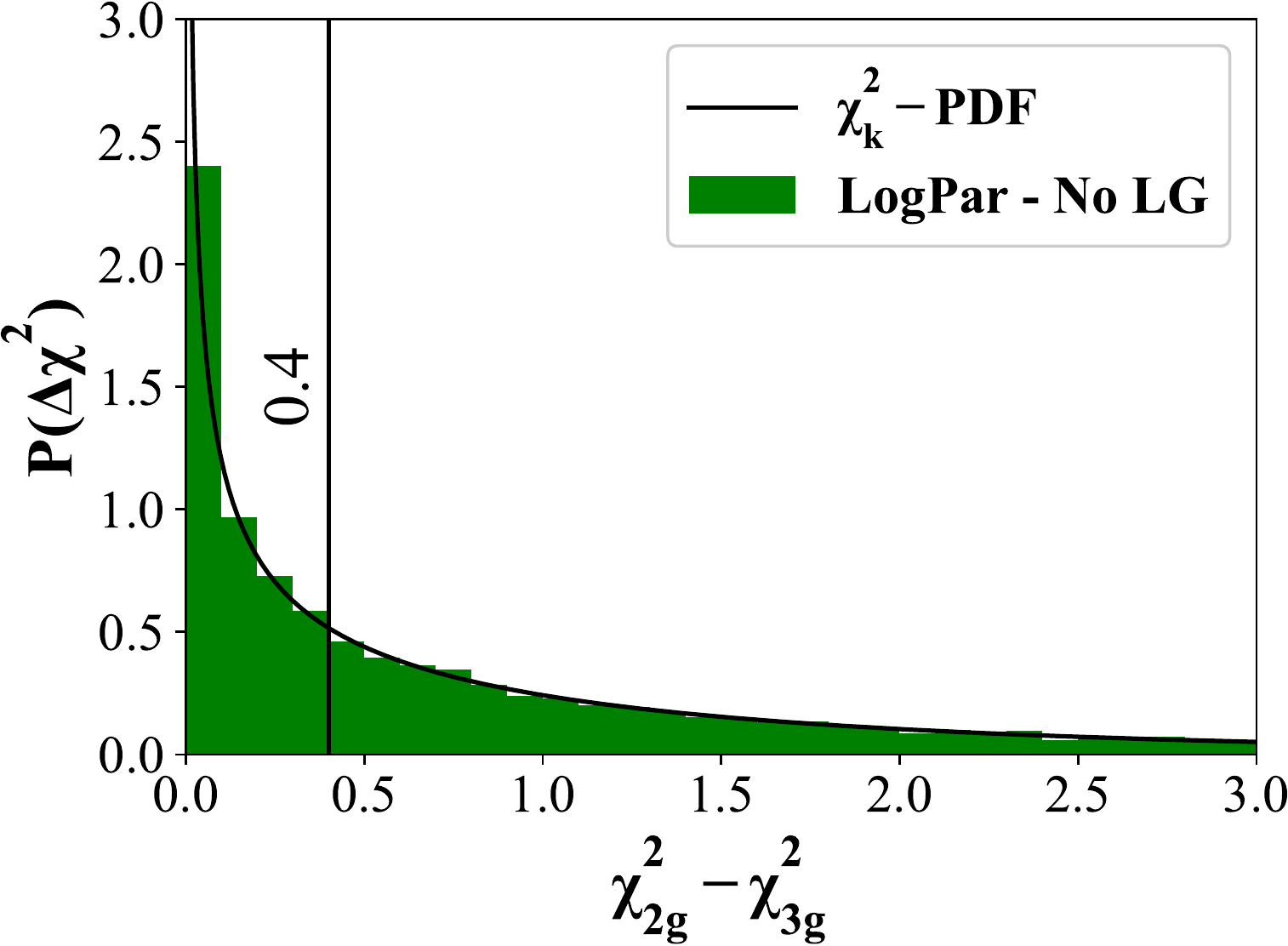}
} \\ 
\subfigure{
  \includegraphics[width=.33\textwidth]{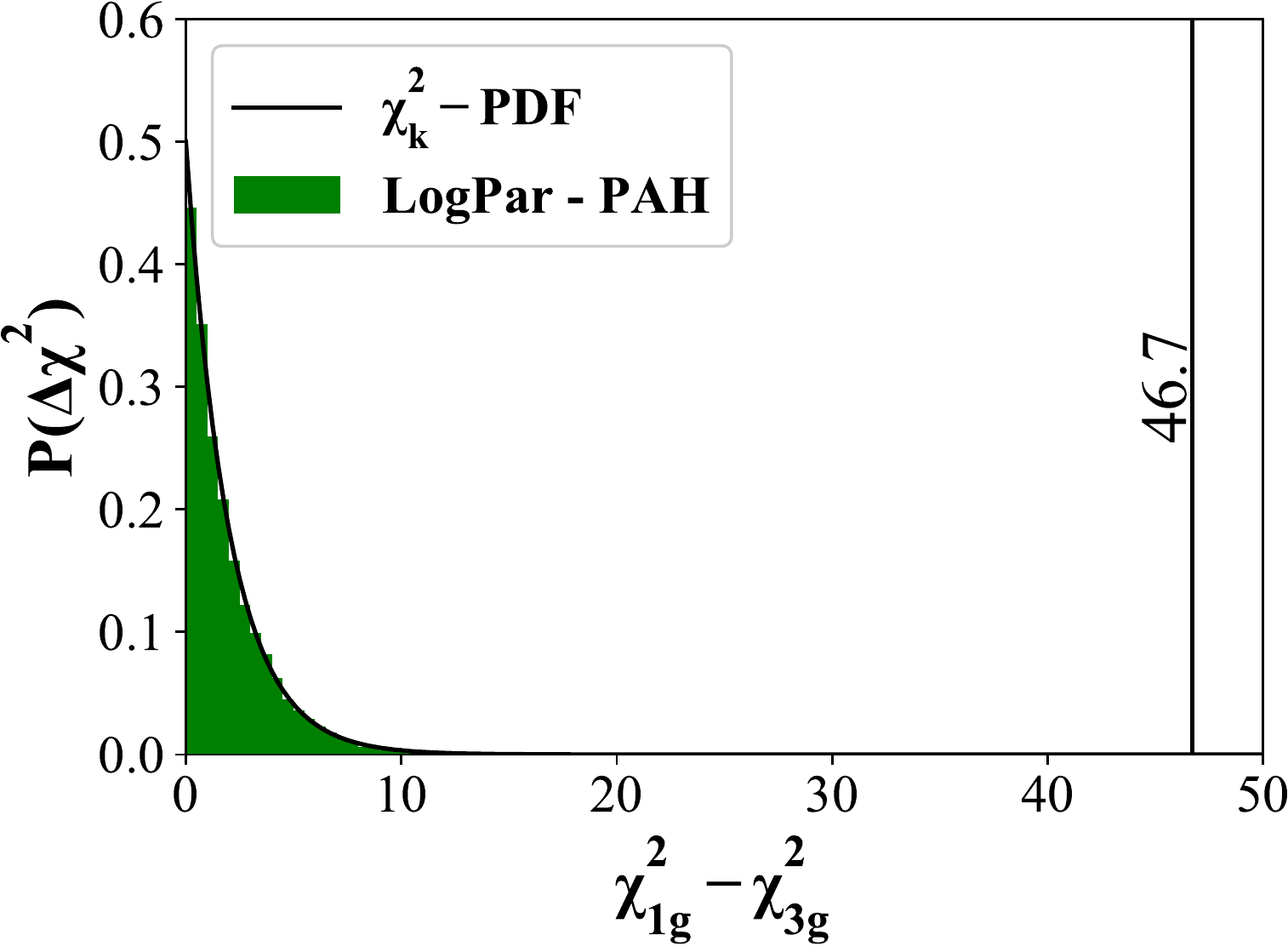}
}%
\subfigure{
  \includegraphics[width=.33\textwidth]{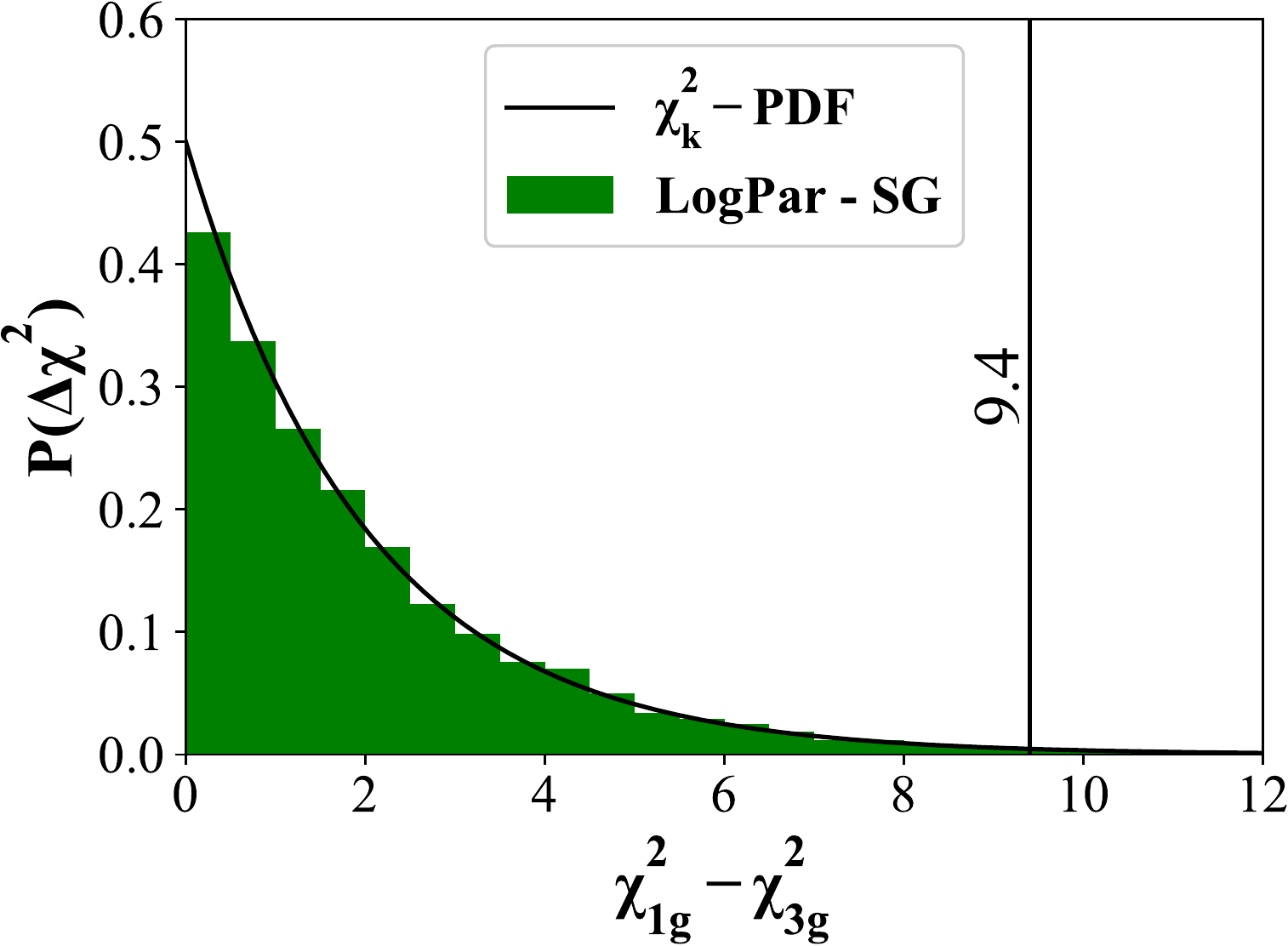}
}%
\subfigure{
  \includegraphics[width=.33\textwidth]{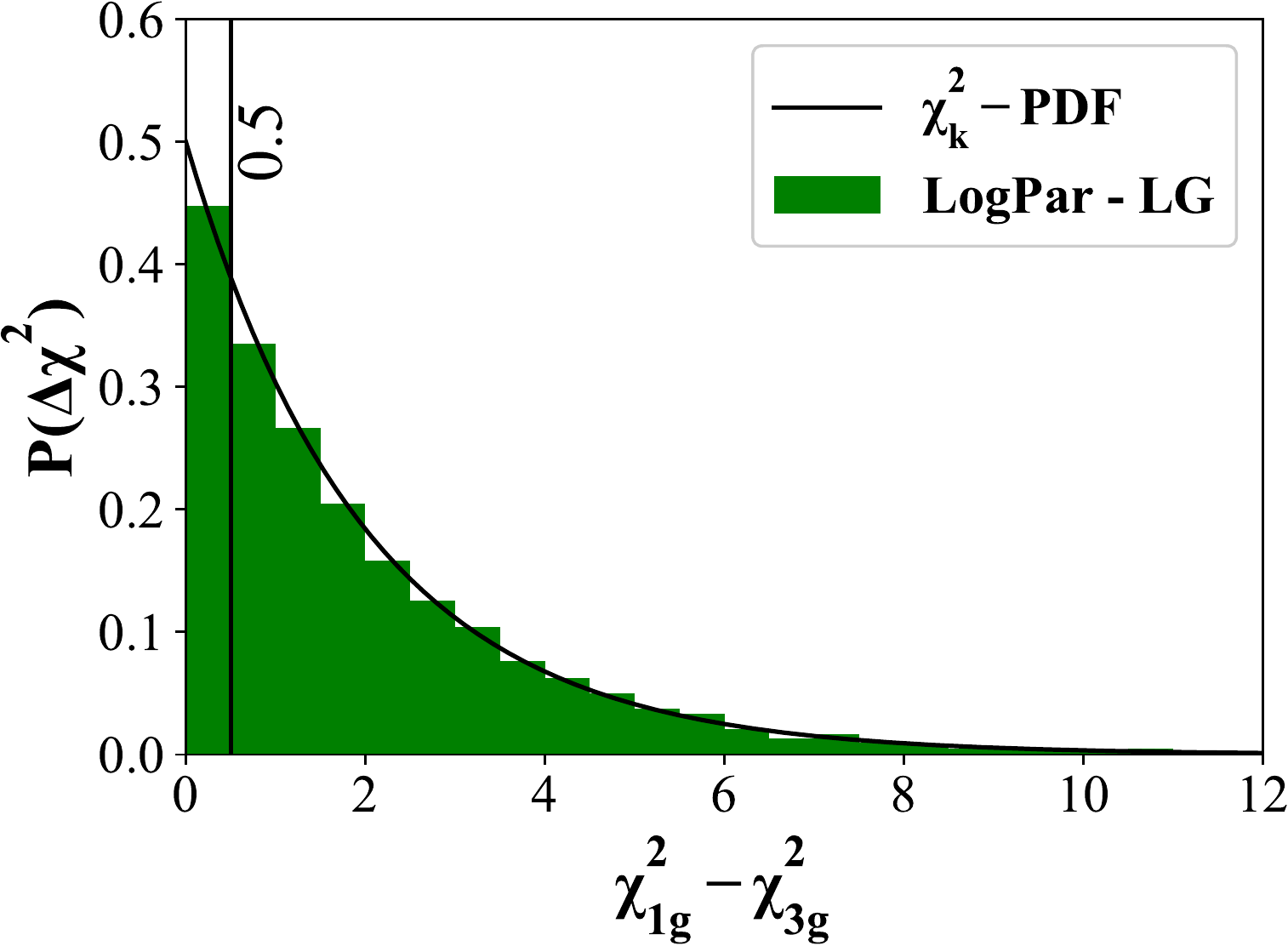}
} \\ 
\end{center}
  \caption{
Distribution of test statistic ($-2\,\textrm{ln}({\cal L}_0/{\cal L}_1) = \Delta\chi^2$) for a log-parabola intrinsic spectrum, 1- and 2-grain models as null hypotheses ($H_0$) and 3-grain model as alternative hypothesis ($H_1$). Top: distribution for effective 2-grain models superimposed to the $\chi^{2}_{k}$ pdf with $k=1$ dof. Bottom: distribution for single grain models superimposed to the $\chi^{2}_{k}$ pdf with $k=2$ dofs. The vertical lines correspond to the test statistic value for the best fits of table \ref{table:results_eblmodels+log-par}.}
  \label{fig:likeli_ratio_logpar}
\end{figure*}

\begin{figure*}[ht]
\begin{center}
\subfigure{
  \includegraphics[width=.33\textwidth]{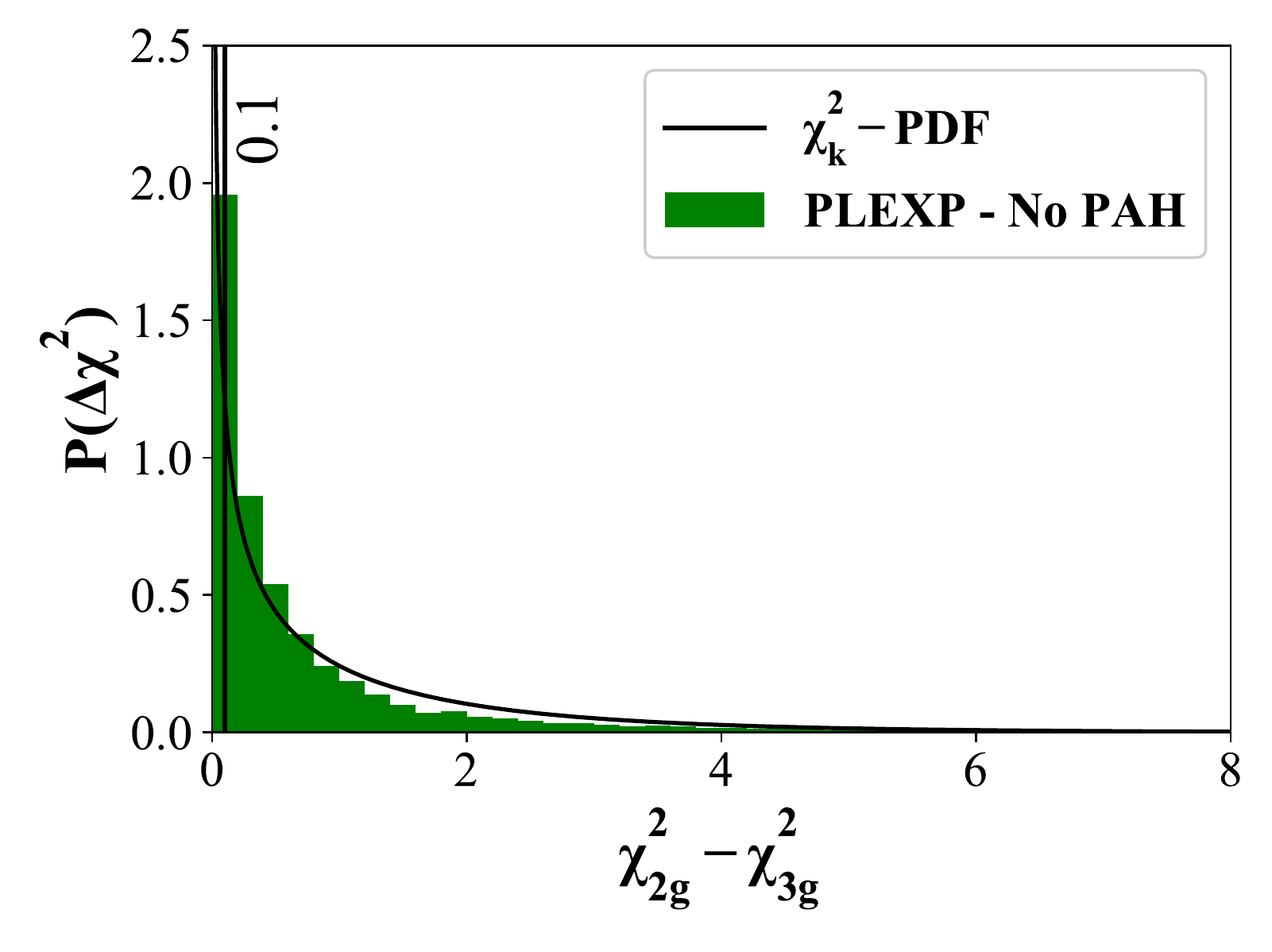}
}%
\subfigure{
  \includegraphics[width=.33\textwidth]{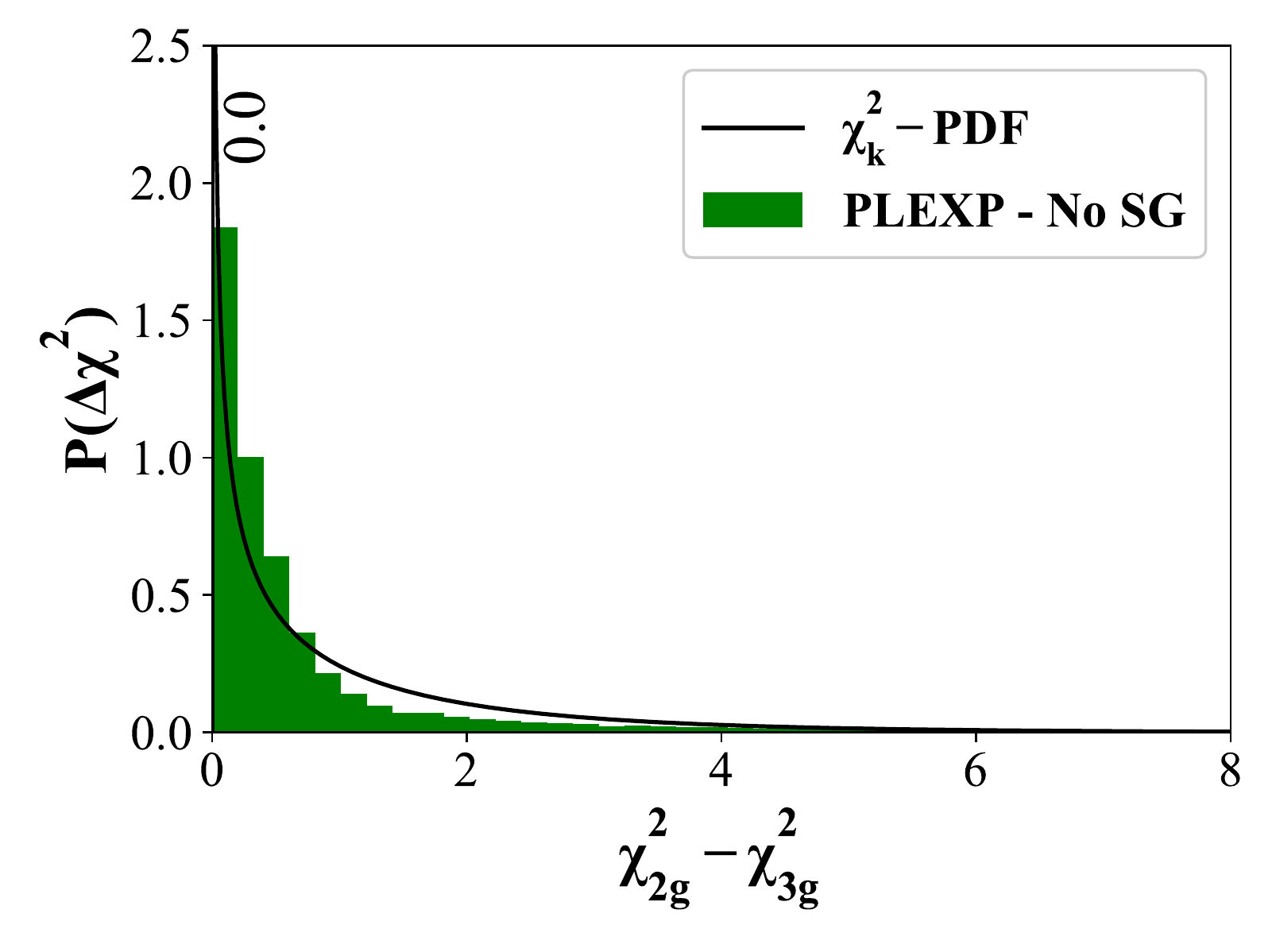}
}%
\subfigure{
  \includegraphics[width=.33\textwidth]{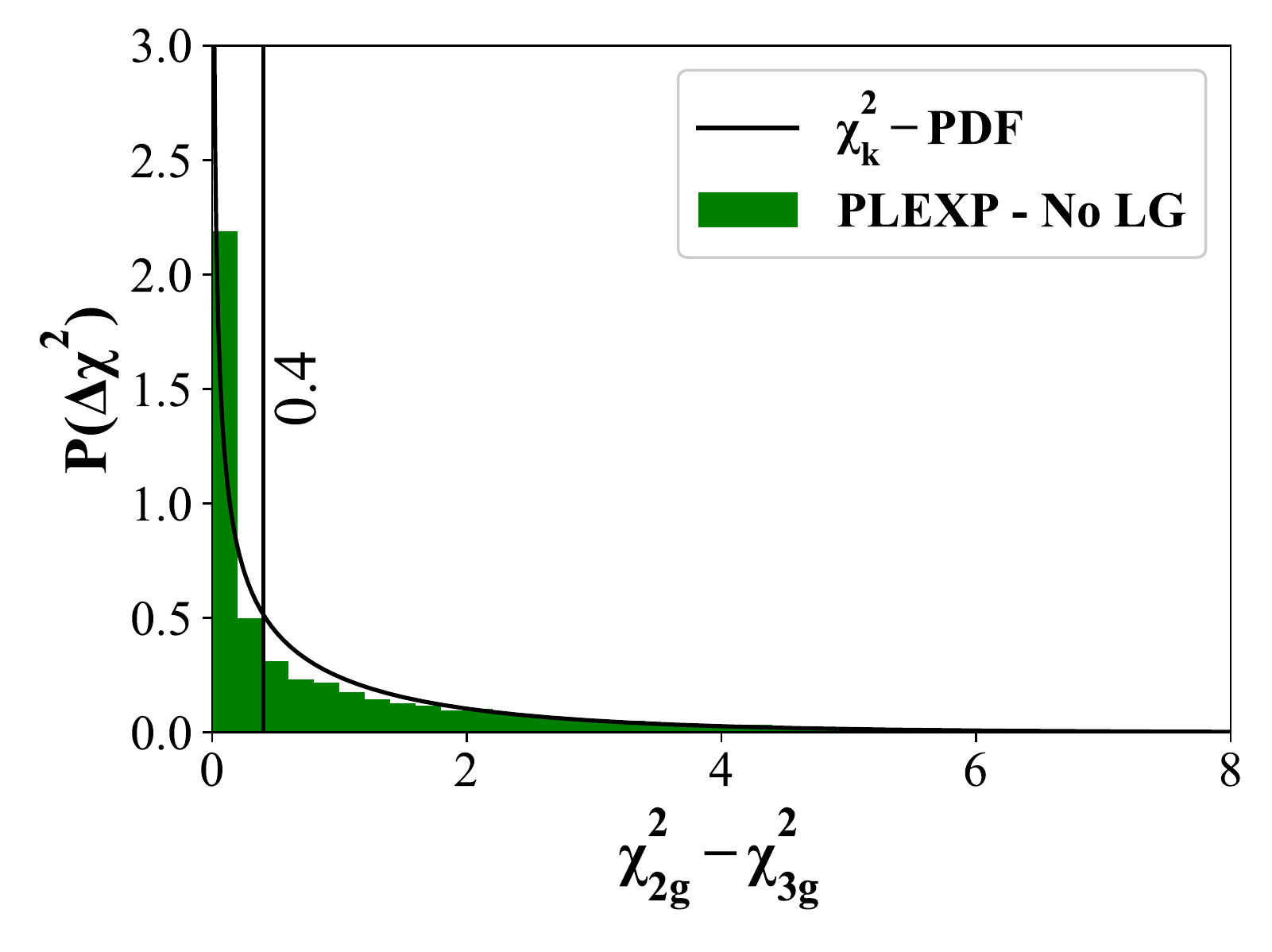}
} \\ 
\subfigure{
  \includegraphics[width=.33\textwidth]{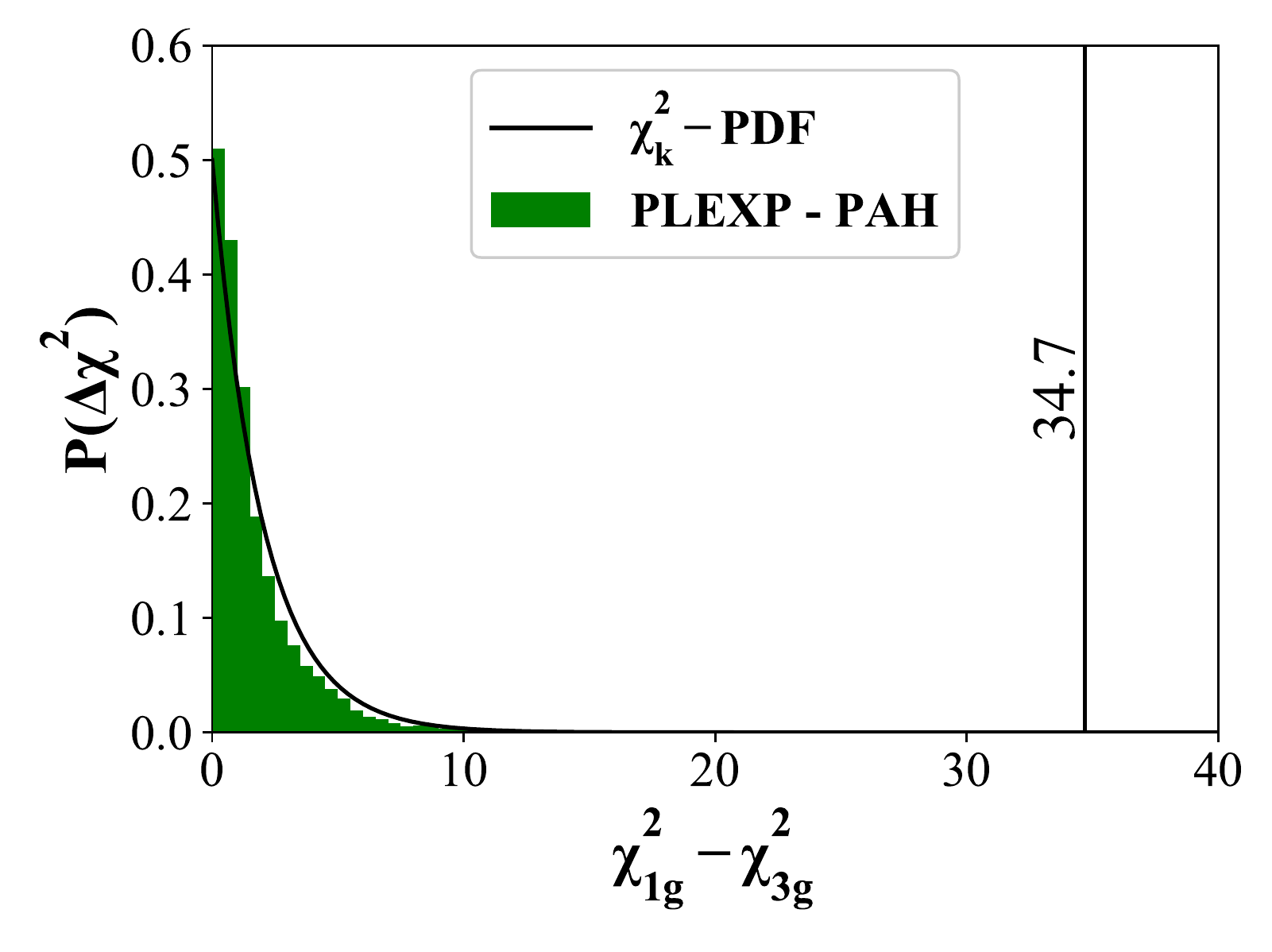}
}%
\subfigure{
  \includegraphics[width=.33\textwidth]{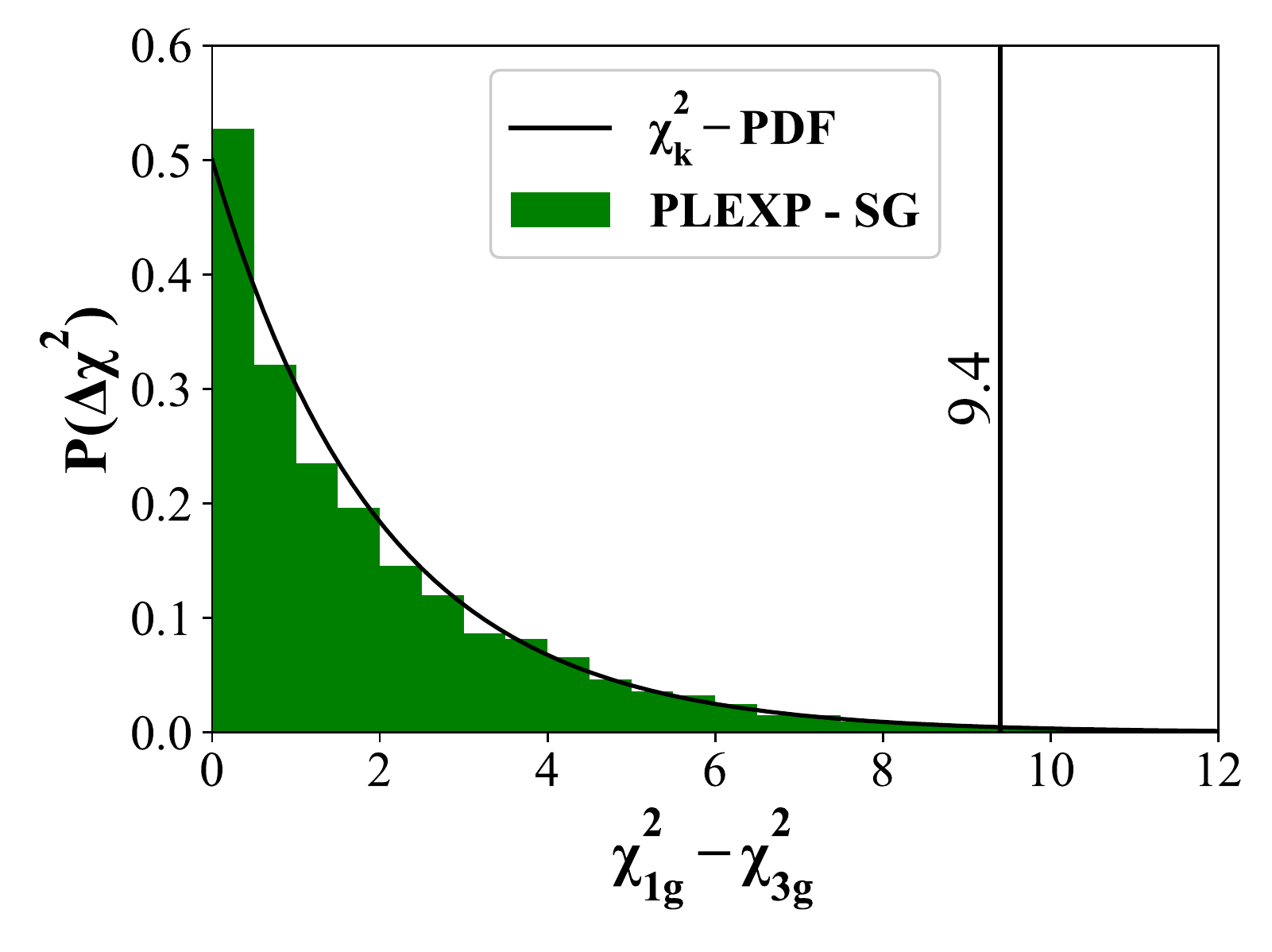}
}%
\subfigure{
  \includegraphics[width=.33\textwidth]{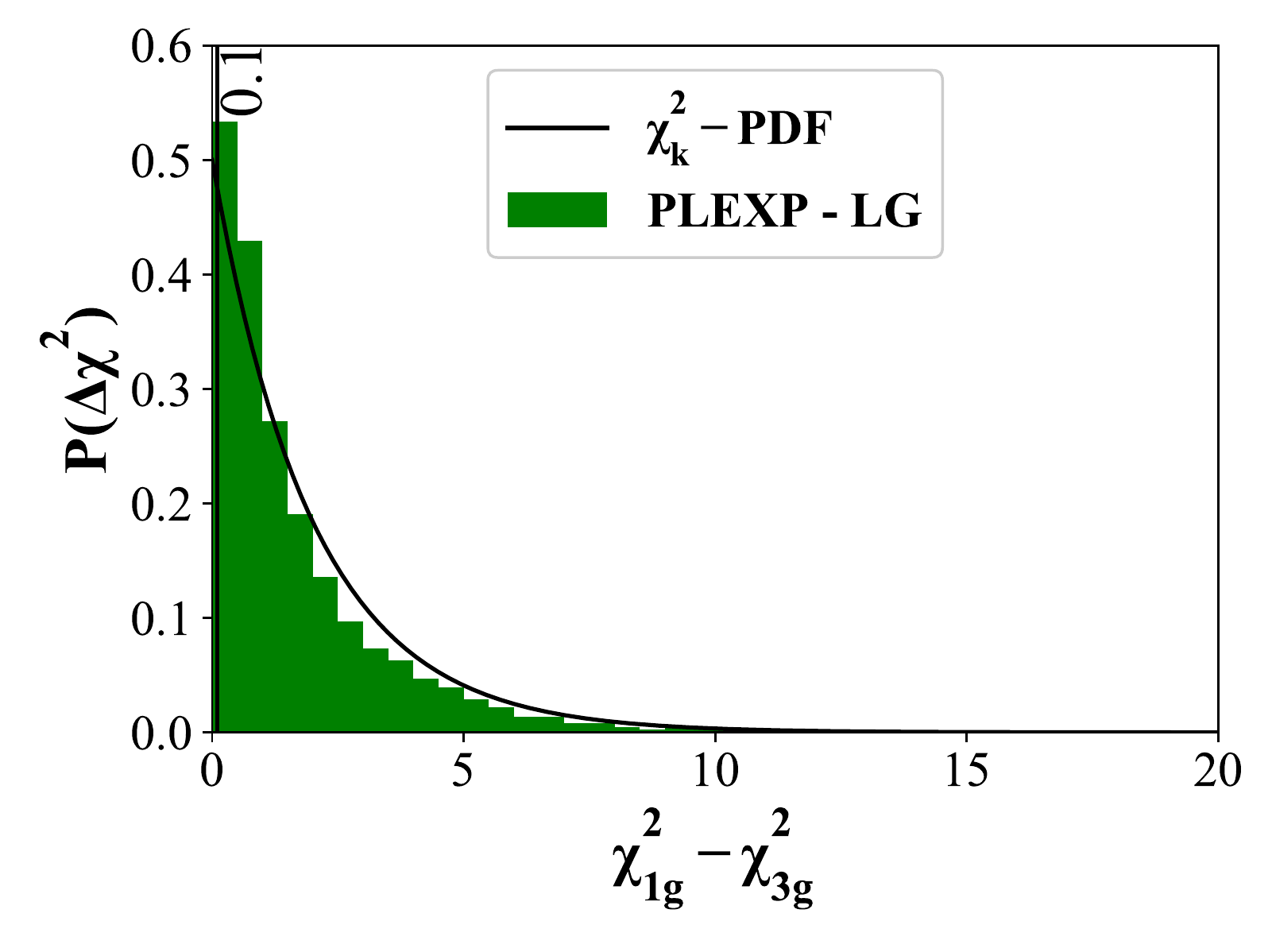}
} \\ 
\end{center}
  \caption{
Distribution of test statistic ($-2\,\textrm{ln}({\cal L}_0/{\cal L}_1) = \Delta\chi^2$) for a power-law with cutoff intrinsic spectrum, 1- and 2-grain models as null hypotheses ($H_0$) and 3-grain model as alternative hypothesis ($H_1$). Top: distribution for effective 2-grain models superimposed to the $\chi^{2}_{k}$ pdf with $k=1$ dof. Bottom: distribution for single grain models superimposed to the $\chi^{2}_{k}$ pdf with $k=2$ dofs. The vertical lines correspond to the test statistic value for the best fits of table \ref{table:results_eblmodels+pl+expcut}.}
  \label{fig:likeli_ratio_plcutoff}
\end{figure*}

Finally, we perform a hypothesis test by comparing the 1- and 2-grain scenarios (the null hypotheses $H_0$) against the 3-grain one (the alternative $H_1$), using a nested log-likelihood ratio. The test statistic will be $-2\,\textrm{ln}({\cal L}_0/{\cal L}_1) = \Delta\chi^2$. According to Wilks' theorem \cite{wilks1938}, in the limit of a large data sample, the asymptotic pdf of this statistic (when $H_0$ holds true) should be a $\chi_k^2$ distribution with a number of degrees of freedom $k$ equal to the difference in dimensionally of the corresponding parameter spaces. Therefore, $k=1$ ($H_0=$ ``two grains'') or $k=2$ ($H_0=$ ``single grain''), for the tests performed here.

\begin{table*}[h!]
\setlength{\tabcolsep}{6pt}
\begin{center}
\begin{tabular}{cccccccc}
\hline
\hline
& \multicolumn{2}{ |c }{power-law} & \multicolumn{2}{|c}{log parabola} & \multicolumn{2}{|c}{power-law $\times$ cutoff}\\
\hline
\hline
null hypothesis & $\Delta\chi^2$ & $P(\ge \Delta \chi^2)$ & $\Delta\chi^2$ & $P(\ge \Delta \chi^2)$ & $\Delta\chi^2$ & $P(\ge \Delta \chi^2)$ \\
\hline
PAH+small  & 0.4 & $0.53$  & 0.4 & $ 0.53 $ & 0.4 & 0.53 \\
PAH+large   & 31.8 & $ 1.7 \times 10^{-8} $ & 0.4 & $ 0.53 $ & 0.0 & 1.0 \\
small+large & 3.6 & $ 0.06 $  & 0.1 & $ 0.75 $ & 0.1 & 0.75 \\
PAH             & 82.3 & $ 1.4 \times 10^{-18} $ & 46.7 & $ 7.2 \times 10^{-11} $ & 34.7 & $ 2.9 \times 10^{-8} $ \\
small           & 9.4 & $ 0.01 $ & 9.4 & $ 0.01 $ & 9.4 & 0.01 \\
large            & 33.2 & $ 6.2 \times 10^{-8} $ & 0.5 & $ 0.78 $ & 0.1 & 0.95 \\
\hline
\hline
\end{tabular}
\end{center}
\caption{Summary of the nested likelihood ratio test for 2-grain and 1-grain models with power-law, log-parabola and power-law with an exponential cutoff intrinsic spectra. The 3-grains model was used as alternative hypothesis.}
\label{table:results_likeli+pl}
\end{table*}

Figures \ref{fig:likeli_ratio_pl}, \ref{fig:likeli_ratio_logpar} and \ref{fig:likeli_ratio_plcutoff} show the distributions of $\Delta\chi^2$ for the three spectra, using as null hypotheses the 1- and 2-grain best fits of tables \ref{table:results_eblmodels+pl}, \ref{table:results_eblmodels+log-par} and \ref{table:results_eblmodels+pl+expcut}. The expected asymptotic pdf of $\Delta\chi^2$ is also superimposed and shows that for the size of Mkn\,501 flare state SED, it is already an excellent approximation to the exact pdf. The $p$-values of table \ref{table:results_likeli+pl} were, therefore, calculated using the asymptotic formula. We see that the single grain scenario represented by PAHs can be excluded at more than 5$\sigma$ ($p=2.9 \times 10^{-8}$), regardless of the intrinsic spectrum used. It is clear from the 2-grain fits of figure \ref{fig:two_grains_models_bestfits} that a PAH-only attenuation is unable to account for the strong flux drop of Mkn\,501 SED above 10 TeV.

\begin{figure}[h]
  \centering
  \includegraphics[width=\textwidth]{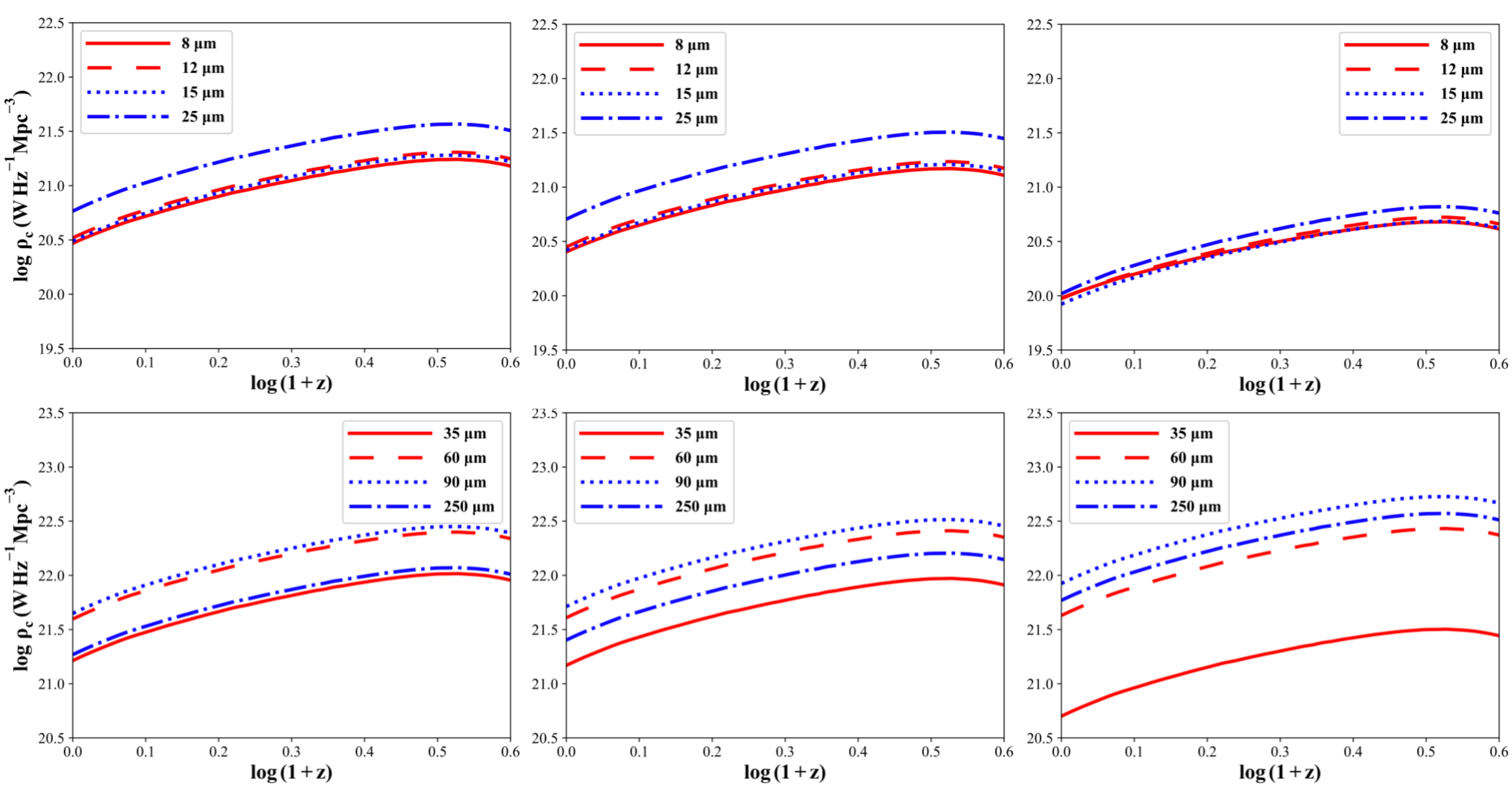}
  \caption{EBL luminosity density (i.e. emissivity) as a function of redshift predicted by equation \ref{eq:j_dust}, using the best-fit fractions for the 3-grain cases. Left: fit with power-law intrinsic spectrum; middle: log-parabola; right: power-law with cutoff.}
  \label{fig:emissivity}
\end{figure}

We would like to mention that the bolometric intensities associated to the best-fits of tables \ref{table:results_eblmodels+pl}, \ref{table:results_eblmodels+log-par} and \ref{table:results_eblmodels+pl+expcut} are around $I_{bol}=48.0$ nW m$^{-2}$ sr$^{-1}$, with variations in the first digit, since the stellar component is fixed and the broad range of redshifts over which the integration is performed dilutes the temperature dependence of $I_{bol}$. For comparison, Finke et al. has $I_{bol}=46.8$ nW m$^{-2}$ sr$^{-1}$. Therefore, the best-fits found here correspond to conservative estimates of the EBL contribution, since the bolometric intensities mentioned are very close the direct galaxy counts lower bounds (see figure \ref{fig:ebl_spectrum}). It is also interesting to compare measurements of the luminosity density with the predictions from formula \ref{eq:j_dust} in the IR range using the best-fit fractions obtained here. In \cite{Scully:2014wpa} and \cite{Stecker:2016fsg}, for example, empirical methods were developed to extract the EBL luminosity density as a function of redshift over a broad range of wavelengths, all the way from the Lyman limit to the far-IR (850 $\mu$m). Figure \ref{fig:emissivity} shows the redshift evolution of the luminosity density at different wavelengths for the 3-grain scenarios obtained with different Mkn 501 intrinsic spectrum parameterizations. For the 68\% confidence level bands presented in \cite{Scully:2014wpa,Stecker:2016fsg}, the curves agree within 1-2$\sigma$ with the measurements.

\section{Global fit properties for an extended sample of gamma-ray sources}
\label{global}

\begin{figure}[h]
  \centering
  \includegraphics[width=12cm]{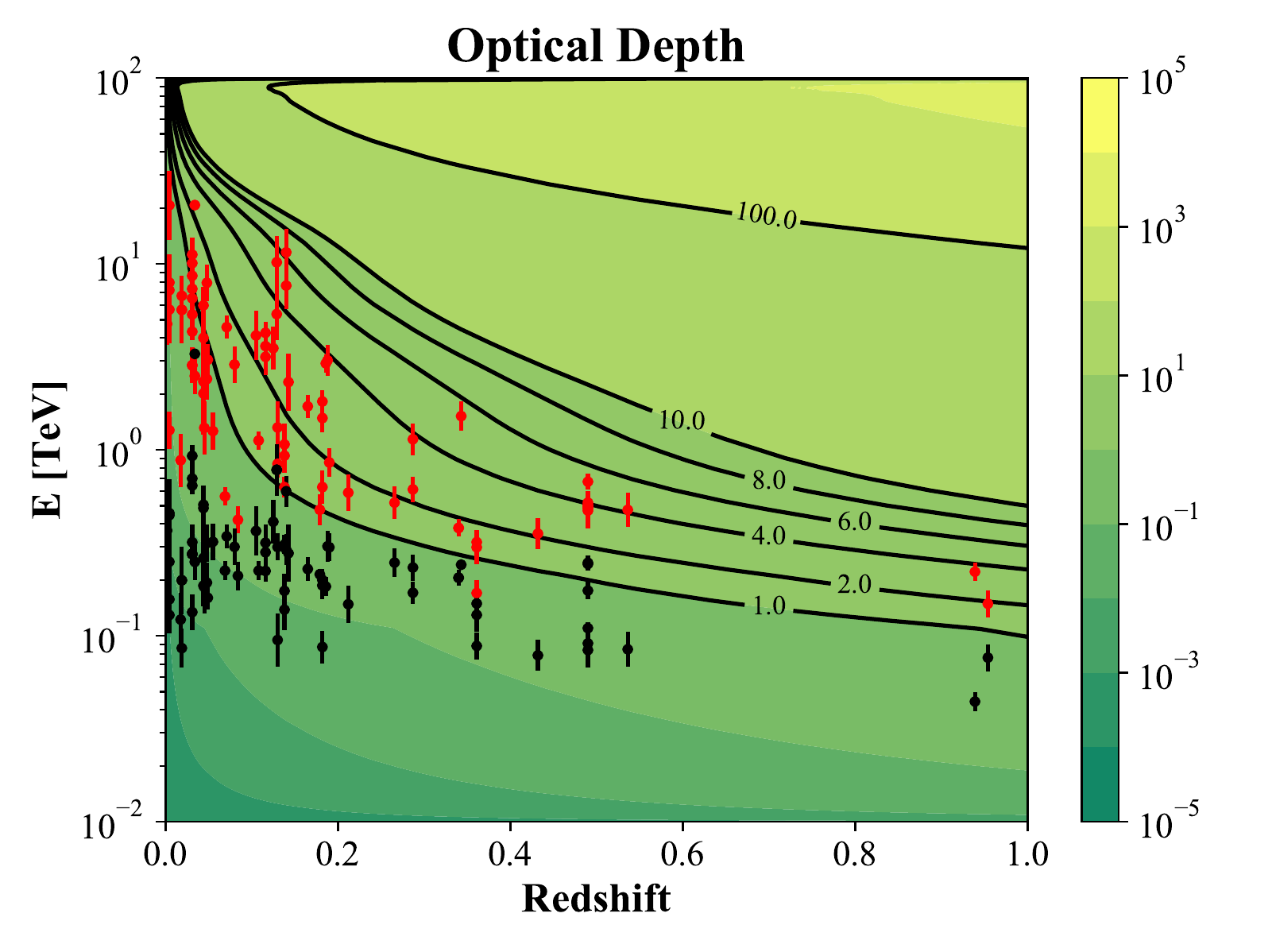}
  \caption{Heat map showing the optical depth to gamma-rays according to Finke et al. model in the $E_\gamma \times z$ parameter space. The lowest (black) and highest (red) energy bins for each observation shown in tables \ref{table:gamma-ray-srcs-1} and  \ref{table:gamma-ray-srcs-2} are superimposed to the plot. Black curves at specific values of $\tau$ are shown. The curve corresponding to $\tau=1$ is defined as the cosmic gamma-ray horizon (CGRH).}
  \label{fig:cgrh_finke_all_srcs}
\end{figure}

In this section, we describe some tests performed to compare two EBL scenarios: the model by Finke et al. with its nominal dust fractions and the same star+dust model with dust fractions tuned using Mkn\,501 measured SED presented in the last section. The procedure adopted also had as an objective to go from a single source analysis, as presented in the previous sections, to an extended sample of gamma-ray sources. The fits performed in this section will only vary the parameters of the intrinsic spectra, the dust fractions being either the Finke et al. nominal ones or those tuned to Mkn 501 in the 3-grain cases. We started by pre-selecting a sample of extragalactic gamma-ray sources from the TeVCat catalog \cite{2008ICRC:tevcat}. From this initial sample, we were able to collect in the literature 78 spectra of 41 different sources, all of them observed by IACTs. Tables \ref{table:gamma-ray-srcs-1} and \ref{table:gamma-ray-srcs-2} summarize important information on the spectra used.  The observations listed in this table were made at energies where current EBL models indicate non-negligible attenuation effects. This can be better appreciated when the lowest and highest energy bins of each observation are superimposed to the optical depth map of Finke et al. model in the 2D-parameter space $E_\gamma \times z$, as shown in figure \ref{fig:cgrh_finke_all_srcs}. One can see for a large number of observations, the highest energy bin lying above the curve corresponding to optical depth $\tau=1$, the so called cosmic gamma-ray horizon (CGRH) \cite{Dominguez:2013lfa}.

\begin{table*}[h]
\scriptsize
\setlength{\tabcolsep}{9pt}
\begin{center}
\begin{tabular}{ccccccc}
\hline
\hline
Name & Redshift & Type & Survey & Period of Observ. & Reference \\
\hline
1ES 0229+200 & 0.14 & BL Lac & HESS & 2005-2006 &\cite{aharonian2007new} \\
& & & VERITAS & 2010-2012 &  \cite{aliu2014three} \\
1ES 0347-121 & 0.188 & BL Lac & HESS & 2006 &  \cite{aharonian2007discovery} \\
1ES 0414+009 & 0.287 & BL Lac & HESS & 2005-2009 &  \cite{abramowski2012discovery_1es0414} \\
& & & VERITAS & 2008-2011 &  \cite{aliu2012multiwavelength} \\
1ES 0806+524 & 0.138 & BL Lac & MAGIC & 2011 & \cite{aleksic2015magic} \\
& & & VERITAS & 2006-2008 & \cite{acciari2008discovery} \\
1ES 1011+496 & 0.212 & BL Lac & MAGIC & 2007 & \cite{albert2007discovery} \\
1ES 1101-232 & 0.186 & BL Lac & HESS & 2004-2005 & \cite{aharonian2006low} \\
1ES 1215+303 & 0.13 & BL Lac & MAGIC & 2011 & \cite{aleksic2015magic} \\
& & & VERITAS & 2011 & \cite{aliu2013long} \\
1ES 1218+304 & 0.182 & BL Lac & VERITAS & 2008-2009 & \cite{acciari2010discovery_1es1218} \\
& & & VERITAS & 2007 & \cite{acciari2009veritas} \\
& & & MAGIC & 2005 & \cite{albert2006discovery_1es1218} \\
1ES 1312-423 & 0.105 & BL Lac & HESS & 2004-2010 & \cite{hess2013hess} \\
1ES 1727+502 & 0.055 & BL Lac & VERITAS & 2013 &  \cite{archambault2015veritas} \\
1ES 1741+196 & 0.084 & BL Lac & VERITAS & 2009-2014 & \cite{abeysekara2016veritas} \\
1ES 1959+650 & 0.048 & BL Lac & VERITAS & 2007-2011 & \cite{aliu2013multiwavelength} \\
& & & MAGIC & 2006 & \cite{tagliaferri2008simultaneous} \\
1ES 2344+514 & 0.044 & BL Lac & VERITAS & 2007-2008 & \cite{acciari2011multiwavelength} \\
& & & & 2007 &  \cite{acciari2011multiwavelength} \\
& & & MAGIC & 2008 & \cite{aleksic2013simultaneous} \\
& & & MAGIC & 2005-2006 & \cite{albert2007observation} \\
1RXS J101015.9 & 0.142639 & BL Lac & HESS & 2006-2010 &  \cite{abramowski2012discovery_1rxsj101015} \\
3C 279 & 0.5362 & FSRQ & MAGIC & 2008 &  \cite{albert2008very_3c279} \\
3C66A & 0.34 & BL Lac & VERITAS & 2008 & \cite{abdo2010multi} \\
4C+2135 & 0.432 & FSRQ & MAGIC & 2010 & \cite{aleksic2011magic} \\
AP Librae & 0.049 & BL Lac & HESS & 2010-2011 &  \cite{abramowski2015high} \\
BL Lacertae & 0.069 & BL Lac & VERITAS & 2011 &  \cite{arlen2012rapid} \\
Centaurus A & 0.00183 & FR I & HESS & 2004-2008 &  \cite{aharonian2009discovery} \\
H 1426+428 & 0.129 & BL Lac & HEGRA & 1999-2000  & \cite{aharonian2003observations} \\
& & & & 2002 &  \cite{aharonian2003observations} \\
H 2356-309 & 0.165 & BL Lac & HESS & 2004-2007 &  \cite{abramowski2010multi} \\
IC 310 & 0.0189 & BL Lac & MAGIC & 2012 &  \cite{aleksic2014black} \\
& & & & 2009-2010 &  \cite{aleksic2014rapid} \\
M87 & 0.0044 & FR I & HESS & 2005 & \cite{aharonian2006fast} \\
& & & & 2004 & \cite{aharonian2006fast} \\
& & & MAGIC & 2005-2007 & \cite{aleksic2012magic} \\
& & & & 2008 &  \cite{albert2008very_m87} \\
& & & VERITAS & 2007 &  \cite{acciari2008observation} \\
Markarian 180 & 0.045 & BL Lac & MAGIC & 2006 &  \cite{albert2006discovery_mkn180} \\
Markarian 421 & 0.031 & BL Lac & MAGIC & 2004-2005 & \cite{albert2007observations} \\
& & & & 2006 & \cite{acciari2009simultaneous} \\
& & & VERITAS & 2008 &  \cite{acciari2011tev} \\
Markarian 501 & 0.034 & BL Lac & HEGRA & 1997 & \cite{aharonian2001reanalysis} \\
& & & VERITAS & 2009 & \cite{acciari2011spectral} \\
NGC 1275 & 0.017559 & FR I & MAGIC & 2009-2014 &  \cite{ahnen2016deep} \\
PG 1553+113 & 0.49 & BL Lac & VERITAS & 2010-2012  & \cite{aliu2015veritas} \\
& & & MAGIC & 2008 &  \cite{aleksic2010simultaneous} \\
& & & & 2006 &  \cite{albert2009magic} \\
& & & HESS & 2013-2014 &  \cite{abdalla2017gamma} \\
& & & HESS & 2005-2006 &  \cite{abramowski20152012} \\
& & & HESS & 2012 &  \cite{abramowski20152012} \\
\hline
\hline
\end{tabular}
\end{center}
\caption{Gamma-ray sources selected from TeVCat \cite{2008ICRC:tevcat}.}
\label{table:gamma-ray-srcs-1}
\end{table*}

\begin{table*}[h]
\scriptsize
\setlength{\tabcolsep}{10pt}
\begin{center}
\begin{tabular}{ccccccc}
\hline
\hline
Name & Redshift & Type & Survey & Period of Observ. & Reference \\
\hline
PKS 0301-243 & 0.2657 & BL Lac & HESS & 2009-2011 &  \cite{abramowski2013discovery} \\
PKS 0447-439 & 0.343 & BL Lac & HESS & 2009 &  \cite{abramowski2013discoveryPKS0447} \\
PKS 1441+25 & 0.939 & FSRQ & MAGIC & 2015 &  \cite{ahnen2015very} \\
PKS 1510-089 & 0.361 & FSRQ & HESS & 2009 &  \cite{abramowski2013hess} \\
& & & MAGIC & 2015-PeriodA &  \cite{ahnen2017multiwavelength} \\
& & & & 2015-PeriodB &  \cite{ahnen2017multiwavelength} \\
PKS 2005-489 & 0.071 & BL Lac & HESS & 2004-2007 &  \cite{acero2010pks} \\
PKS 2155-304 & 0.116 & BL Lac & HESS & 2006 &  \cite{abramowski2013constraints} \\
& & & & 2005-2007 &  \cite{abramowski2010vhe} \\
& & & MAGIC & 2006  & \cite{aleksic2012high} \\
RBS 0413 & 0.19 & BL Lac & VERITAS & 2009 &  \cite{aliu2012discovery} \\
RGB J0152+017 & 0.08 & BL Lac & HESS & 2007 &  \cite{aharonian2008discovery} \\
RGB J0710+591 & 0.125 & BL Lac & VERITAS & 2008-2009 &  \cite{acciari2010discovery_rgbj0710} \\
RX J0648.7+1516 &	0.179 & BL Lac & VERITAS & 2010 &  \cite{aliu2011multiwavelength} \\
S3 0218+35 & 0.954 & FSRQ & MAGIC & 2014 &  \cite{ahnen2016detection} \\
VER J0521+211 & 0.108 & BL Lac & VERITAS & 2009-2010 &  \cite{archambault2013discovery} \\
\hline
\hline
\end{tabular}
\end{center}
\caption{Continuation of Table \ref{table:gamma-ray-srcs-1}.}
\label{table:gamma-ray-srcs-2}
\end{table*}

\begin{figure}[h]
  \centering
  \includegraphics[width=12cm]{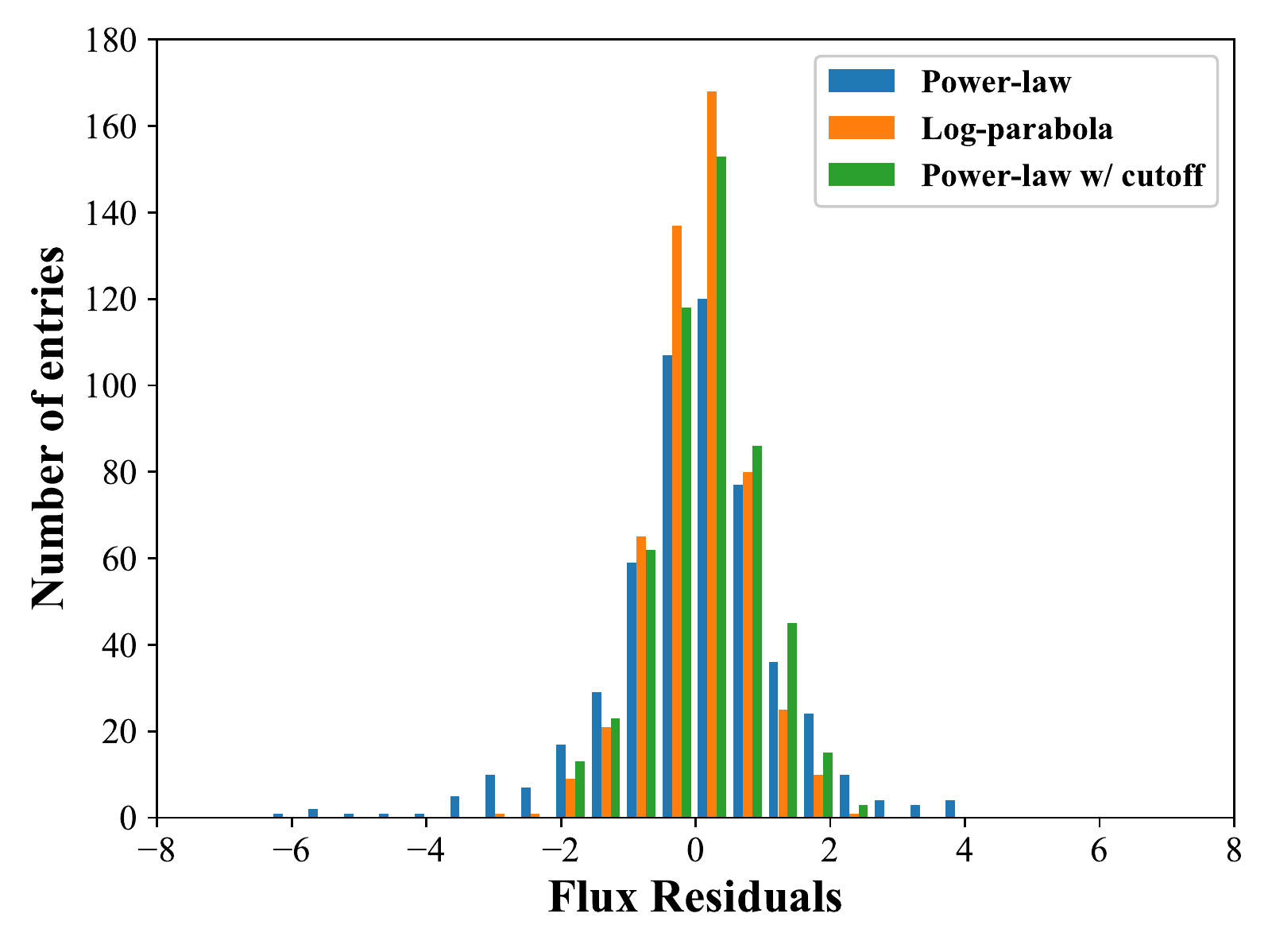}
  \caption{Distributions of flux residuals for an EBL attenuation model based on the nominal dust fractions of Finke et al.. Each histogram correspond to a different intrinsic spectrum parameterization.}
  \label{fig:fluxres_3intr+F10}
\end{figure}

\begin{figure*}[ht]
\begin{center}
\subfigure{
  \includegraphics[width=.33\textwidth]{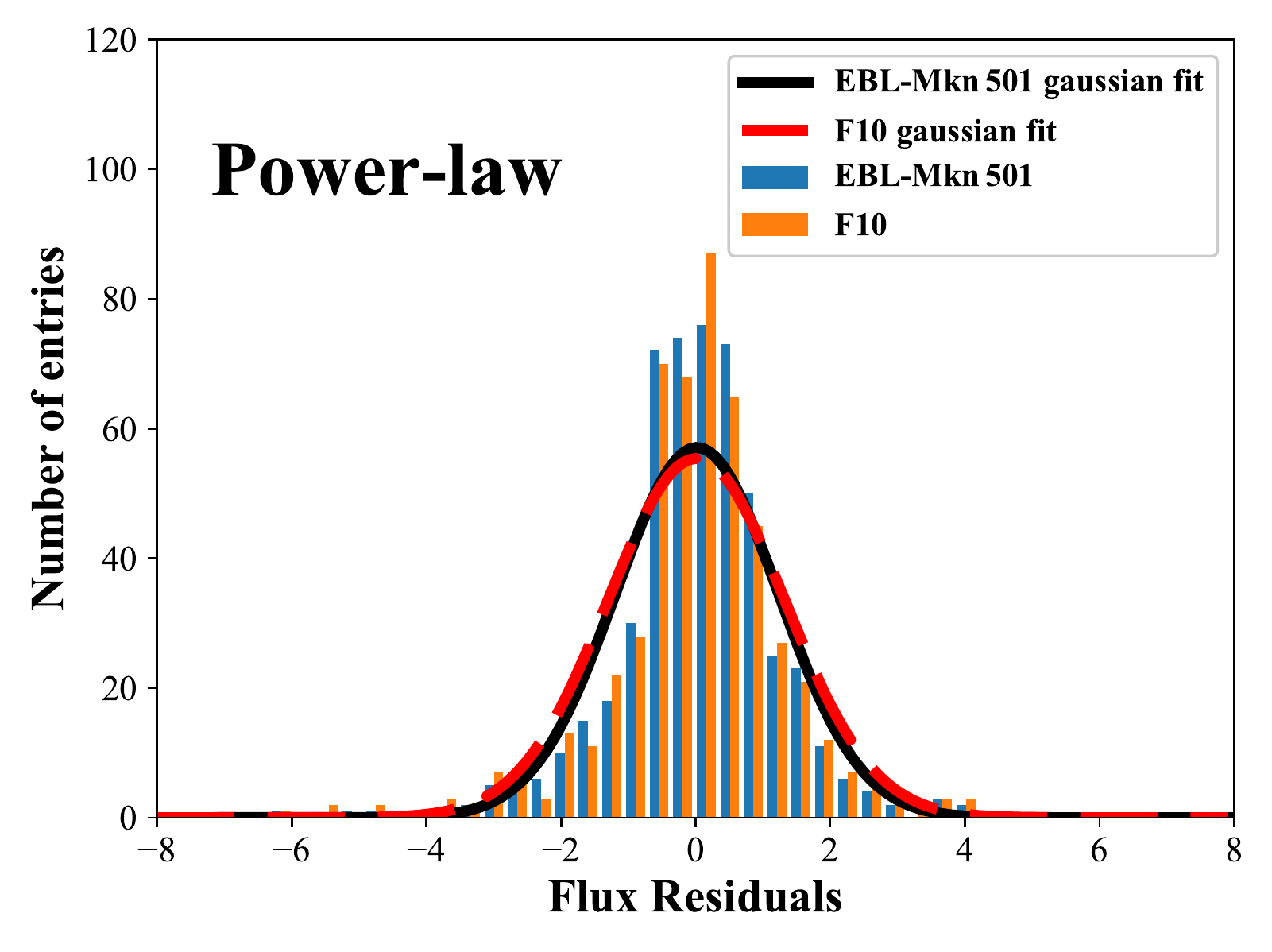}
}%
\subfigure{
  \includegraphics[width=.33\textwidth]{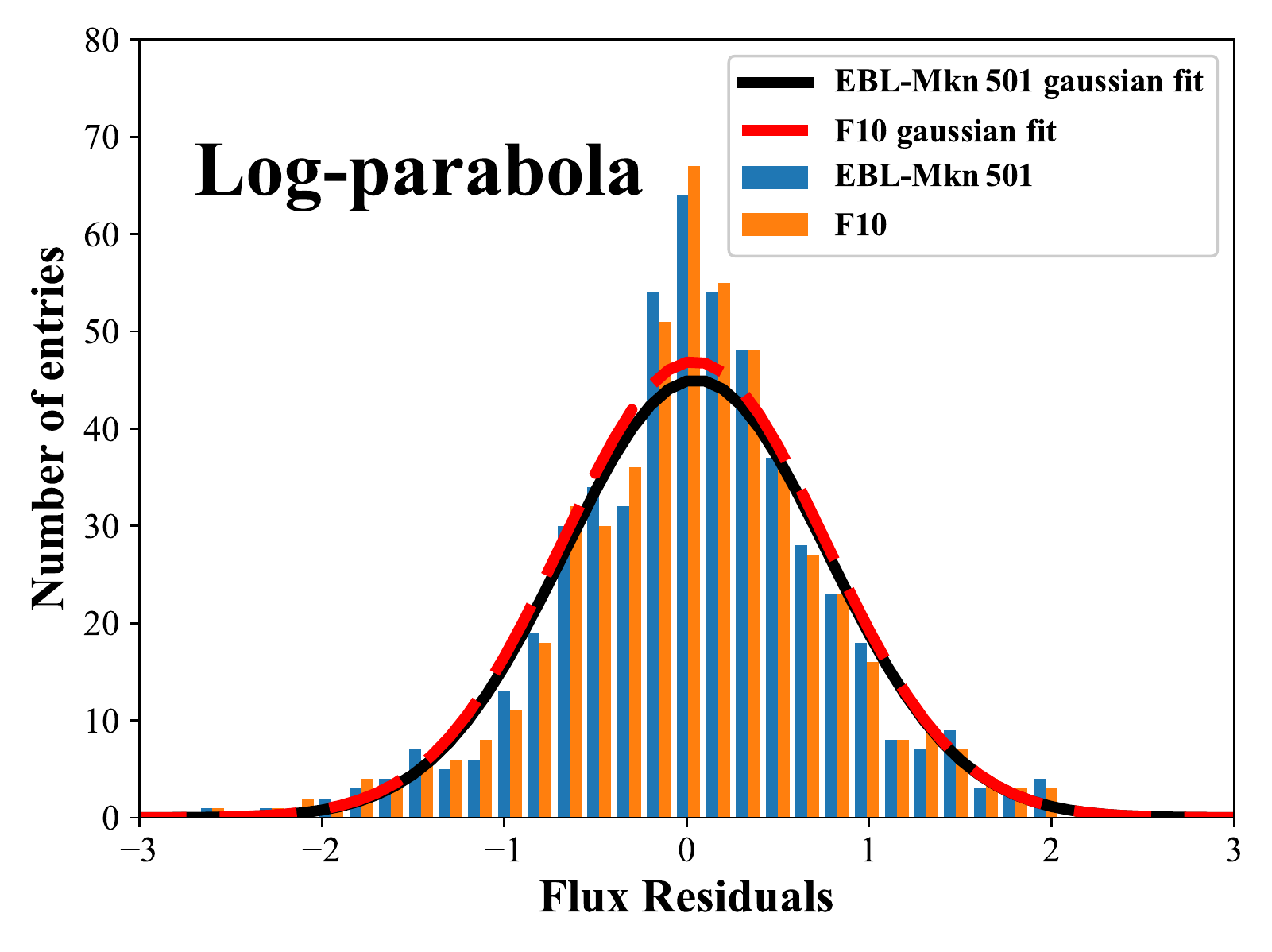}
}%
\subfigure{
  \includegraphics[width=.33\textwidth]{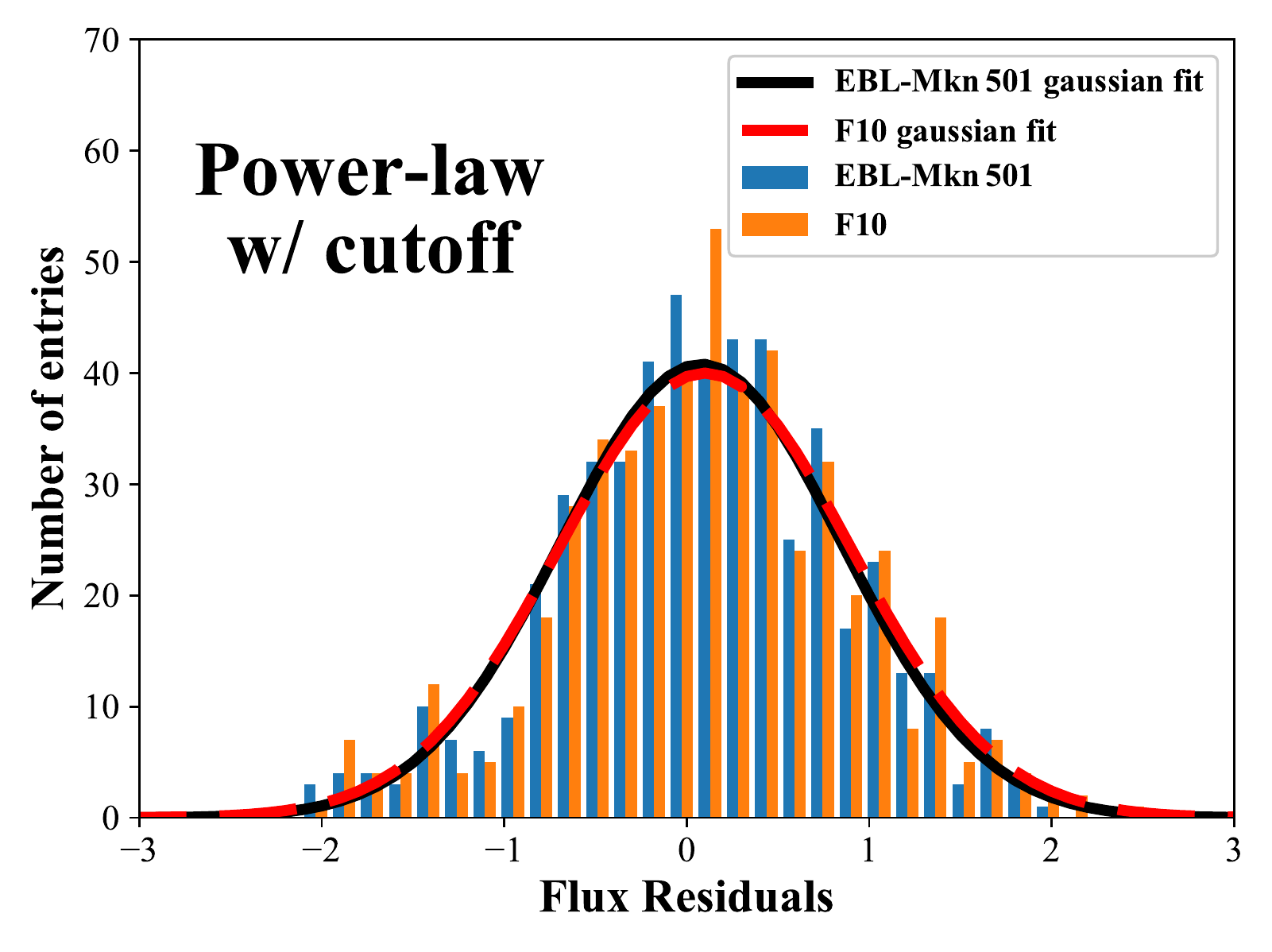}
}%
\end{center}
 \caption{Distributions of flux residuals for different combinations of EBL attenuation and intrinsic spectrum. At each plot, two distributions of residuals are shown: one for the nominal fractions scenario and another for the tuned fractions case. (Left) power-law; (Center) log-parabola; (Right) power-law with exponential cutoff.}
  \label{fig:res_fluxes}
\end{figure*}

With this sample of 78 SEDs, we have studied the distribution of residuals obtained when the measured spectrum ($\Phi_i$) at energy bin $i$ is compared to the predicted flux at Earth ($\Phi(E)=e^{-\tau}\Phi_0 $, for different combinations of intrinsic spectrum convoluted with an EBL attenuation factor as given by equation \ref{eq:flux_params}), taking into account the uncertainties on the measured flux $\Phi_i$ ($\sigma_i$):
\begin{equation}
\textrm{Flux residual}_i \equiv \frac{\Phi_i - \Phi(E_i)}{\sigma_i},
\end{equation}
which, defined in this way, are expected to follow a normal distribution with zero mean and unit variance when the errors $\sigma_i$ are Gaussian and the model $\Phi(E)$ appropriately describes the measurements $\Phi_i$. Figure \ref{fig:fluxres_3intr+F10} shows the distributions of the residuals (one entry for each energy bin of the 72 \footnote{In order to ensure that all the fits have at least one degree of freedom, six spectra with just three energy bins measured were excluded from the sample.} independent SED observations) for different intrinsic spectrum parameterizations and an EBL attenuation using Finke et al. nominal fractions. One can see a clear improvement in the description of the measurements when log-parabola or a spectrum with a cutoff is used due, of course, to the extra parameter present in these parameterizations which can even absorb part of the EBL attenuation effects that could be imprinted in the SED. The improvement can be seen even by eye in the reduction of differential flux outliers (compared the the power-law case) when these two spectra are used. For a more quantitative analysis, Gaussian fits to the residual distributions were performed and the results are summarized in table \ref{table:mean_sigma} (columns labeled as ``nominal fractions''). The distributions of figure \ref{fig:fluxres_3intr+F10}, when fitted with a log-parabola or a spectrum with a cutoff, have reduced $\chi^2$ closer to unity when compared to the PL case. Additional tests were made by fixing the intrinsic spectrum and comparing the nominal fractions scenario with the Mkn 501-tuned one. The corresponding distribution of residuals can be seen in figure \ref{fig:res_fluxes} for all 3 intrinsic spectra. The Gaussian fit results in table \ref{table:mean_sigma} show again that, after tuning the fractions, power-law gives the worse reduced $\chi^2$ among the three spectra. However, except for the power-law case, the differences in the reduced $\chi^2$ between the two sets of dust fractions analyzed are small.
\renewcommand{\arraystretch}{1.2}
\begin{table*}[h!]
\small
\setlength{\tabcolsep}{6pt}
\begin{center}
\begin{tabular}{ccccccc}
\hline
\hline
& \multicolumn{3}{ |c }{nominal fractions} & \multicolumn{3}{|c}{tuned fractions} \\
\hline
\hline
& $\mu$ & $\sigma$ & $\chi^2$/dof & $\mu$  & $\sigma$ & $\chi^2$/dof   \\
\hline
PL  & $-0.01 \pm 0.08$ & $1.31 \pm 0.06$ & 61.23 & $0.02 \pm 0.08$ & $ 1.21 \pm 0.05 $ & 97.79 \\
LP   & $0.04 \pm 0.05$ & $ 0.72 \pm 0.03 $ & 1.91 & $0.05 \pm 0.05 $ & $ 0.72 \pm 0.03 $ & 1.96 \\
PLC & $0.10 \pm 0.05$ & $ 0.80 \pm 0.04 $ & 1.87  & $0.08 \pm 0.05 $ & $ 0.77 \pm 0.04 $ & 1.47 \\
best spec. (PL) & \multirow{ 3}{*}{$0.05 \pm 0.04$}  & \multirow{ 3}{*}{$0.71 \pm 0.03$} & \multirow{ 3}{*}{2.46}   & $0.06 \pm 0.04 $ & $ 0.71 \pm 0.03 $ & 2.25 \\
best spec. (LP) &  &  &   & $0.05 \pm 0.04 $ & $ 0.71 \pm 0.03 $ & 2.39 \\
best spec. (PLC) &  &  &   & $0.05 \pm 0.04 $ & $ 0.71 \pm 0.03 $ & 1.31 \\
$r>0.8$ (PL) & $0.11 \pm 0.09$ & $ 1.24 \pm 0.06 $  & 70.17 & $0.08 \pm 0.08 $ & $ 1.18 \pm 0.06 $ & 44.74 \\
$r\le0.8$ (PL) & $-0.40 \pm 0.40$ & $ 1.80 \pm 0.30 $ & 2.12 & $-0.30 \pm 0.30 $ & $ 1.50 \pm 0.20 $ & 9.18 \\
$r>0.8$ (LP) & $0.03 \pm 0.05$ & $ 0.70 \pm 0.04 $  & 1.65 & $0.03 \pm 0.05 $ & $ 0.70 \pm 0.04 $ & 1.91 \\
$r\le0.8$ (LP) & $0.10 \pm 0.20$ & $ 0.90 \pm 0.10 $ & 1.22 & $0.10 \pm 0.10 $ & $ 0.90 \pm 0.10 $ & 1.35 \\
$r>0.8$ (PLC) & $0.06 \pm 0.06$ & $ 0.79 \pm 0.04 $  & 1.77 & $0.06 \pm 0.05 $ & $ 0.77 \pm 0.04 $ & 1.35 \\
$r\le0.8$ (PLC) & $0.40 \pm 0.20$ & $ 1.00 \pm 0.20 $ & 1.13 & $0.30 \pm 0.20 $ & $ 0.90 \pm 0.10 $ & 0.45 \\
\hline
\hline
\end{tabular}
\end{center}
\caption{Mean, standard deviation and reduced $\chi^2$ of Gaussian fits to the distributions of residuals for SED fits performed with different dust fractions and blazar intrinsic spectra: PL (power-law), LP (log-parabola) and PLC (power-law with cutoff).}
\label{table:mean_sigma}
\end{table*}
\renewcommand{\arraystretch}{1}

In order to disentangle, at least partially, intrinsic spectrum effects from the EBL attenuation ones, we finally performed two additional tests. Firstly, a comparison was made based on the approach adopted in \cite{biteau2015extragalactic}, where the fit residuals for the nominal and Mkn\,501-tuned fractions were calculated for each source using the intrinsic spectrum that lead to the best fit quality (more precisely, the largest $P(>\chi^2)$ for a given ndof). The same two scenarios for the set of dust fractions are compared. The three lines of table \ref{table:mean_sigma}  identified by the label ``best spec.'' summarize the fit results in these cases, each line representing one of the three sets of dust fractions, depending on the spectrum parameterization used during the tuning procedure. The reduced $\chi^2$s for the scenario of tuned fractions are slightly smaller than the nominal fractions case.
\begin{figure}[h]
  \centering
  \includegraphics[width=10cm]{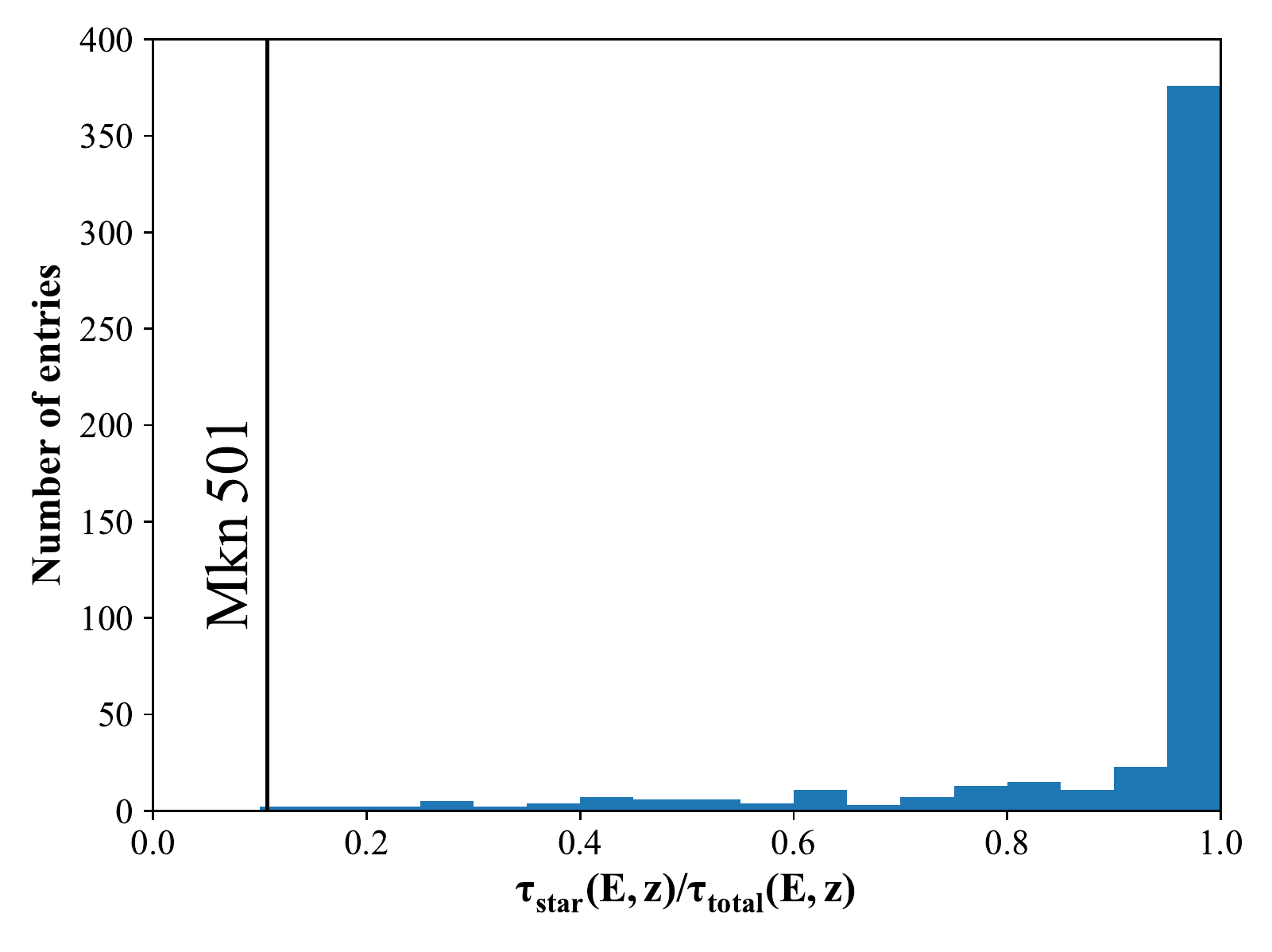}
  \caption{Distribution of the optical depth estimator $\tau_{star}/\tau_{total}$ (see text) for all the SED bins of the sample of blazar spectra shown in tables \ref{table:gamma-ray-srcs-1} and \ref{table:gamma-ray-srcs-2}. The vertical line correspond to the highest energy bin of Mkn$\,$501.}
  \label{fig:dust_estimator}
\end{figure}

\begin{figure*}[ht]
\begin{center}
\subfigure{
  \includegraphics[width=.49\textwidth]{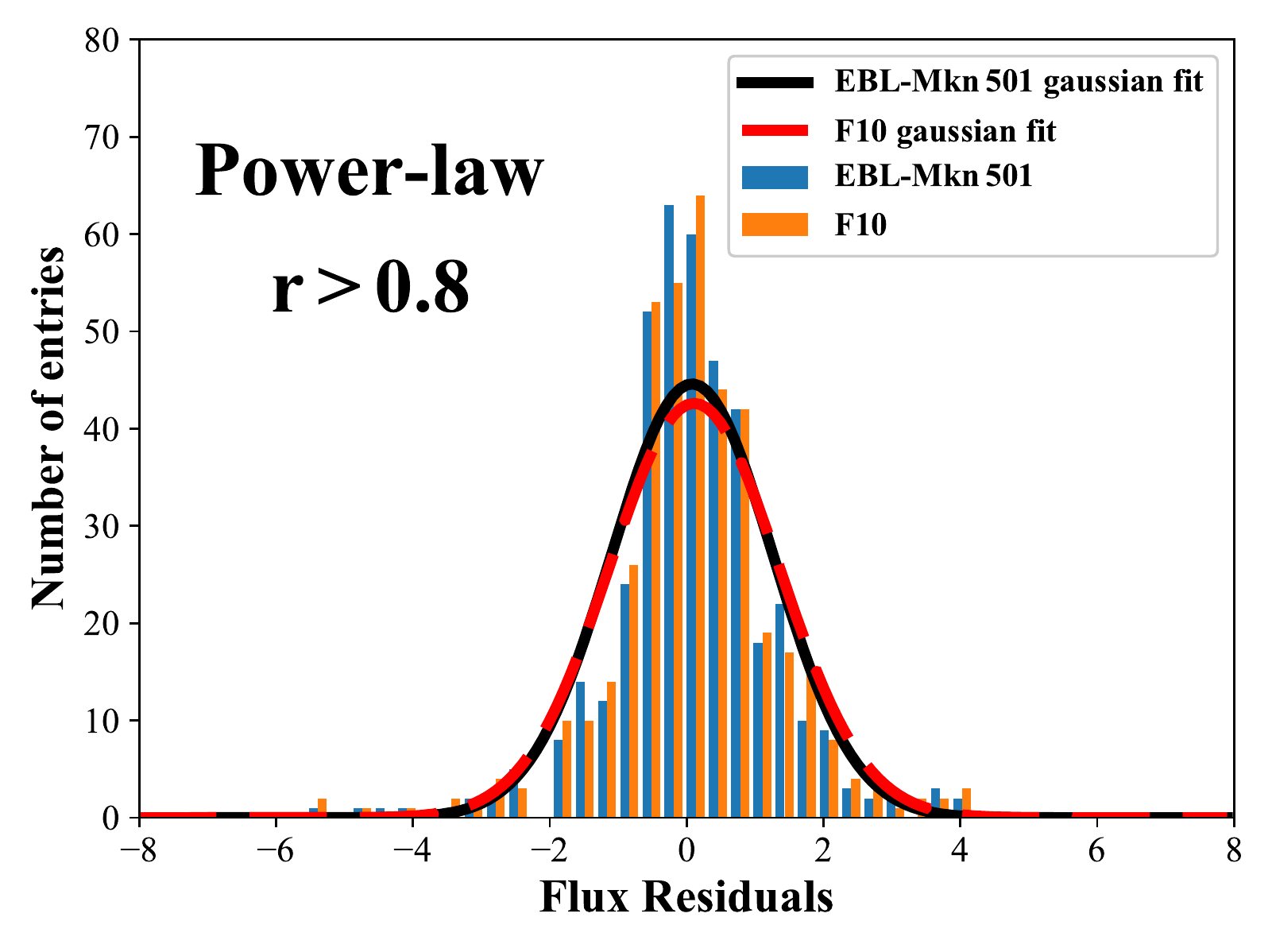}
}%
\subfigure{
  \includegraphics[width=.49\textwidth]{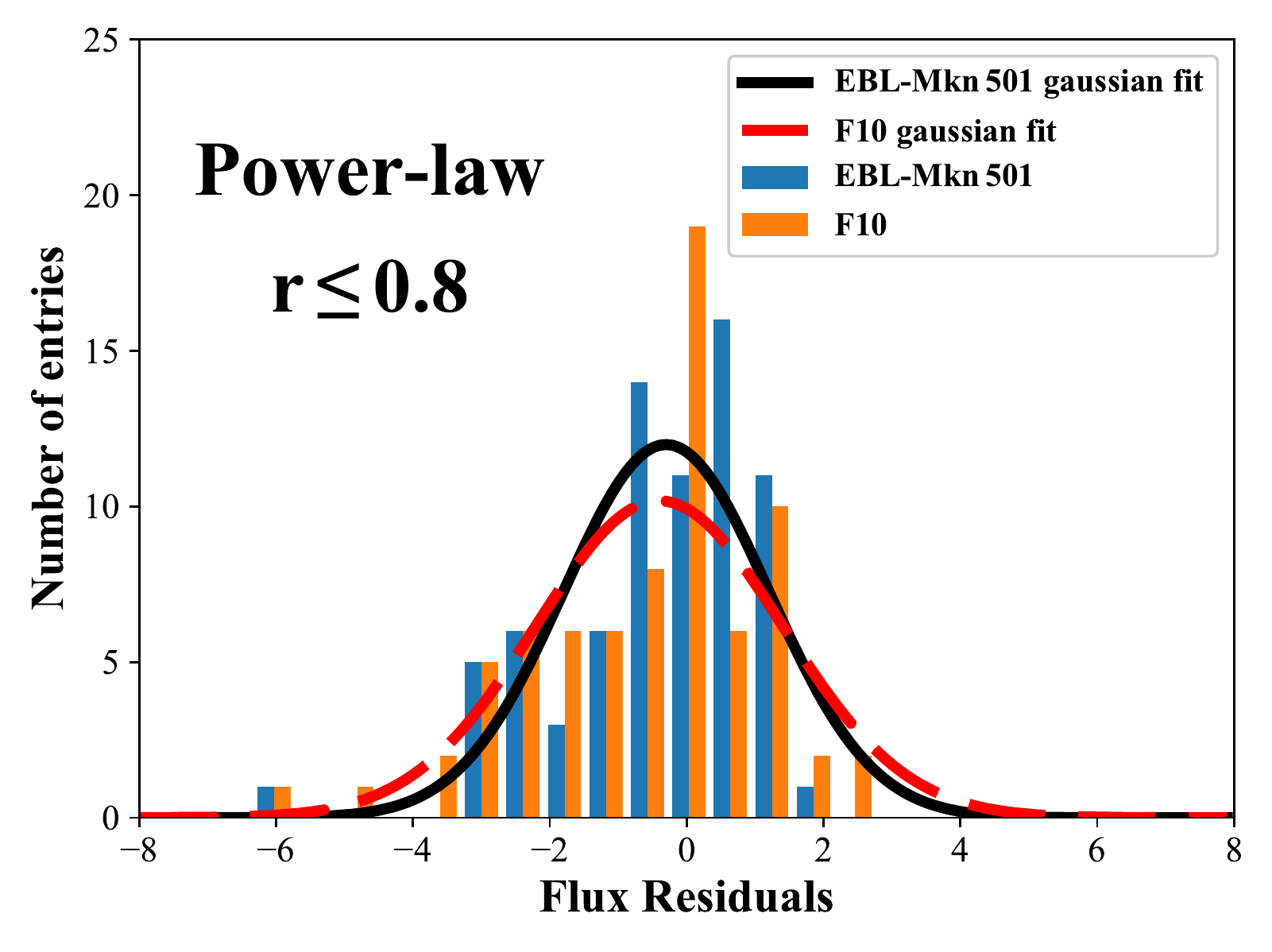}
} \\ 
\subfigure{
  \includegraphics[width=.49\textwidth]{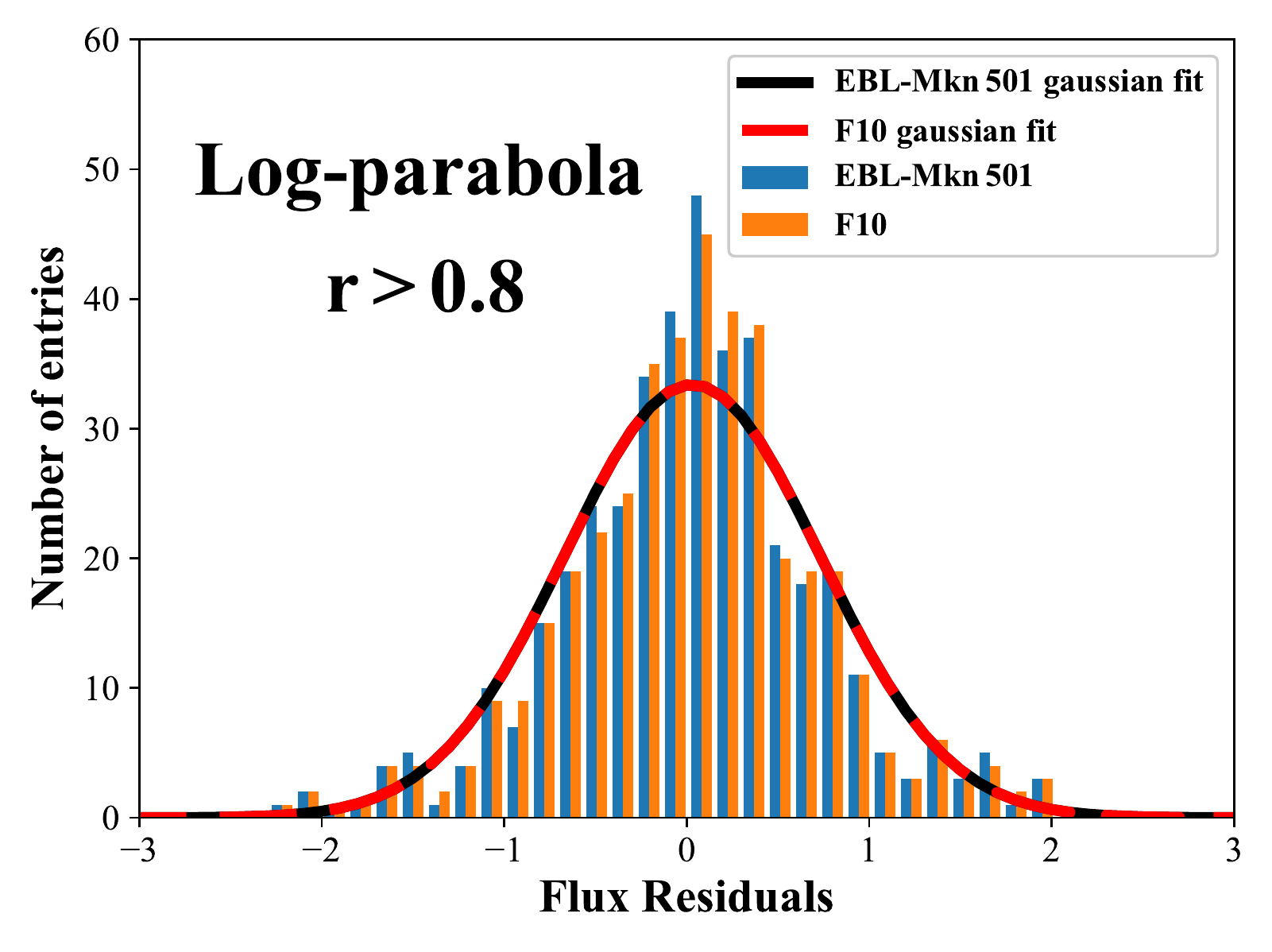}
}%
\subfigure{
  \includegraphics[width=.49\textwidth]{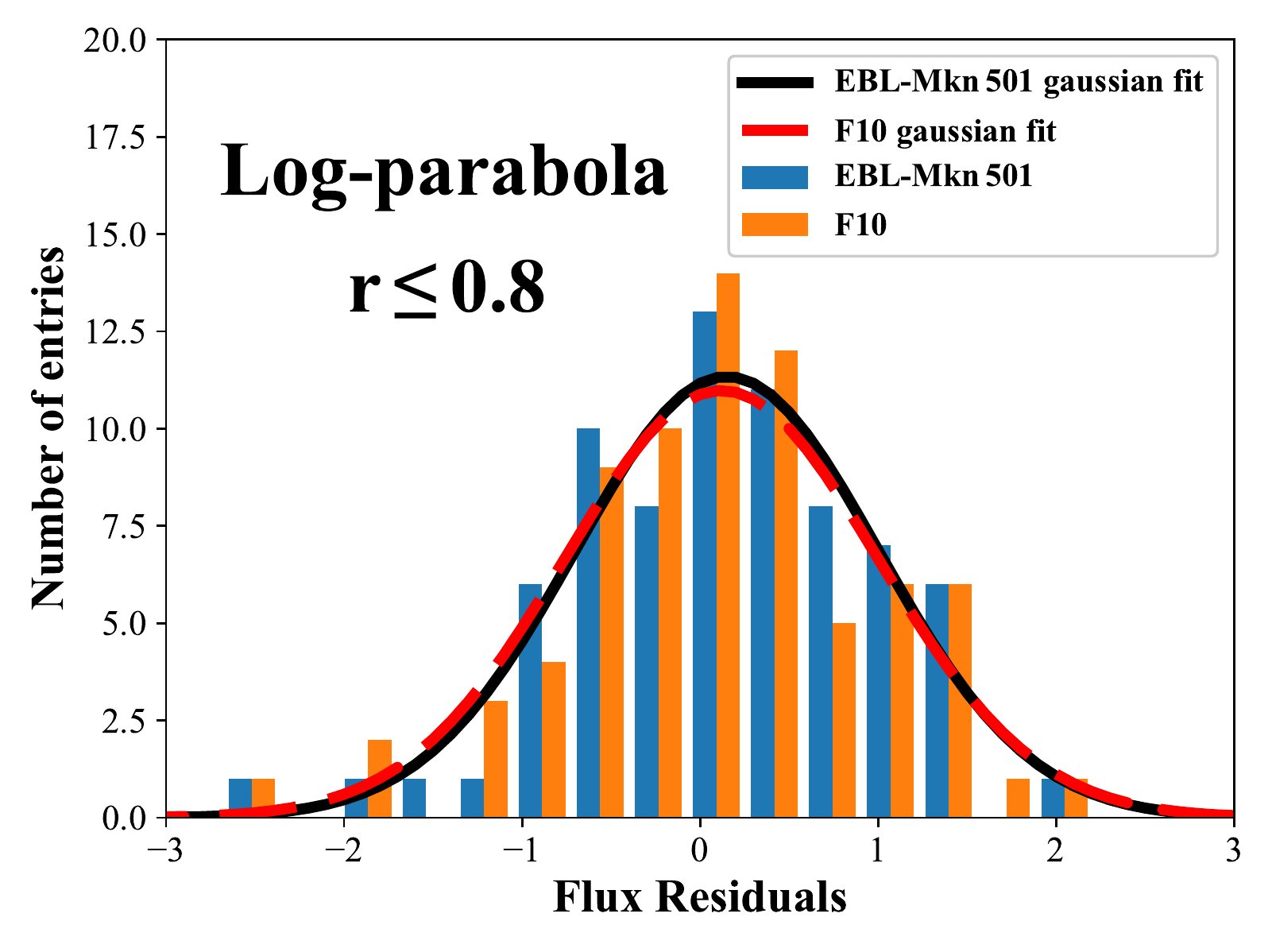}
} \\ 
\subfigure{
  \includegraphics[width=.49\textwidth]{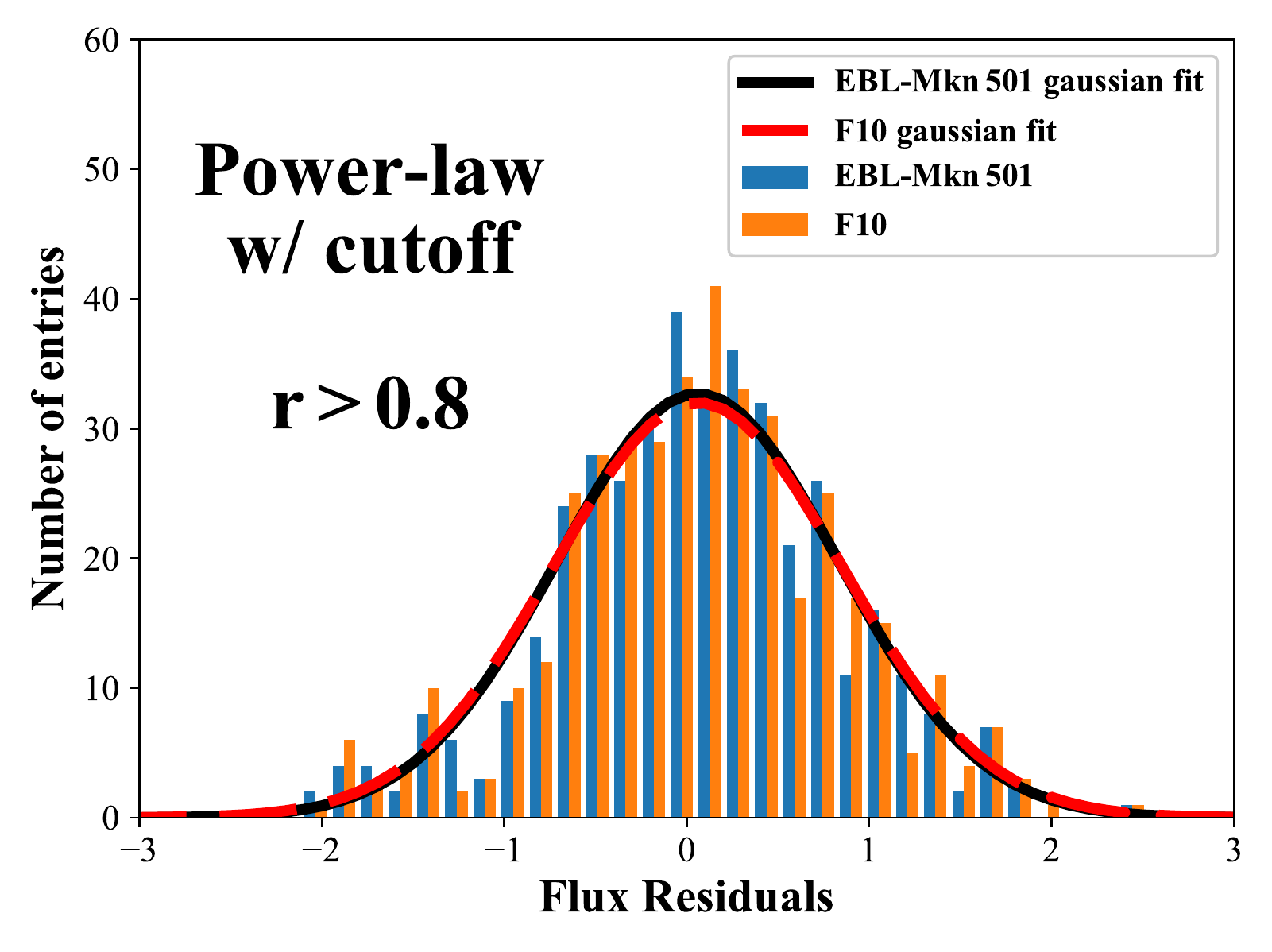}
}%
\subfigure{
  \includegraphics[width=.49\textwidth]{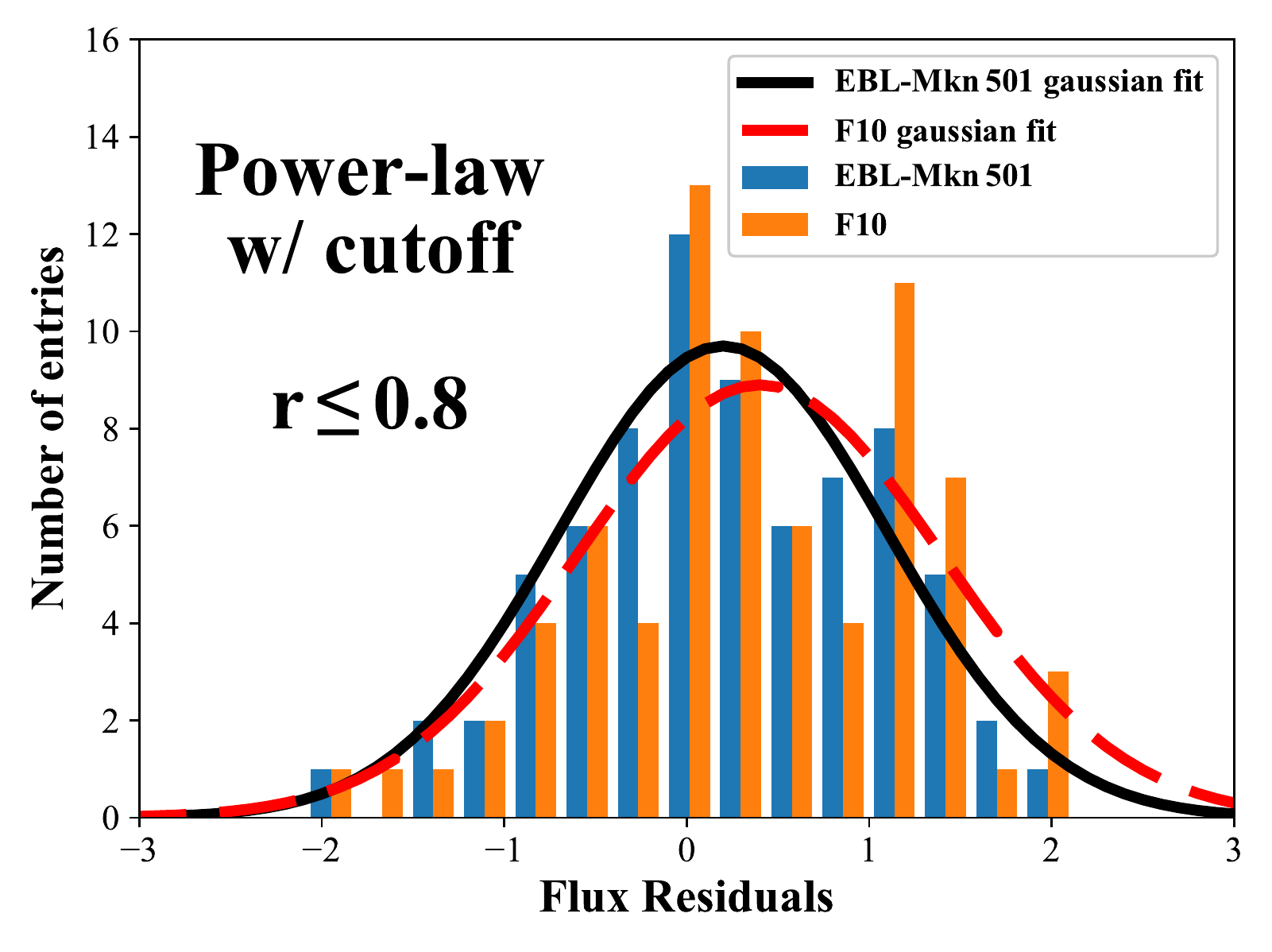}
} \\ 
\end{center}
 \caption{Distributions of flux residuals for two different populations of SED bins and different blazar intrinsic spectra. Left: plots corresponding to the bins dominated by dust attenuation ($\tau_{star}/\tau_{total} \leq 0.8$) according to the estimator of equation  \ref{eq:estimator}. Right: bins dominated by stellar attenuation ($\tau_{star}/\tau_{total}>0.8$). At each plot, two histograms are compared: nominal fractions and tuned fractions.}
  \label{fig:res_fluxes_subset_star_dust}
\end{figure*}

In the second test, we have subdivided the blazar sample using an attenuation estimator to produce stellar- and dust-dominated SED bins. The distribution of the estimator, taken as the following predicted ratio of optical depths for a source at the same redshift $z$ of the blazar emitting photons of energy equal to the corresponding central value of the energy bin
\begin{equation}
r(E,z)=\frac{\tau_{star}(E,z)}{\tau_{star}(E,z)+\tau_{dust}(E,z)},
\label{eq:estimator}
\end{equation}
is shown in figure \ref{fig:dust_estimator} for the sample of 39 blazars \footnote{Two of the blazars in the input sample (CenA and M87) are too close that their optical depths due to starlight are negligible.}. In order to get the optical depths for equation \ref{eq:estimator}, we used the nominal dust fractions of Finke et al. model. We also verified that the use of the Mkn\,501 tuned fractions did not change the classification of the subsamples. The plot clearly shows that the current sample of blazars detected by IACTs is dominated by star attenuation. When the maximum energy of the measured SED is used, this estimator shows that Mkn 501 is the source with the highest expected level of dust attenuation in the sample. The corresponding residual distributions for the two subsamples ($r>0.8$ and $r\le 0.8$) are shown in figure \ref{fig:res_fluxes_subset_star_dust} once again for the three spectra and the two sets of dust fractions. The numbers summarized in table \ref{table:mean_sigma} do not indicate a uniform systematic change in the quality of fit when one goes from the nominal fractions to the tuned ones. The changes in the reduced $\chi^2$ show positive and negative variations depending on the intrinsic spectrum and on the range of $r$ analyzed.

Regarding the mean ($\mu$) and standard deviation ($\sigma$) of the Gaussian fits, excluding the cases with bad fit quality ($\chi^2/$ndof $\gtrsim 10$), we see that the mean values are consistent with zero at the 1- to 2-sigma level, whereas $\sigma$ can be up to 30\% smaller than the ideal case of unit variance. Such an effect could be due to an overestimation of the flux uncertainties, but investigating this is beyond the scope of this paper, since it would require extra information at the telescope and data processing levels.

\section{Conclusions}
\label{conclusions}

We have addressed here the issue of the contribution of dust emission at IR wavelengths to the opacity of the extragalactic medium. Using an existing EBL model based on the blackbody emission of stars and dust grains of different sizes and temperatures, we have been able to study separately the contribution of each grain type to the attenuation of TeV gamma-rays. With a single TeV source at redshift $z=0.034$, Mkn\,501, we showed that its measured SED has already some sensitivity to the relative contributions of different dust grains. The fit was performed for three different intrinsic spectrum parameterizations while the temperatures of the grains were kept fixed during all the fit procedure.

For this single source fit, some residual degeneracy is still present between the amount of some grains (small and large) and the curvature of the intrinsic spectrum. More specifically, when the intrinsic spectrum lacks curvature (power-law) or has an energy independent curvature (log-parabola), the competition between small and large grains at the very end of Mkn\,501 SED is won by the small ones due to their slightly harder attenuation factor. However, the flux suppression due to this dust component can be mimicked by some extra curvature of the blazar spectrum (like in the power-law with cutoff case). A nested likelihood ratio test was able to exclude the PAH-only scenario (with the temperature of the dust grains fixed a priori) at more than 5$\sigma$, for the attenuation of this dust component has an effective energy dependence in the form of a single spectral index over a broad energy range, therefore, being unable to account for the strong flux suppression of Mkn\,501 flare state SED seen above 10 TeV, even when there is an energy cutoff in the source spectrum. On the other hand, in the region just below 10 TeV, the presence of PAH molecules is essential for the attenuated spectrum to have an energy dependence consistent with the measured SED. Therefore, by separating the attenuation due to each EBL component, we can clearly see the potential of a precisely measured SED to constrain both the spectrum and EBL parameters.

The extension of this procedure from the single source level to a sample of well measured AGNs at different redshifts has the potential to put much stronger constraints on the same parameters, since many (if not all) of the mentioned degeneracies could be broken in that case, due to the increase in the number of degrees of freedom. A first step towards that goal was given here by checking the consistency of the EBL parameters tuned to Mkn\,501 in describing the attenuated spectra of a set of 78 SEDs from 41 different blazars selected from TeVCat. We have studied the distribution of SED fit residuals for several combinations of dust fraction sets and intrinsic spectra. By splitting the sample of blazar SED bins into stellar and dust attenuation dominated subsamples, we could not identify a uniform systematic change in the quality of Gaussian fits performed on these residual distributions when going from the nominal fractions to the tuned ones. This result is consistent with the fact that the current sample of blazars detected with IACTs is still dominated by starlight attenuation as shown by an apropriate estimator.

The next generation of IACTs, represented by CTA, is expected to discover a whole new sample of extragalactic AGNs at high redshifts due to its $\sim$10 factor enhancement in sensitivity. Its extended energy range, covering almost four decades from below 100 GeV to 100 TeV, will provide SEDs where the attenuation effects of all EBL components are expected to play some role: from the stellar one at visible and UV wavelengths (affecting the tens to hundreds of GeV region of the AGN spectrum) to the mid-IR and PAH-dominated range (attenuating the $\sim$ TeV region of the spectrum) to the far-IR region of small and large grains (important for the attenuation at very high energy tail of the spectrum around tens of TeV).

\ack

The authors acknowledge the CTA GPROPA task members for fruitful discussions, in special David Williams and Jonathan Biteau for the extremely relevant comments and suggestions. We are grateful for the valuable comments made by the referee that certainly improved the paper. We also thank the
use of computer time at CENAPAD-SP (Centro Nacional de Processamento de Alto Desempenho em S\~ao Paulo), projeto UNICAMP/FINEP-MCT. D.R.M.P. thanks Coordenadoria de Aperfei\c{c}oamento de Pessoal de Ensino Superior (CAPES). E.M.S. acknowledges Funda\c{c}\~ao de Amparo \`a Pesquisa do Estado de S\~ao Paulo (FAPESP) for BPE grant number 2016/19171-8 and Projeto Tem\'atico grant number 2015/15897-1. E.M.S. is also grateful to Conselho Nacional de Desenvolvimento Cient\'ifico e Tecnol\'ogico (CNPq) for grant number 310565/2015-4.

\section*{References}
\bibliography{bibliography}{}

\providecommand{\newblock}{}
\begin{thebibliography}{100}
\expandafter\ifx\csname url\endcsname\relax
  \def\url#1{{\tt #1}}\fi
\expandafter\ifx\csname urlprefix\endcsname\relax\def\urlprefix{URL }\fi
\providecommand{\eprint}[2][]{\url{#2}}

\bibitem{Gould:1967zza}
Gould R~J and Schreder G~P 1967 {\em Phys. Rev.\/} {\bf 155} 1408--1411

\bibitem{Gould:1967zzb}
Gould R~J and Schreder G~P 1967 {\em Phys. Rev.\/} {\bf 155} 1404--1407

\bibitem{Hauser:2001xs}
Hauser M~G and Dwek E 2001 {\em Ann. Rev. Astron. Astrophys.\/} {\bf 39}
  249--307 (\textit{Preprint} \eprint{astro-ph/0105539})

\bibitem{Acharya:2017ttl}
Acharya B~S {\em et~al.\/} (Cherenkov Telescope Array Consortium) 2017
  (\textit{Preprint} \eprint{1709.07997})

\bibitem{Primack:1998wn}
Primack J~R, Bullock J~S, Somerville R~S and MacMinn D 1999 {\em Astropart.
  Phys.\/} {\bf 11} 93--102 (\textit{Preprint} \eprint{astro-ph/9812399})

\bibitem{Somerville:2011kq}
Somerville R~S, Gilmore R~C, Primack J~R and Dominguez A 2012 {\em Mon. Not.
  Roy. Astron. Soc.\/} {\bf 423} 1992 (\textit{Preprint} \eprint{1104.0669})

\bibitem{Gilmore:2011ks}
Gilmore R~C, Somerville R~S, Primack J~R and Dominguez A 2012 {\em Mon. Not.
  Roy. Astron. Soc.\/} {\bf 422} 3189 (\textit{Preprint} \eprint{1104.0671})

\bibitem{Malkan:1997yd}
Malkan M~A and Stecker F~W 1998 {\em Astrophys. J.\/} {\bf 496} 13--16
  (\textit{Preprint} \eprint{astro-ph/9710072})

\bibitem{Stecker:2005qs}
Stecker F~W, Malkan M~A and Scully S~T 2006 {\em Astrophys. J.\/} {\bf 648}
  774--783 (\textit{Preprint} \eprint{astro-ph/0510449})

\bibitem{Franceschini:2008tp}
Franceschini A, Rodighiero G and Vaccari M 2008 {\em Astron. Astrophys.\/} {\bf
  487} 837 (\textit{Preprint} \eprint{0805.1841})

\bibitem{Finke:2009xi}
Finke J~D, Razzaque S and Dermer C~D 2010 {\em Astrophys. J.\/} {\bf 712}
  238--249 (\textit{Preprint} \eprint{0905.1115})

\bibitem{aharonian1999time}
Aharonian F (HEGRA) 1999 {\em Astron. Astrophys.\/} {\bf 349} 11--28
  (\textit{Preprint} \eprint{astro-ph/9903386})

\bibitem{eggleton1989distribution}
{Eggleton} P~P, {Fitchett} M~J and {Tout} C~A 1989 {\em Astrophys. J.\/} {\bf
  347} 998--1011

\bibitem{Bruzual:2003tq}
Bruzual G and Charlot S 2003 {\em Mon. Not. Roy. Astron. Soc.\/} {\bf 344} 1000
  (\textit{Preprint} \eprint{astro-ph/0309134})

\bibitem{baldry03}
Baldry I~K and Glazebrook K 2003 {\em The Astrophysical Journal\/} {\bf 593}
  258

\bibitem{Salpeter:1955it}
Salpeter E~E 1955 {\em Astrophys. J.\/} {\bf 121} 161--167

\bibitem{Krugel2003}
Kr\"ugel E 2003 {\em The Physics of Interstellar Dust\/} (Institute of Physics)

\bibitem{biteau2015extragalactic}
Biteau J and Williams D~A 2015 {\em The Astrophysical Journal\/} {\bf 812} 60

\bibitem{Peebles1971}
Peebles P~J~E 1971 {\em Physical Cosmology\/} (Princeton University Press)

\bibitem{Razzaque:2008te}
Razzaque S, Dermer C~D and Finke J~D 2009 {\em Astrophys. J.\/} {\bf 697}
  483--492 (\textit{Preprint} \eprint{0807.4294})

\bibitem{aharonian2001reanalysis}
Aharonian F, Akhperjanian A, Barrio J, Bernl{\"o}hr K, Bolz O, B{\"o}rst H,
  Bojahr H, Contreras J, Cortina J, Denninghoff S {\em et~al.\/} 2001 {\em
  Astronomy \& Astrophysics\/} {\bf 366} 62--67

\bibitem{blanton:2003galaxy}
Blanton M~R, Hogg D~W, Bahcall N~A, Brinkmann J, Britton M, Connolly A~J,
  Csabai I, Fukugita M, Loveday J, Meiksin A {\em et~al.\/} 2003 {\em The
  Astrophysical Journal\/} {\bf 592} 819

\bibitem{cole:20012df}
Cole S, Norberg P, Baugh C~M, Frenk C~S, Bland-Hawthorn J, Bridges T, Cannon R,
  Colless M, Collins C, Couch W {\em et~al.\/} 2001 {\em Monthly Notices of the
  Royal Astronomical Society\/} {\bf 326} 255--273

\bibitem{kochanek:2001k}
Kochanek C, Pahre M, Falco E, Huchra J, Mader J, Jarrett T, Chester T, Cutri R
  and Schneider S 2001 {\em The Astrophysical Journal\/} {\bf 560} 566

\bibitem{budavari:2005ultraviolet}
Budav{\'a}ri T, Szalay A~S, Charlot S, Seibert M, Wyder T~K, Arnouts S, Barlow
  T~A, Bianchi L, Byun Y~I, Donas J {\em et~al.\/} 2005 {\em The Astrophysical
  Journal Letters\/} {\bf 619} L31

\bibitem{tresse:2007cosmic}
Tresse L, Ilbert O, Zucca E, Zamorani G, Bardelli S, Arnouts S, Paltani S,
  Pozzetti L, Bottini D, Garilli B {\em et~al.\/} 2007 {\em Astronomy \&
  Astrophysics\/} {\bf 472} 403--419

\bibitem{sawicki:2006keck}
Sawicki M and Thompson D 2006 {\em The Astrophysical Journal\/} {\bf 648} 299

\bibitem{dahlen:2007evolution}
Dahlen T, Mobasher B, Dickinson M, Ferguson H~C, Giavalisco M, Kretchmer C and
  Ravindranath S 2007 {\em The Astrophysical Journal\/} {\bf 654} 172

\bibitem{smith:2009luminosity}
Smith A~J, Loveday J and Cross N~J 2009 {\em Monthly Notices of the Royal
  Astronomical Society\/} {\bf 397} 868--882

\bibitem{magnelli:20090}
Magnelli B, Elbaz D, Chary R, Dickinson M, Le~Borgne D, Frayer D and Willmer C
  2009 {\em Astronomy \& Astrophysics\/} {\bf 496} 57--75

\bibitem{takeuchi:2006iso}
Takeuchi T~T, Ishii T~T, Dole H, Dennefeld M, Lagache G and Puget J~L 2006 {\em
  Astronomy \& Astrophysics\/} {\bf 448} 525--534

\bibitem{babbedge:2006luminosity}
Babbedge T~S, Rowan-Robinson M, Vaccari M, Surace J, Lonsdale C, Clements D,
  Fang F, Farrah D, Franceschini A, Gonzalez-Solares E {\em et~al.\/} 2006 {\em
  Monthly Notices of the Royal Astronomical Society\/} {\bf 370} 1159--1180

\bibitem{huang:2007local}
Huang J~S, Ashby M, Barmby P, Brodwin M, Brown M, Caldwell N, Cool R,
  Eisenhardt P, Eisenstein D, Fazio G {\em et~al.\/} 2007 {\em The
  Astrophysical Journal\/} {\bf 664} 840

\bibitem{le:2005infrared}
Le~Floc’h E, Papovich C, Dole H, Bell E~F, Lagache G, Rieke G~H, Egami E,
  P{\'e}rez-Gonz{\'a}lez P~G, Alonso-Herrero A, Rieke M~J {\em et~al.\/} 2005
  {\em The Astrophysical Journal\/} {\bf 632} 169

\bibitem{flores:199915}
Flores H, Hammer F, Thuan T, C{\'e}sarsky C, Desert F, Omont A, Lilly S, Eales
  S, Crampton D and Le~Fevre O 1999 {\em The Astrophysical Journal\/} {\bf 517}
  148

\bibitem{cirasuolo:2007evolution}
Cirasuolo M, McLure R, Dunlop J, Almaini O, Foucaud S, Smail I, Sekiguchi K,
  Simpson C, Eales S, Dye S {\em et~al.\/} 2007 {\em Monthly Notices of the
  Royal Astronomical Society\/} {\bf 380} 585--595

\bibitem{faucher:2008flat}
Faucher-Giguere C~A, Lidz A, Hernquist L and Zaldarriaga M 2008 {\em The
  Astrophysical Journal Letters\/} {\bf 682} L9

\bibitem{reddy:2008multiwavelength}
Reddy N~A, Steidel C~C, Pettini M, Adelberger K~L, Shapley A~E, Erb D~K and
  Dickinson M 2008 {\em The Astrophysical Journal Supplement Series\/} {\bf
  175} 48

\bibitem{caputi:2007infrared}
Caputi K~I, Lagache G, Yan L, Dole H, Bavouzet N, Le~Floc’h E, Choi P, Helou
  G and Reddy N 2007 {\em The Astrophysical Journal\/} {\bf 660} 97

\bibitem{perez:2005spitzer}
P{\'e}rez-Gonz{\'a}lez P~G, Rieke G~H, Egami E, Alonso-Herrero A, Dole H,
  Papovich C, Blaylock M, Jones J, Rieke M, Rigby J {\em et~al.\/} 2005 {\em
  The Astrophysical Journal\/} {\bf 630} 82

\bibitem{Driver:2008sd}
Driver S~P, Popescu C~C, Tuffs R~J, Graham A~W, Liske J and Baldry I 2008 {\em
  Astrophys. J.\/} {\bf 678} L101 (\textit{Preprint} \eprint{0803.4164})

\bibitem{wilks1938}
Wilks S~S 1938 {\em Ann. Math. Statist.\/} {\bf 9} 60--62
  \urlprefix\url{https://doi.org/10.1214/aoms/1177732360}

\bibitem{Scully:2014wpa}
Scully S~T, Malkan M~A and Stecker F~W 2014 {\em Astrophys. J.\/} {\bf 784} 138
  (\textit{Preprint} \eprint{1401.4435})

\bibitem{Stecker:2016fsg}
Stecker F~W, Scully S~T and Malkan M~A 2016 {\em Astrophys. J.\/} {\bf 827} 6
  [Erratum: Astrophys. J.863,no.1,112(2018)] (\textit{Preprint}
  \eprint{1605.01382})

\bibitem{2008ICRC:tevcat}
{Wakely} S~P and {Horan} D 2008 {\em International Cosmic Ray Conference\/}
  {\bf 3} 1341--1344

\bibitem{Dominguez:2013lfa}
Domínguez A, Finke J~D, Prada F, Primack J~R, Kitaura F~S, Siana B and Paneque
  D 2013 {\em Astrophys. J.\/} {\bf 770} 77 (\textit{Preprint}
  \eprint{1305.2162})

\bibitem{aharonian2007new}
Aharonian F, Akhperjanian A, De~Almeida U~B, Bazer-Bachi A, Behera B, Beilicke
  M, Benbow W, Bernl{\"o}hr K, Boisson C, Bolz O {\em et~al.\/} 2007 {\em
  Astronomy \& Astrophysics\/} {\bf 475} L9--L13

\bibitem{aliu2014three}
Aliu E, Archambault S, Arlen T, Aune T, Behera B, Beilicke M, Benbow W, Berger
  K, Bird R, Bouvier A {\em et~al.\/} 2014 {\em The Astrophysical Journal\/}
  {\bf 782} 13

\bibitem{aharonian2007discovery}
Aharonian F, Akhperjanian A, De~Almeida U~B, Bazer-Bachi A, Behera B, Beilicke
  M, Benbow W, Bernl{\"o}hr K, Boisson C, Bolz O {\em et~al.\/} 2007 {\em
  Astronomy \& Astrophysics\/} {\bf 473} L25--L28

\bibitem{abramowski2012discovery_1es0414}
Abramowski A, Acero F, Aharonian F, Akhperjanian A, Anton G, Balzer A, Barnacka
  A, De~Almeida U~B, Becherini Y, Becker J {\em et~al.\/} 2012 {\em Astronomy
  \& Astrophysics\/} {\bf 538} A103

\bibitem{aliu2012multiwavelength}
Aliu E, Archambault S, Arlen T, Aune T, Beilicke M, Benbow W, B{\"o}ttcher M,
  Bouvier A, Bugaev V, Cannon A {\em et~al.\/} 2012 {\em The Astrophysical
  Journal\/} {\bf 755} 118

\bibitem{aleksic2015magic}
Aleksi{\'c} J, Ansoldi S, Antonelli L, Antoranz P, Babic A, Bangale P, Barrio
  J, Becerra~Gonz{\'a}lez J, Bednarek W, Bernardini E {\em et~al.\/} 2015 {\em
  Monthly Notices of the Royal Astronomical Society\/} {\bf 451} 739--750

\bibitem{acciari2008discovery}
Acciari V, Aliu E, Arlen T, Bautista M, Beilicke M, Benbow W, B{\"o}ttcher M,
  Bradbury S, Buckley J, Bugaev V {\em et~al.\/} 2008 {\em The Astrophysical
  Journal Letters\/} {\bf 690} L126

\bibitem{albert2007discovery}
Albert J, Aliu E, Anderhub H, Antoranz P, Armada A, Baixeras C, Barrio J,
  Bartko H, Bastieri D, Becker J {\em et~al.\/} 2007 {\em The Astrophysical
  Journal Letters\/} {\bf 667} L21

\bibitem{aharonian2006low}
Aharonian F, Akhperjanian A, Bazer-Bachi A, Beilicke M, Benbow W, Berge D,
  Bernl{\"o}hr K, Boisson C, Bolz O, Borrel V {\em et~al.\/} 2006 {\em
  Nature\/} {\bf 440} 1018

\bibitem{aliu2013long}
Aliu E, Archambault S, Arlen T, Aune T, Beilicke M, Benbow W, Bird R, Bouvier
  A, Buckley J, Bugaev V {\em et~al.\/} 2013 {\em The Astrophysical Journal\/}
  {\bf 779} 92

\bibitem{acciari2010discovery_1es1218}
Acciari V, Aliu E, Beilicke M, Benbow W, Boltuch D, B{\"o}ttcher M, Bradbury S,
  Bugaev V, Byrum K, Cesarini A {\em et~al.\/} 2010 {\em The Astrophysical
  Journal Letters\/} {\bf 709} L163

\bibitem{acciari2009veritas}
Acciari V, Aliu E, Arlen T, Beilicke M, Benbow W, Bradbury S, Buckley J, Bugaev
  V, Butt Y, Byrum K {\em et~al.\/} 2009 {\em The Astrophysical Journal\/} {\bf
  695} 1370

\bibitem{albert2006discovery_1es1218}
Albert J, Aliu E, Anderhub H, Antoranz P, Armada A, Asensio M, Baixeras C,
  Barrio J, Bartelt M, Bartko H {\em et~al.\/} 2006 {\em arXiv preprint
  astro-ph/0603529\/}

\bibitem{hess2013hess}
Collaboration H, Abramowski A, Acero F, Aharonian F, Akhperjanian A,
  Ang{\"u}ner E, Anton G, Balenderan S, Balzer A, Barnacka A {\em et~al.\/}
  2013 {\em Monthly Notices of the Royal Astronomical Society\/} {\bf 434}
  1889--1901

\bibitem{archambault2015veritas}
Archambault S, Archer A, Beilicke M, Benbow W, Bird R, Biteau J, Bouvier A,
  Bugaev V, Cardenzana J, Cerruti M {\em et~al.\/} 2015 {\em The Astrophysical
  Journal\/} {\bf 808} 110

\bibitem{abeysekara2016veritas}
Abeysekara A, Archambault S, Archer A, Benbow W, Bird R, Biteau J, Buchovecky
  M, Buckley J, Bugaev V, Byrum K {\em et~al.\/} 2016 {\em Monthly Notices of
  the Royal Astronomical Society\/} {\bf 459} 2550--2557

\bibitem{aliu2013multiwavelength}
Aliu E, Archambault S, Arlen T, Aune T, Beilicke M, Benbow W, Bird R,
  B{\"o}ttcher M, Bouvier A, Bugaev V {\em et~al.\/} 2013 {\em The
  Astrophysical Journal\/} {\bf 775} 3

\bibitem{tagliaferri2008simultaneous}
Tagliaferri G, Foschini L, Ghisellini G, Maraschi L, Tosti G, Albert J, Aliu E,
  Anderhub H, Antoranz P, Baixeras C {\em et~al.\/} 2008 {\em The Astrophysical
  Journal\/} {\bf 679} 1029

\bibitem{acciari2011multiwavelength}
Acciari V, Aliu E, Arlen T, Aune T, Beilicke M, Benbow W, Boltuch D, Bugaev V,
  Cannon A, Ciupik L {\em et~al.\/} 2011 {\em The Astrophysical Journal\/} {\bf
  738} 169

\bibitem{aleksic2013simultaneous}
Aleksi{\'c} J, Antonelli L~A, Antoranz P, Asensio M, Backes M, De~Almeida U~B,
  Barrio J~A, Bednarek W, Berger K, Bernardini E {\em et~al.\/} 2013 {\em
  Astronomy \& Astrophysics\/} {\bf 556} A67

\bibitem{albert2007observation}
Albert J, Aliu E, Anderhub H, Antoranz P, Armada A, Baixeras C, Barrio J,
  Bartko H, Bastieri D, Becker J {\em et~al.\/} 2007 {\em The Astrophysical
  Journal\/} {\bf 662} 892

\bibitem{abramowski2012discovery_1rxsj101015}
Abramowski A, Acero F, Aharonian F, Akhperjanian A, Anton G, Balzer A, Barnacka
  A, Becherini Y, Becker J, Bernl{\"o}hr K {\em et~al.\/} 2012 {\em Astronomy
  \& Astrophysics\/} {\bf 542} A94

\bibitem{albert2008very_3c279}
Albert J, Aliu E, Anderhub H, Antonelli L, Antoranz P, Backes M, Baixeras C,
  Barrio J, Bartko H, Bastieri D {\em et~al.\/} 2008 {\em Science\/} {\bf 320}
  1752--1754

\bibitem{abdo2010multi}
Abdo A, Ackermann M, Ajello M, Baldini L, Ballet J, Barbiellini G, Bastieri D,
  Bechtol K, Bellazzini R, Berenji B {\em et~al.\/} 2010 {\em The Astrophysical
  Journal\/} {\bf 726} 43

\bibitem{aleksic2011magic}
Aleksi{\'c} J, Antonelli L, Antoranz P, Backes M, Barrio J, Bastieri D,
  Gonz{\'a}lez J~B, Bednarek W, Berdyugin A, Berger K {\em et~al.\/} 2011 {\em
  The Astrophysical Journal Letters\/} {\bf 730} L8

\bibitem{abramowski2015high}
Abramowski A, Aharonian F, Benkhali F~A, Akhperjanian A, Ang{\"u}ner E, Anton
  G, Backes M, Balenderan S, Balzer A, Barnacka A {\em et~al.\/} 2015 {\em
  Astronomy \& Astrophysics\/} {\bf 573} A31

\bibitem{arlen2012rapid}
Arlen T, Aune T, Beilicke M, Benbow W, Bouvier A, Buckley J, Bugaev V, Cesarini
  A, Ciupik L, Connolly M {\em et~al.\/} 2012 {\em The Astrophysical Journal\/}
  {\bf 762} 92

\bibitem{aharonian2009discovery}
Aharonian F, Akhperjanian A, Anton G, De~Almeida U~B, Bazer-Bachi A, Becherini
  Y, Behera B, Benbow W, Bernl{\"o}hr K, Boisson C {\em et~al.\/} 2009 {\em The
  Astrophysical Journal Letters\/} {\bf 695} L40

\bibitem{aharonian2003observations}
Aharonian F, Akhperjanian A, Beilicke M, Bernl{\"o}hr K, B{\"o}rst H~G, Bojahr
  H, Bolz O, Coarasa T, Contreras J, Cortina J {\em et~al.\/} 2003 {\em
  Astronomy \& Astrophysics\/} {\bf 403} 523--528

\bibitem{abramowski2010multi}
Abramowski A, Acero F, Aharonian F, Akhperjanian A, Anton G, De~Almeida U~B,
  Bazer-Bachi A, Becherini Y, Behera B, Benbow W {\em et~al.\/} 2010 {\em
  Astronomy \& Astrophysics\/} {\bf 516} A56

\bibitem{aleksic2014black}
Aleksi{\'c} J, Ansoldi S, Antonelli L, Antoranz P, Babic A, Bangale P, Barrio
  J, Gonz{\'a}lez J~B, Bednarek W, Bernardini E {\em et~al.\/} 2014 {\em
  Science\/} {\bf 346} 1080--1084

\bibitem{aleksic2014rapid}
Aleksi{\'c} J, Antonelli L, Antoranz P, Babic A, de~Almeida U~B, Barrio J,
  Gonz{\'a}lez J~B, Bednarek W, Berger K, Bernardini E {\em et~al.\/} 2014 {\em
  Astronomy \& Astrophysics\/} {\bf 563} A91

\bibitem{aharonian2006fast}
Aharonian F, Akhperjanian A, Bazer-Bachi A, Beilicke M, Benbow W, Berge D,
  Bernl{\"o}hr K, Boisson C, Bolz O, Borrel V {\em et~al.\/} 2006 {\em
  Science\/} {\bf 314} 1424--1427

\bibitem{aleksic2012magic}
Aleksi{\'c} J, Alvarez E, Antonelli L, Antoranz P, Asensio M, Backes M, Barrio
  J, Bastieri D, Gonz{\'a}lez J~B, Bednarek W {\em et~al.\/} 2012 {\em
  Astronomy \& Astrophysics\/} {\bf 544} A96

\bibitem{albert2008very_m87}
Albert J, Aliu E, Anderhub H, Antonelli L, Antoranz P, Backes M, Baixeras C,
  Barrio J, Bartko H, Bastieri D {\em et~al.\/} 2008 {\em The Astrophysical
  Journal Letters\/} {\bf 685} L23

\bibitem{acciari2008observation}
Acciari V, Beilicke M, Blaylock G, Bradbury S, Buckley J, Bugaev V, Butt Y,
  Celik O, Cesarini A, Ciupik L {\em et~al.\/} 2008 {\em The Astrophysical
  Journal\/} {\bf 679} 397

\bibitem{albert2006discovery_mkn180}
Albert J, Aliu E, Anderhub H, Antoranz P, Armada A, Asensio M, Baixeras C,
  Barrio J, Bartko H, Bastieri D {\em et~al.\/} 2006 {\em The Astrophysical
  Journal Letters\/} {\bf 648} L105

\bibitem{albert2007observations}
Albert J, Aliu E, Anderhub H, Antoranz P, Armada A, Asensio M, Baixeras C,
  Barrio J, Bartko H, Bastieri D {\em et~al.\/} 2007 {\em The Astrophysical
  Journal\/} {\bf 663} 125

\bibitem{acciari2009simultaneous}
Acciari V, Aliu E, Aune T, Beilicke M, Benbow W, B{\"o}ttcher M, Bradbury S,
  Buckley J, Bugaev V, Butt Y {\em et~al.\/} 2009 {\em The Astrophysical
  Journal\/} {\bf 703} 169

\bibitem{acciari2011tev}
Acciari V, Aliu E, Arlen T, Aune T, Beilicke M, Benbow W, Boltuch D, Bradbury
  S, Buckley J, Bugaev V {\em et~al.\/} 2011 {\em The Astrophysical Journal\/}
  {\bf 738} 25

\bibitem{acciari2011spectral}
Acciari V, Arlen T, Aune T, Beilicke M, Benbow W, B{\"o}ttcher M, Boltuch D,
  Bradbury S, Buckley J, Bugaev V {\em et~al.\/} 2011 {\em The Astrophysical
  Journal\/} {\bf 729} 2

\bibitem{ahnen2016deep}
Ahnen M~L, Ansoldi S, Antonelli L, Antoranz P, Babic A, Banerjee B, Bangale P,
  De~Almeida U~B, Barrio J, Gonz{\'a}lez J~B {\em et~al.\/} 2016 {\em Astronomy
  \& Astrophysics\/} {\bf 589} A33

\bibitem{aliu2015veritas}
Aliu E, Archer A, Aune T, Barnacka A, Behera B, Beilicke M, Benbow W, Berger K,
  Bird R, Buckley J {\em et~al.\/} 2015 {\em The Astrophysical Journal\/} {\bf
  799} 7

\bibitem{aleksic2010simultaneous}
Aleksi{\'c} J, Anderhub H, Antonelli L, Antoranz P, Backes M, Baixeras C,
  Balestra S, Barrio J, Bastieri D, Gonz{\'a}lez J~B {\em et~al.\/} 2010 {\em
  Astronomy \& Astrophysics\/} {\bf 515} A76

\bibitem{albert2009magic}
Albert J, Aliu E, Anderhub H, Antoranz P, Baixeras C, Barrio J, Bartko H,
  Bastieri D, Becker J, Bednarek W {\em et~al.\/} 2009 {\em Astronomy \&
  Astrophysics\/} {\bf 493} 467--469

\bibitem{abdalla2017gamma}
Abdalla H, Abramowski A, Aharonian F, Benkhali F~A, Akhperjanian A, Andersson
  T, Ang{\"u}ner E, Arrieta M, Aubert P, Backes M {\em et~al.\/} 2017 {\em
  Astronomy \& Astrophysics\/} {\bf 600} A89

\bibitem{abramowski20152012}
Abramowski A, Aharonian F, Benkhali F~A, Akhperjanian A, Ang{\"u}ner E, Backes
  M, Balenderan S, Balzer A, Barnacka A, Becherini Y {\em et~al.\/} 2015 {\em
  The Astrophysical Journal\/} {\bf 802} 65

\bibitem{abramowski2013discovery}
Abramowski A, Acero F, Aharonian F, Benkhali F~A, Akhperjanian A, Ang{\"u}ner
  E, Anton G, Balenderan S, Balzer A, Barnacka A {\em et~al.\/} 2013 {\em
  Astronomy \& Astrophysics\/} {\bf 559} A136

\bibitem{abramowski2013discoveryPKS0447}
Abramowski A, Acero F, Akhperjanian A, Anton G, Balenderan S, Balzer A,
  Barnacka A, Becherini Y, Tjus J~B, Behera B {\em et~al.\/} 2013 {\em
  Astronomy \& Astrophysics\/} {\bf 552} A118

\bibitem{ahnen2015very}
Ahnen M~L, Ansoldi S, Antonelli L, Antoranz P, Babic A, Banerjee B, Bangale P,
  De~Almeida U~B, Barrio J, Bednarek W {\em et~al.\/} 2015 {\em The
  Astrophysical Journal Letters\/} {\bf 815} L23

\bibitem{abramowski2013hess}
Abramowski A, Acero F, Aharonian F, Akhperjanian A, Anton G, Balenderan S,
  Balzer A, Barnacka A, Becherini Y, Tjus J~B {\em et~al.\/} 2013 {\em
  Astronomy \& Astrophysics\/} {\bf 554} A107

\bibitem{ahnen2017multiwavelength}
Ahnen M~L, Ansoldi S, Antonelli L, Arcaro C, Babi{\'c} A, Banerjee B, Bangale
  P, De~Almeida U~B, Barrio J, Bednarek W {\em et~al.\/} 2017 {\em Astronomy \&
  Astrophysics\/} {\bf 603} A29

\bibitem{acero2010pks}
Acero F, Aharonian F, Akhperjanian A, Anton G, De~Almeida U~B, Bazer-Bachi A,
  Becherini Y, Behera B, Benbow W, Bernl{\"o}hr K {\em et~al.\/} 2010 {\em
  Astronomy \& Astrophysics\/} {\bf 511} A52

\bibitem{abramowski2013constraints}
Abramowski A, Acero F, Aharonian F, Benkhali F~A, Akhperjanian A, Ang{\"u}ner
  E, Anton G, Balenderan S, Balzer A, Barnacka A {\em et~al.\/} 2013 {\em
  Physical Review D\/} {\bf 88} 102003

\bibitem{abramowski2010vhe}
Abramowski A, Acero F, Aharonian F, Akhperjanian A, Anton G, De~Almeida U~B,
  Bazer-Bachi A, Becherini Y, Behera B, Benbow W {\em et~al.\/} 2010 {\em
  Astronomy \& Astrophysics\/} {\bf 520} A83

\bibitem{aleksic2012high}
Aleksi{\'c} J, Alvarez E, Antonelli L, Antoranz P, Asensio M, Backes M,
  de~Almeida U~B, Barrio J, Bastieri D, Gonz{\'a}lez J~B {\em et~al.\/} 2012
  {\em Astronomy \& Astrophysics\/} {\bf 544} A75

\bibitem{aliu2012discovery}
Aliu E, Archambault S, Arlen T, Aune T, Beilicke M, Benbow W, B{\"o}ttcher M,
  Bouvier A, Bradbury S, Buckley J {\em et~al.\/} 2012 {\em The Astrophysical
  Journal\/} {\bf 750} 94

\bibitem{aharonian2008discovery}
Aharonian F, Akhperjanian A, de~Almeida U~B, Bazer-Bachi A, Behera B, Beilicke
  M, Benbow W, Bernl{\"o}hr K, Boisson C, Borrel V {\em et~al.\/} 2008 {\em
  Astronomy \& Astrophysics\/} {\bf 481} L103--L107

\bibitem{acciari2010discovery_rgbj0710}
Acciari V, Aliu E, Arlen T, Aune T, Bautista M, Beilicke M, Benbow W,
  B{\"o}ttcher M, Boltuch D, Bradbury S {\em et~al.\/} 2010 {\em The
  Astrophysical Journal Letters\/} {\bf 715} L49

\bibitem{aliu2011multiwavelength}
Aliu E, Aune T, Beilicke M, Benbow W, B{\"o}ttcher M, Bouvier A, Bradbury S,
  Buckley J, Bugaev V, Cannon A {\em et~al.\/} 2011 {\em The Astrophysical
  Journal\/} {\bf 742} 127

\bibitem{ahnen2016detection}
Ahnen M~L, Ansoldi S, Antonelli L, Antoranz P, Arcaro C, Babic A, Banerjee B,
  Bangale P, de~Almeida U~B, Barrio J {\em et~al.\/} 2016 {\em Astronomy \&
  Astrophysics\/} {\bf 595} A98

\bibitem{archambault2013discovery}
Archambault S, Arlen T, Aune T, Behera B, Beilicke M, Benbow W, Bird R, Bouvier
  A, Buckley J, Bugaev V {\em et~al.\/} 2013 {\em The Astrophysical Journal\/}
  {\bf 776} 69

\end{thebibliography}

\end{document}